\documentclass[11pt,titlepage,a4paper]{article}
\usepackage{bozhomac}  
\usepackage{bozhlogo}  
\usepackage{cite}	
\usepackage{varioref}	

\title{\bfseries	\vspace*{-1.7in}
{\huge Lagrangian quantum field theory\\[1ex] in momentum picture}
 \\[1.3ex]
{\LARGE III.\ Free vector fields}
}

\vspace{1.7ex}

\author{
Bozhidar Z.\ Iliev
\thanks{Laboratory of Mathematical Modeling in Physics,
Institute for Nuclear Research and \mbox{Nuclear} Energy,
Bulgarian Academy of Sciences,
Boul.\ Tzarigradsko chauss\'ee~72, 1784 Sofia, Bulgaria}
\thanks{E-mail address: bozho@inrne.bas.bg}
\thanks{URL: http://theo.inrne.bas.bg/$\sim$bozho/}
}

\date{	
 \vspace{2.27ex}\ShortTitle{QFT in momentum picture: III}\\[0.27ex]
 \vspace{3.27ex}
\small
	\begin{tabular}{r@{$\colon\to~$}l}
 \vspace{0.09ex} Last update	& May 1, 2005	\\[0.09ex]
 \vspace{0.27ex} Produced	& \fbox{\today}	\\[0.27ex]
	\end{tabular} \\[1.27ex]
\normalsize
	\begin{tabular}{r@{$\colon~$}l}
 \vspace{0.27ex} http://www.arXiv.org e-Print archive No. & hep-th/0505007
	\end{tabular} \\[-0.27ex]
 \vspace{4.27ex}{\Huge\BOZHO}	\\[4.27ex]
 \vspace{0.27ex}\Subject{Quantum field theory}
								\\[2.27ex]
	\begin{tabular}{r@{\hspace{0.512em}}|@{\hspace{0.512em}}l}
 \vspace{0.27ex}\MSC[2000]{81Q99, 81T99\\\hspace{0pt}}
&
 \vspace{0.27ex}\PACS[2003]{03.70.+k, 11.10.Ef, 11.10.-z,\\
				11.90.+t, 12.90.+b}
	\end{tabular} \\[1.27ex]
 \vspace{0.27ex}\KeyWords{Quantum field theory, Pictures of motion\\
Pictures of motion in quantum field theory, Momentum picture\\
Free vector field, Free neutral vector field, Free charged vector field\\
Free massive vector fields, Free massless vector fields satisfying the
Lorenz condition\\
Free electromagnetic field, Equations of motion for free vector field\\
Klein-Gordon equation, Proca equation, Proca equation in momentum picture\\
Maxwell equations, Maxwell-Lorentz equations,
Maxwell equations in momentum picture\\
Lorenz condition, Lorenz gauge, Coulomb gauge\\
Commutation relations for free vector field,
State vectors of free vector field}\\[0.27ex]
}

\newcommand{\U}{{\ope{U}}}  		
\newcommand{\tU}{{\tope{U}}}		
\newcommand{\tu}{{\Tilde{u}}}		
\newcommand{\ta}{{\Tilde{a}}}		
\newcommand{\bk}{\boldsymbol{k}}  	

 \newcommand{\Hil}{\mathcal{F}}		
	\newcommand{\base}{\mathit{M}}	

\newcommand{\ope}[2][{}]{\lindex[\mathcal{#2}]{}{#1}} 
\newcommand{\tope}[2][{}]{\ope[#1]{\Tilde{#2}}} 
\newcommand{\ocross}{\overset{\circ}{\times}}

\begin{document}		

\renewcommand{\thepage}{\roman{page}}

\renewcommand{\thefootnote}{\fnsymbol{footnote}} 
\maketitle				
\renewcommand{\thefootnote}{\arabic{footnote}}   

\tableofcontents		


\begin{abstract}

	Free vector fields, satisfying the Lorenz condition, are
investigated in details in the momentum picture of motion in Lagrangian
quantum field theory. The field equations are equivalently written in terms
of creation and annihilation operators and on their base the commutation
relations are derived. Some problems concerning the vacuum and state vectors
of free vector field are discussed. Special attention is paid to
peculiarities of the massless case; in particular, the electromagnetic field
is explored. Several Lagrangians, describing free vector fields, are
considered and the basic consequences of them are pointed and compared.

\end{abstract}

\renewcommand{\thepage}{\arabic{page}}


\section {Introduction}
\label{Introduction}

	This paper is devoted to an exploration of two types of free vector
fields in the momentum picture of Lagrangian quantum field theory%
\footnote{~%
In this paper we considered only the Lagrangian (canonical) quantum field
theory in which the quantum fields are represented as operators, called field
operators, acting on some Hilbert space, which in general is unknown if
interacting fields are studied. These operators are supposed to satisfy some
equations of motion, from them are constructed conserved quantities
satisfying conservation laws, etc. From the view\ndash point of present\ndash
day quantum field theory, this approach is only a preliminary stage for more
or less rigorous formulation of the theory in which the fields are
represented via operator\ndash valued distributions, a fact required even for
description of free fields. Moreover, in non\ndash perturbative directions,
like constructive and conformal field theories, the main objects are the
vacuum mean (expectation) values of the fields and from these are
reconstructed the Hilbert space of states and the acting on it fields.
Regardless of these facts, the Lagrangian (canonical) quantum field theory is
an inherent component of the most of the ways of presentation of quantum
field theory adopted explicitly or implicitly in books
like~\cite{Bogolyubov&Shirkov,Bjorken&Drell,Roman-QFT,Ryder-QFT,Schweber,
Akhiezer&Berestetskii,Ramond-FT,Bogolyubov&et_al.-AxQFT,Bogolyubov&et_al.-QFT}.
Besides, the Lagrangian approach is a source of many ideas for other
directions of research, like the axiomatic quantum field
theory~\cite{Roman-QFT,Bogolyubov&et_al.-AxQFT,Bogolyubov&et_al.-QFT}.%
}:
massive vector fields and massless vector fields, the latter satisfying the
Lorenz%
\footnote{~%
The Lorenz condition and gauge (see below equation~\eref{3.6b})
are named in honor of the Danish theoretical physicist
Ludwig Valentin Lorenz (1829--1891),
who has first published it in 1867~\cite{Lorenz/1867}
(see also~\cite[pp.\ 268-269, 291]{Whittaker-History});
however this condition was first introduced in lectures by Bernhard
G.~W.~Riemann in 1861 as pointed in~\cite[p.~291]{Whittaker-History}. It
should be noted that the \emph{Lorenz} condition/gauge is quite often
erroneously referred to as the Loren\emph{t}z condition/gauge after the name of
the Dutch theoretical physicist Hendrik Antoon Lorentz (1853--1928) as, e.g.,
in~\cite[p.~18]{Roman-QFT} and in~\cite[p.~45]{Gockeler&Schucker}.%
}
condition as addition to the Lagrangian formalism. Since the massive
free vector fields satisfy the Lorenz condition as a consequence of the
Euler\ndash Lagrange equations, both kinds of fields are treated on almost
equal footing in the present work. However, the massless case has its own
peculiarities to which special attention is paid. Most of the known results,
concerning the mentioned fields in Heisenberg picture, are reproduced in
momentum picture of motion. New results are obtained too. For example, the
field equations in terms of creation and annihilation operators and the
(second) quantization of electromagnetic field in Lorenz gauge (imposed on
the fields potentials as operators), a special case of which is the
quantization in Coulomb gauge.

	The work may be regarded as a continuation
of~\cite{bp-QFTinMP-scalars,bp-QFTinMP-spinors}, where free scalar and
spinor, respectively, fields are studied in momentum picture.

	The basic moments of the method, we will follow in this work, are the
following ones:
\\\indent
	(i) In Heisenberg picture is fixed a (second) non\ndash quantized and
non\ndash normally ordered  operator\ndash valued Lagrangian, which is
supposed to be polynomial (or convergent power series) in the field operators
and their first partial derivatives;
\\\indent
	(ii) As conditions additional to the Lagrangian formalism are
postulated the commutativity between the components of the momentum operator
(see~\eref{2.1} below) and the Heisenberg relations between the field
operators and momentum operator (see~\eref{2.28} below);
\\\indent
	(iii) Following the Lagrangian formalism in momentum picture, the
creation and annihilation operators are introduced and the dynamical
variables and field equations are written in their terms;
\\\indent
	(iv) From the last equations, by imposing some additional restrictions
on the creation and annihilation operators, the (anti)commutation relations
for these operators are derived;
\\\indent
	(v) At last, the vacuum and normal ordering procedure are defined, by
means of which the theory can be developed to a more or less complete form.

	The main difference of the above scheme from the standard one is that
we postulate the below\ndash written relations~\eref{2.1} and~\eref{2.28}
and, then, we look for compatible with them and the field equations
(anti)commutation relations. (Recall, ordinary the (anti)commutation
relations are postulated at first and the validity of the
equations~\eref{2.1} and~\eref{2.28} is explored after
that~\cite{Bjorken&Drell-2}.)

	The organization of the material is as follows.

	In Sect.~\ref{Sect2} are reviewed, for reference purposes, the basic
aspects of momentum picture of motion of quantum field theory. The
description of free vector fields in this picture is presented in
Sect.~\ref{Sect3}.

	The structure of the solutions of the field equations is analyzed in
Sect.~\ref{Sect4}. Decompositions of these solutions, equivalent to the
Fourier decompositions in Heisenberg picture, are established. A suitably
normalized system of classical solutions of the field equations is
constructed. The creation and annihilation operators for the fields
considered are introduced in Sect.~\ref{Sect5} on a base of the
decompositions and system of classical solutions just mentioned. A physical
interpretation of these operators is derived from the Heisenberg relations,
which are external to the Lagrangian formalism. At this point, the first
problem with the massless case, concerning the angular momentum operator,
appears. In Sect.~\ref{Sect6}, the operators of the dynamical variables of
free vector fields (satisfying the Lorenz condition) are calculated in
Heisenberg picture of motion in terms of creation and annihilation operators
in momentum picture. Special attention is paid to the spin angular momentum
operator and the above mentioned problem is analyzed further.

	In Sect.~\ref{Sect7}, the field equations are equivalently rewritten
in terms of creation and annihilation operators. As a consequence of them,
the dynamical variables in momentum picture are found. It should be
mentioned, in the massless case, the creation and annihilation operators
corresponding to the degrees of freedom, `parallel' to the 4\ndash momentum
variable, do not enter in the field equations. In Sect.~\ref{Sect8}, the
commutation relations for free vector fields satisfying the Lorenz
conditions are derived. They also do not include the just\ndash mentioned
operators. The commutators between the components of spin angular momentum
operator and between them and the charge operator are calculate on the base
of the established commutation relations. It is pointed that these relations
play a role of field equations under the hypotheses they are derived. To the
normal ordering procedure and definition of vacuum is devoted
Sect.~\ref{Sect9}. Problems, regarding state vectors and physical
interpretation of creation and annihilation operators in the Lagrangian
formalism, are considered in Sect.~\ref{Sect10}.

	Some peculiarities of the massless case are explored in
Sect.~\ref{Sect11}. It is pointed that, generally, new suppositions are
required for the treatment of creation and annihilation operators, connected
with the degrees of freedom `parallel' to the 4\ndash momentum, which are the
cause for the problems arising in the massless case. Two such hypotheses are
analyzed. The obtained formalism is applied to a description of the
electromagnetic field. In fact, it provides a new quantization of this field
in which the Lorenz conditions is imposed directly on the field operators,
which is completely different with respect to the one used in Gupta\ndash
Bleuler quantization. It is shown that for an electromagnetic field no
problems arise, due to a suitable definition of the normal ordering
procedure. The basic relation of quantum field theory of free electromagnetic
field are written explicitly.

	Sect.~\ref{Sect12} contain a discussion of some Lagrangians suitable
for description of free vector fields satisfying the Lorenz condition. The
basic consequences of these Lagrangians are pointed and compared. As a `best'
Lagrangian is pointed the one which is charge\ndash symmetric and, hence, in
which the spin\ndash statistics theorem is encoded. It is proved that the
quantum field theories, arising from the considered Lagrangians, became
identical after the normal ordering procedure is applied.

	In Sect.~\ref{Sect13} is analyzed the role of the Lorenz condition,
when studying massless free vector fields. This is done by investigating a
massless vector field with a Lagrangian equal to the one of a massive vector
field with vanishing mass and without imposing the Lorenz condition as a
subsidiary condition on the field operators. Sect.~\ref{Conclusion} closes
the paper.

\vspace{1.2ex}

	The books~\cite{Bogolyubov&Shirkov,Roman-QFT,Bjorken&Drell} will be
used as standard reference works on quantum field theory. Of course, this is
more or less a random selection between the great number of (text)books and
papers on the theme to which the reader is referred for more details or other
points of view. For this end, e.g.,~\cite{Itzykson&Zuber,Ryder-QFT,Schweber} or
the literature cited
in~\cite{Bogolyubov&Shirkov,Roman-QFT,Bjorken&Drell,Itzykson&Zuber,
Ryder-QFT,Schweber} may be helpful.

	Throughout this paper $\hbar$ denotes the Planck's constant (divided
by $2\pi$), $c$ is the velocity of light in vacuum, and $\iu$ stands for the
imaginary unit. The superscripts $\dag$ and $\top$ means respectively
Hermitian conjugation and transposition of operators or matrices, the
superscript $\ast$ denotes complex conjugation, and the symbol $\circ$
denotes compositions of mappings/operators.

	By $\delta_{fg}$, or $\delta_f^g$ or $\delta^{fg}$  ($:=1$ for $f=g$,
$:=0$ for $f=g$) is denoted the Kronecker $\delta$\ndash symbol, depending on
arguments $f$ and $g$, and $\delta^n(y)$, $y\in\field[R]^n$, stands for the
$n$\ndash dimensional Dirac $\delta$\ndash function; $\delta(y):=\delta^1(y)$
for $y\in\field[R]$.

	The Minkowski spacetime is denoted by $\base$. The Greek indices run
from 0 to $\dim\base-1=3$. All Greek indices will be raised and lowered by
means of the standard 4\ndash dimensional Lorentz metric tensor
$\eta^{\mu\nu}$ and its inverse $\eta_{\mu\nu}$ with signature
$(+\,-\,-\,-)$. The Latin indices $a,b,\dots$ run from 1 to $\dim\base-1=3$
and, usually, label the spacial components of some object. The Einstein's
summation convention over indices repeated on different levels is assumed
over the whole range of their values.

	At the end, a technical remark is in order. The derivatives with
respect to operator\ndash valued (non\ndash commuting) arguments will be
calculated according to the rules of the classical analysis of commuting
variables, which is an everywhere silently accepted
practice~\cite{Bogolyubov&Shirkov,Bjorken&Drell-2}. As it is demonstrated
in~\cite{bp-QFT-action-principle}, this is not quite correct but does not
lead to incorrect results when free vector fields are concerned.


\section{The momentum picture}
	\label{Sect2}

	In this section, we present a summary of the momentum picture in
quantum field theory, introduce in~\cite{bp-QFT-pictures} and developed
in~\cite{bp-QFT-MP}.

	Let us consider a system of quantum fields, represented in Heisenberg
picture of motion by field operators $\tope{\varphi}_i(x)\colon\Hil\to\Hil$,
$i=1,\dots,n\in\field[N]$, acting on the system's Hilbert space $\Hil$ of
states and depending on a point $x$ in Minkowski spacetime $\base$. Here and
henceforth, all quantities in Heisenberg picture will be marked by a tilde
(wave) ``$\tope{\mspace{6mu}}\mspace{3mu}$'' over their kernel symbols. Let
$\tope{P}_\mu$ denotes the system's (canonical) momentum vectorial operator,
defined via the energy\ndash momentum tensorial operator $\tope{T}^{\mu\nu}$
of the system, viz.
	\begin{equation}
			\label{2.0}
\tope{P}_\mu
:=
\frac{1}{c}\int\limits_{x^0=\const} \tope{T}_{0\mu}(x) \Id^3\bs x .
	\end{equation}
Since this operator is Hermitian, $\tope{P}_\mu^\dag=\tope{P}_\mu$, the
operator
	\begin{equation}	\label{12.112}
\ope{U}(x,x_0)
 =
\exp\Bigl( \iih \sum_\mu (x^\mu-x_0^\mu)\tope{P}_{\mu}  \Bigr) ,
	\end{equation}
where $x_0\in\base$ is arbitrarily fixed and $x\in\base$,%
\footnote{~%
The notation $x_0$, for a fixed point in $\base$, should not be confused with
the zeroth covariant coordinate $\eta_{0\mu}x^\mu$ of $x$ which, following
the convention $x_\nu:=\eta_{\nu\mu}x^\mu$, is denoted by the same symbol
$x_0$. From the context, it will always be clear whether $x_0$ refers to a
point in $\base$ or to the zeroth covariant coordinate of a point
$x\in\base$.%
}
is unitary, \ie
\(
\ope{U}^\dag(x_0,x)
:= (\ope{U}(x,x_0))^\dag
 = \ope{U}^{-1}(x,x_0)
 = (\ope{U}(x,x_0))^{-1}
\)
and, via the formulae
	\begin{align}	\label{12.113}
\tope{X}\mapsto \ope{X}(x)
	&= \ope{U}(x,x_0) (\tope{X})
\\			\label{12.114}
\tope{A}(x)\mapsto \ope{A}(x)
	&= \ope{U}(x,x_0)\circ (\tope{A}(x)) \circ \ope{U}^{-1}(x,x_0) ,
	\end{align}
realizes the transition to the \emph{momentum picture}. Here $\tope{X}$ is a
state vector in system's Hilbert space of states $\Hil$ and
$\tope{A}(x)\colon\Hil\to\Hil$ is (observable or not) operator\ndash valued
function of $x\in\base$ which, in particular, can be polynomial or convergent
power series in the field operators $\tope{\varphi}_i(x)$; respectively
$\ope{X}(x)$ and $\ope{A}(x)$ are the corresponding quantities in momentum
picture.
	In particular, the field operators transform as
	\begin{align}	\label{12.115}
\tope{\varphi}_i(x)\mapsto \ope{\varphi}_i(x)
     = \ope{U}(x,x_0)\circ \tope{\varphi}_i(x) \circ \ope{U}^{-1}(x,x_0) .
	\end{align}
	Notice, in~\eref{12.112} the multiplier $(x^\mu-x_0^\mu)$ is regarded
as a real parameter (in which $\tope{P}_\mu$ is linear). Generally,
$\ope{X}(x)$ and $\ope{A}(x)$ depend also on the point $x_0$ and, to be
quite correct, one should write $\ope{X}(x,x_0)$ and $\ope{A}(x,x_0)$ for
$\ope{X}(x)$ and $\ope{A}(x)$, respectively. However, in the most situations
in the present work, this dependence is not essential or, in fact, is not
presented at all. For that reason, we shall \emph{not} indicate it explicitly.

	As it was said above, we consider quantum field theories in which the
components $\tope{P}_\mu$ of the momentum operator commute between themselves
and satisfy the Heisenberg relations/equations with the field operators, \ie
we suppose that $\tope{P}_\mu$ and $\tope{\varphi}_i(x)$ satisfy the
relations:
	\begin{align}	\label{2.1}
& [\tope{P}_\mu, \tope{P}_\nu ]_{\_} = 0
\\			\label{2.28}
& [\tope{\varphi}_i(x), \tope{P}_\mu]_{\_} = \ih\pd_\mu \tope{\varphi}_i(x).
	\end{align}
Here $[A,B]_{\pm}:=A\circ B \pm B\circ A$, $\circ$ being the composition of
mappings sign, is the commutator/anticommutator of operators (or matrices)
$A$ and $B$. The momentum operator $\tope{P}_\mu$ commutes with the
`evolution' operator $\ope{U}(x,x_0)$ (see below~\eref{12.118}) and its
inverse,
	\begin{equation}	\label{2.2}
	[ \tope{P}_\mu, \ope{U}(x,x_0) ]_{\_} =0
\qquad
	[ \tope{P}_\mu, \ope{U}^{-1}(x,x_0) ]_{\_} =0 ,
	\end{equation}
due to ~\eref{2.1} and~\eref{12.112}. So, the momentum operator remains
unchanged in momentum picture, \viz we have (see~\eref{12.114}
and~\eref{2.2})
	\begin{equation}	\label{2.3}
	 \ope{P}_\mu = \tope{P}_\mu.
	\end{equation}

	Since from~\eref{12.112} and~\eref{2.1} follows
	\begin{equation}	\label{12.116}
\ih \frac{\pd\ope{U}(x,x_0)}{\pd x^\mu}
=
\ope{P}_{\mu} \circ \ope{U}(x,x_0)
\qquad
\ope{U}(x_0,x_0) = \id_\Hil ,
	\end{equation}
we see that, due to~\eref{12.113}, a state vector $\ope{X}(x)$ in
momentum picture is a solution of the initial\ndash value problem
	\begin{equation}	\label{12.117}
\ih \frac{\pd\ope{X}(x)}{\pd x^\mu}
=
\ope{P}_{\mu}  (\ope{X}(x))
\qquad
\ope{X}(x)|_{x=x_0}=\ope{X}(x_0) = \tope{X}
	\end{equation}
which is a 4-dimensional analogue of a similar problem for the Schr{\"o}dinger
equation in quantum mechanics~\cite{Messiah-QM,Dirac-PQM,Prugovecki-QMinHS}.

	By virtue of~\eref{12.112}, or in view of the independence of
$\ope{P}_{\mu} $ of $x$, the solution of~\eref{12.117} is
	\begin{equation}	\label{12.118}
\ope{X}(x)
= \ope{U}(x,x_0) (\ope{X}(x_0))
= \e^{\iih(x^\mu-x_0^\mu)\ope{P}_{\mu} } (\ope{X}(x_0)).
	\end{equation}
Thus, if $\ope{X}(x_0)=\tope{X}$  is an eigenvector of
$\ope{P}_{\mu} $ ($=\tope{P}_{\mu} $)
with eigenvalues $p_\mu$,
	\begin{equation}	\label{12.119}
\ope{P}_{\mu}  (\ope{X}(x_0)) = p_\mu \ope{X}(x_0)
\quad
( =p_\mu \tope{X} = \tope{P}_{\mu}  (\tope{X}) ) ,
	\end{equation}
we have the following \emph{explicit} form of the state vectors
	\begin{equation}	\label{12.120}
\ope{X}(x)
=
\e^{ \iih(x^\mu-x_0^\mu)p_\mu } (\ope{X}(x_0)).
	\end{equation}
It should clearly be understood, \emph{this is the general form of all state
vectors} as they are eigenvectors of all (commuting)
observables~\cite[p.~59]{Roman-QFT}, in particular, of the momentum operator.

	In momentum picture, all of the field operators happen to be
constant in spacetime, \ie
	\begin{equation}	\label{2.4}
\varphi_i(x)
= \ope{U}(x,x_0)\circ \tope{\varphi}_i(x) \circ \ope{U}^{-1}(x,x_0)
= \varphi_i(x_0)
= \tope{\varphi}_i(x_0)
=: \varphi_{(0)\, i} .
	\end{equation}
Evidently, a similar result is valid for any (observable or not such)
function of the field operators which is polynomial or convergent power series
in them and/or their first partial derivatives. However, if $\tope{A}(x)$ is an
arbitrary operator or depends on the field operators in a different way, then
the corresponding to it operator $\ope{A}(x)$ according to~\eref{12.114} is,
generally, not spacetime\ndash constant and depends on the both points $x$
and $x_0$. As a rules, if $\ope{A}(x)=\ope{A}(x,x_0)$ is independent of $x$,
we, usually, write $\ope{A}$ for $\ope{A}(x,x_0)$, omitting all arguments.

	It should be noted, the Heisenberg relations~\eref{2.28} in momentum
picture transform into the identities $\pd_\mu\varphi_i=0$ meaning that the
field operators $\varphi_i$ in momentum picture are spacetime constant
operators (see~\eref{2.4}). So, in momentum picture, the Heisenberg
relations~\eref{2.28} are incorporated in the constancy of the field
operators.

	Let $\tope{L}$ be the system's Lagrangian (in Heisenberg picture). It
is supposed to be polynomial or convergent power series in the field
operators and their first partial derivatives, \ie
$\tope{L}=\tope{L}(\varphi_i(x),\pd_\nu\varphi_i(x))$ with $\pd_\nu$ denoting
the partial derivative operator relative to the $\nu^{\mathrm{th}}$
coordinate $x^\nu$. In momentum picture it transforms into
	\begin{equation}	\label{2.5}
\ope{L} =\tope{L}(\varphi_i(x), y_{j\nu})
\qquad
y_{j\nu}=\iih[\varphi_j,\ope{P}_{\nu} ]_{\_}  ,
	\end{equation}
\ie in momentum picture one has simply to replace the field operators in
Heisenberg picture with their values at a fixed point $x_0$ and the partial
derivatives  $\pd_\nu\tope{\varphi}_j(x)$ in Heisenberg picture with the
above\ndash defined quantities $y_{j\nu}$.
	The (constant) field operators $\varphi_i$ satisfy the following
\emph{algebraic Euler\ndash Lagrange equations in momentum picture}:%
\footnote{~%
In~\eref{12.129} and similar expressions appearing further, the derivatives
of functions of operators with respect to operator arguments are calculated
in the same way as if the operators were ordinary (classical)
fields/functions, only the order of the arguments should not be changed.
This is a silently accepted practice in the
literature~\cite{Roman-QFT,Bjorken&Drell}. In the most cases such a procedure
is harmless, but it leads to the problem of non\ndash unique definitions of
the quantum analogues of the classical conserved quantities, like the
energy\ndash momentum and charge operators. For some details on this range of
problems in quantum field theory, see~\cite{bp-QFT-action-principle}; in
particular, the \emph{loc.\ cit.}\ contains an example of a Lagrangian whose
field equations are \emph{not} the Euler\ndash Lagrange
equations~\eref{12.129} obtained as just described.%
}
	\begin{equation}	\label{12.129}
\Bigl\{
\frac{\pd\tope{L}(\varphi_j,y_{l\nu})} {\pd \varphi_i}
-
\iih
\Bigl[
\frac{\pd\tope{L}(\varphi_j,y_{l\nu})} {y_{i\mu}}
,
\ope{P}_{\mu}
\Bigr]_{\_}
\Bigr\}
\Big|_{ y_{j\nu}=\iih[\varphi_j,\ope{P}_{\nu} ]_{\_} }
= 0 .
	\end{equation}
Since $\ope{L}$ is supposed to be polynomial or convergent power series  in
its arguments, the equations~\eref{12.129}  are \emph{algebraic}, not
differential, ones. This result is a natural one in view of~\eref{2.4}.

	Suppose a quantum system under consideration possesses a charge (\eg
electric one) and angular momentum, described by respectively the current
operator $\tope{J}_\mu(x)$ and (total) angular momentum tensorial density
operator
	\begin{equation}	\label{2.9}
\tope{M}_{\mu\nu}^{\lambda} (x)
=
-\tope{M}_{\nu\mu}^{\lambda} (x)
=
  x_\mu \Sprindex[\tope{T}]{\nu}{\lambda}
- x_\nu \Sprindex[\tope{T}]{\mu}{\lambda}
+ \tope{S}_{\mu\nu}^{\lambda} (x)
	\end{equation}
with $x_\nu:=\eta_{\nu\mu}x^\mu$ and
 $\tope{S}_{\mu\nu}^{\lambda}(x)= - \tope{S}_{\nu\mu}^{\lambda}(x)$
being the spin angular momentum (density) operator. The (constant, time\ndash
independent) conserved quantities corresponding to them, the charge operator
$\tope{Q}$ and total angular momentum operator $\tope{M}_{\mu\nu}$,
respectively are
	\begin{gather}
			\label{2.10}
\tope{Q} := \frac{1}{c} \int\limits_{x^0=\const} \tope{J}_0(x) \Id^3\bs x
\\			\label{2.11}
\tope{M}_{\mu\nu} = \tope{L}_{\mu\nu}(x) + \tope{S}_{\mu\nu}(x) ,
	\end{gather}
where
	\begin{subequations}	\label{2.12}
      \begin{align}
      		\label{2.12a}
\tope{L}_{\mu\nu}(x)
& :=
\frac{1}{c} \int\limits_{x^0=\const}
\{
  x_\mu \Sprindex[\tope{T}]{\nu}{0}(x) - x_\nu \Sprindex[\tope{T}]{\mu}{0}(x)
\} \Id^3\bs x
\\			\label{2.12b}
\tope{S}_{\mu\nu}(x)
& := \frac{1}{c} \int\limits_{x^0=\const} \tope{S}_{\mu\nu}^0(x) \Id^3\bs x
      \end{align}
	\end{subequations}
are the orbital and spin, respectively, angular momentum operators (in
Heisenberg picture). Notice, we write $\tope{L}_{\mu\nu}(x)$ and
$\tope{S}_{\mu\nu}(x)$, but, as a result of~\eref{2.12}, these operators may
depend only on the zeroth (time) coordinate of $x\in\base$. When working in
momentum picture, in view of~\eref{12.114}, the following representations
turn to be useful:
      \begin{gather}
      		\label{2.6}
\ope{P}_\mu = \tope{P}_\mu
=
\frac{1}{c} \int\limits_{x^0=\const}
	\ope{U}^{-1}(x,x_0)\circ \ope{T}_{0\mu} \circ \ope{U}(x,x_0)
	\Id^3\bs{x}
\displaybreak[1]\\      		\label{2.13}
\tope{Q}
=
\frac{1}{c} \int\limits_{x^0=\const}
	\ope{U}^{-1}(x,x_0)\circ \ope{J}_{0} \circ \ope{U}(x,x_0)
	\Id^3\bs{x}
\displaybreak[1]\\      		\label{2.14}
\tope{L}_{\mu\nu} (x)
=
\frac{1}{c} \int\limits_{x^0=\const}
	\ope{U}^{-1}(x,x_0)\circ \{
  x_\mu \Sprindex[\ope{T}]{\nu}{0} - x_\nu \Sprindex[\ope{T}]{\mu}{0}
\} \circ \ope{U}(x,x_0)
	\Id^3\bs{x}
\displaybreak[1]\\      		\label{2.15}
\tope{S}_{\mu\nu} (x)
=
\frac{1}{c} \int\limits_{x^0=\const}
	\ope{U}^{-1}(x,x_0)\circ \ope{S}_{\mu\nu}^0 \circ \ope{U}(x,x_0)
	\Id^3\bs{x} .
      \end{gather}
These expressions will be employed essentially in the present paper.

	The conservation laws $\frac{\od\tope{Q}}{\od x^0}=0$ and
$\frac{\od\tope{M}_{\mu\nu}}{\od x^0}=0$ (or, equivalently,
 $\pd^\mu\tope{J}_\mu=0$ and $\pd_\lambda\tope{M}_{\mu\nu}^{\lambda}=0$),
can be rewritten as
	\begin{equation}	\label{2.16}
\pd_\mu\tope{Q} = 0
\qquad
\pd_\lambda\tope{M}_{\mu\nu} = 0
	\end{equation}
since~\eref{2.10}--\eref{2.12} imply
$\pd_a\tope{Q} = 0$ and $\pd_a\tope{M}_{\mu\nu} = 0$ for $a=1,2,3$.

	As a result of the skewsymmetry of the operators~\eref{2.11}
and~\eref{2.12} in the subscripts $\mu$ and $\nu$, their spacial components
form a (pseudo\ndash)vectorial operators. If $e^{abc}$, $a,b,c=1,2,3$,
denotes the 3\ndash dimensional Levi\ndash Civita (totally) antisymmetric
symbol, we put $\tope{\bs M}:=(\tope{\bs M}^1,\tope{\bs M}^2,\tope{\bs M}^3)$
with $\tope{\bs M}^a:=e^{abc}\tope{M}_{bc}$ and similarly for the orbital and
spin angular momentum operators. Then~\eref{2.11} and the below written
equation~\eref{2.25} imply
	\begin{align}	\label{2.31}
& \tope{\bs M} = \tope{\bs L}(x) + \tope{\bs S}(x)
\\			\label{2.32}
& \ope{\bs M}(x,x_0)
= \tope{\bs L}(x) + (\bs x -\bs x_0)\times \ope{\bs P} + \tope{\bs S}(x),
	\end{align}
where $\bs x:=(x^1,x^2,x^3)=-(x_1,x_2,x_3)$, $\times$ denotes the Euclidean
cross product, and
\(
\ope{\bs P}
:= (\ope{P}^1,\ope{P}^2,\ope{P}^3)=-(\ope{P}_1,\ope{P}_2,\ope{P}_3) .
\)
Obviously, the correction in~\eref{2.32} to $\tope{\bs M}$ can be interpreted
as a one due to an additional orbital angular momentum when the origin, with
respect to which it is determined, is change from $x$ to $x_0$.

	The consideration of $\tope{Q}$ and$\tope{M}_{\mu\nu}$ as generators
of constant phase transformations and 4\ndash rotations, respectively, leads
to the following
relations~\cite{Bjorken&Drell,Bogolyubov&Shirkov,Itzykson&Zuber}
	\begin{align}	\label{2.17}
& [\tope{\varphi}_i(x), \tope{Q}]_{\_}
	= \varepsilon(\tope{\varphi}_i) q_i \tope{\varphi}_i(x)
\\			\label{2.18}
& [\tope{\varphi}_i(x), \tope{M}_{\mu\nu}]_{\_}
=
\ih\{
x_\mu\pd_\nu\tope{\varphi}_i(x) - x_\nu\pd_\mu\tope{\varphi}_i(x)
+ I_{i\mu\nu}^{j} \tope{\varphi}_j(x)
\} .
	\end{align}
Here: $q_i=\const$ is the charge of the $i^\text{th}$ field,
 $q_j=q_i$ if $\tope{\varphi}_j=\tope{\varphi}_i^\dag$,
$\varepsilon(\tope{\varphi}_i) = 0$ if
		$\tope{\varphi}_i^\dag = \tope{\varphi}_i$,
$\varepsilon(\tope{\varphi}_i) = \pm 1$ if
		$\tope{\varphi}_i^\dag \not= \tope{\varphi}_i$
with
$\varepsilon(\tope{\varphi}_i) + \varepsilon(\tope{\varphi}_i^\dag) = 0$,
and the constants $I_{i\mu\nu}^{j} = -I_{i\nu\mu}^{j}$ characterize the
transformation properties of the field operators under 4\ndash rotations.
(If $\varepsilon(\tope{\varphi}_i)\not=0$, it is a convention whether to put
$\varepsilon(\tope{\varphi}_i)=+1$ or $\varepsilon(\tope{\varphi}_i)=-1$ for a
fixed $i$.)
	Besides, the operators~\eref{2.10}--\eref{2.12} are Hermitian,
	\begin{equation}	\label{2.19}
\tope{Q}^\dag = \tope{Q}, \quad
\tope{M}_{\mu\nu}^\dag = \tope{M}_{\mu\nu}, \quad
\tope{L}_{\mu\nu}^\dag = \tope{L}_{\mu\nu}, \quad
\tope{S}_{\mu\nu}^\dag = \tope{S}_{\mu\nu}, \quad
	\end{equation}
and satisfy the relations%
\footnote{~%
The author is completely aware of the fact that in the literature, for
instance in~\cite[p.~77, eq.~(2-87)]{Roman-QFT} or
in~\cite[eq.~(2.187)]{Ryder-QFT}, the relation~\eref{2.21} is written with an
opposite sign, \ie with $+\ih$ instead of $-\ih$ on its r.h.s. (In this
case~\eref{2.21} is part of the commutation relations characterizing the Lie
algebra of the Poincar\'e group --- see,
e.g.,~\cite[pp.~143--147]{Bogolyubov&et_al.-AxQFT}
or~\cite[sect.~7.1]{Bogolyubov&et_al.-QFT}.) However, such a choice of the
sign in~\eref{2.21} contradicts to the
explicit form of $\tope{P}_\mu$ and $\tope{L}_{\mu\nu}$ in terms of creation
and annihilation operators (see sections~\ref{Sect6} and~\ref{Sect7}) in the
framework of Lagrangian formalism. For  this reason and since the
relation~\eref{2.21} is external to the Lagrangian formalism, we
accept~\eref{2.21} as it is written below. In connection with~\eref{2.21} ---
see below equation~\eref{7.9}, \eref{7.10} and~\eref{7.13}.%
}

	\begin{align}	\label{2.20}
& [\tope{Q}, \tope{P}_\mu]_{\_} = 0
\\			\label{2.21}
& [\tope{M}_{\mu\nu}, \tope{P}_\lambda]_{\_}
= 
- \ih\{ \eta_{\lambda\mu}\tope{P}_\nu  - \eta_{\lambda\nu}\tope{P}_\mu \} .
	\end{align}
Combining the last two equalities with~\eref{12.112} and~\eref{2.1}, we,
after a simple algebraic calculations, obtain%
\footnote{~\label{CommutativityWithMomentum}%
To derive equation~\eref{2.23}, notice that~\eref{2.21} implies
\(
[\tope{M}_{\mu\nu}, \tope{P}_{\mu_1}\circ\dots\circ\tope{P}_{\mu_n} ]_{\_}
=
- \sum_{i=1}^{n}
\bigl( \eta_{\mu\mu_i} \tope{P}_{\nu} - \eta_{\nu\mu_i} \tope{P}_{\mu} \bigr)
\tope{P}_{\mu_1}\circ\dots\circ
\tope{P}_{\mu_{i-1}}\circ\tope{P}_{\mu_{i+1}} \circ \dots \circ
\tope{P}_{\mu_n},
\)
due to $[A,B\circ C]_{\_} = [A,B]_{\_}\circ C+ B\circ [A,C]_{\_}$,
and expand the exponent in~\eref{12.112} into a power series.
More generally, if $[A(x),\tope{P}_\mu]_{\_}=B_\mu(x)$ with
$[B_\mu(x),\tope{P}_\nu]_{\_}=0$, then
\(
[A(x),\ope{U}(x,x_0)]_{\_}
= \iih (x^\mu-x_0^\mu)B_\mu(x) \circ \ope{U}(x,x_0);
\)
in particular, $[A(x),\tope{P}_\mu]_{\_}=0$ implies
 $[A(x),\ope{U}(x,x_0)]_{\_}=0$. Notice, we consider $(x^\mu-x_0^\mu)$ as a
real parameter by which the corresponding operators are multiplied and which
operators are supposed to be linear in it.%
}
	\begin{align}	\label{2.22}
& [\tope{Q}, \ope{U}(x,x_0)]_{\_} = 0
\\			\label{2.23}
& [\tope{M}_{\mu\nu}, \ope{U}(x,x_0)]_{\_}
=
- \{ (x_\mu-x_{0\,\mu}) \tope{P}_\nu
   - (x_\nu-x_{0\,\nu}) \tope{P}_\mu \} \circ \ope{U}(x,x_0) .
	\end{align}
Consequently, in accord with~\eref{12.114}, in momentum picture the charge and
angular momentum operators respectively are
	\begin{align}	\label{2.24}
\ope{Q}(x)
& = \tope{Q} := \ope{Q}
\\ \notag
\ope{M}_{\mu\nu}
& = \ope{U}(x,x_0) \circ \tope{M}_{\mu\nu} \circ \ope{U}^{-1}(x,x_0)
  = \tope{M}_{\mu\nu} + [\ope{U}(x,x_0),\tope{M}_{\mu\nu}]_{\_} \circ
  						\ope{U}^{-1}(x,x_0)
\\ \notag
& =
\tope{M}_{\mu\nu}
+ (x_\mu-x_{0\,\mu}) \ope{P}_\nu - (x_\nu-x_{0\,\nu}) \ope{P}_\mu
\\			\label{2.25}
& =
\tope{L}_{\mu\nu}
+ (x_\mu-x_{0\,\mu}) \ope{P}_\nu - (x_\nu-x_{0\,\nu}) \ope{P}_\mu
+ \tope{S}_{\mu\nu}
= \ope{L}_{\mu\nu} + \ope{S}_{\mu\nu} ,
	\end{align}
where
	\begin{equation}	\label{2.25new}
	\begin{split}
\tope{L}_{\mu\nu}(x)
  := \ope{U}(x,x_0) \circ\ope{L}_{\mu\nu}(x)\circ \ope{U}^{-1}(x,x_0)
\\
\tope{S}_{\mu\nu}(x)
  := \ope{U}(x,x_0) \circ\ope{S}_{\mu\nu}(x)\circ \ope{U}^{-1}(x,x_0)
	\end{split}
	\end{equation}
and~\eref{2.3} was taken into account. Notice, the correction to
$\tope{M}_{\mu\nu}$ on the r.h.s.\ of~\eref{2.25} is typical for the one of
classical orbital angular momentum when the origin, with respect to which it
is determined, is changed from $x$ to $x_0$.%
\footnote{~
In Section~\ref{Sect7}, it will be proved that, for massive free vector fields
(and for massless free vector fields under some conditions), holds the
equation
	\begin{align}	\label{2.25-1}
& [\tope{S}_{\mu\nu} , \ope{P}_\lambda]_{\_} = 0,
\intertext{which implies}	\label{2.25-2}
& [\tope{S}_{\mu\nu} , \ope{U}(x,s_0)]_{\_} = 0.
\intertext{Amongst other things, from here follow the equations}
		\label{2.25-3}
& \ope{S}_{\mu\nu} = \tope{S}_{\mu\nu}
\\				\label{2.25-4}
& \ope{L}_{\mu\nu}
= \tope{L}_{\mu\nu}
   + (x_\mu-x_{0\,\mu}) \ope{P}_\nu - (x_\nu-x_{0\,\nu}) \ope{P}_\mu .
	\end{align}%
}

	In momentum picture, by virtue of~\eref{12.114}, the
relations~\eref{2.17} and~\eref{2.18} respectively read
($ \varepsilon(\varphi) := \varepsilon(\tope{\varphi}) $)
	\begin{align}	\label{2.26}
& [\ope{\varphi}_i, \ope{Q}]_{\_}
	= \varepsilon(\ope{\varphi}_i) q \ope{\varphi}_i
\\			\label{2.27}
& [\ope{\varphi}_i, \ope{M}_{\mu\nu}(x,x_0)]_{\_}
=
x_\mu [\varphi_i ,\ope{P}_\nu]_{\_} - x_\nu [\varphi_i ,\ope{P}_\mu]_{\_}
+ \ih I_{i\mu\nu}^{j} \varphi_j .
	\end{align}
The first of these equation is evident. To derive the second one, we notice
that, by virtue of the Heisenberg relations/equations~\eref{2.28},
the equality~\eref{2.18} is equivalent to
	\begin{align}
			\label{2.29}
& [\tope{\varphi}_i(x), \tope{M}_{\mu\nu}]_{\_}
=
  x_\mu [\tope{\varphi}_i(x) ,\tope{P}_\nu]_{\_}
- x_\nu [\tope{\varphi}_i(x) ,\tope{P}_\mu]_{\_}
+ \ih I_{i\mu\nu}^{j} \tope{\varphi}_j(x)
	\end{align}
from where~\eref{2.27} follows.

	It should be emphasized, the Heisenberg relations~\eref{2.17}
and~\eref{2.18}, as well as the commutation relations~\eref{2.20}
and~\eref{2.21}, are external to the Lagrangian formalism. For  this reason,
one should be quite careful when applying them unless they are explicitly
proved in the framework of Lagrangian scheme.


\section
[Description of free vector field in momentum picture]
{Description of free vector field in momentum picture}
\label{Sect3}

	A vector field $\U$ is described by four operators
$\tU_\mu:=\tU_\mu(x)$, called its components, which transform as components
of a 4\ndash vector under Poincar\'e transformations. The operators $\tU_\mu$
are Hermitian, $\tU_\mu^\dag=\tU_\mu$, for a neutral field and non\ndash
Hermitian, $\tU_\mu^\dag\not=\tU_\mu$, for a charged one. Since the
consideration of $\tU_0,\dots,\tU_3$ as independent scalar fields meets as an
obstacle the non\ndash positivity of the energy
(see, e.g.,~\cite[\S~4.1]{Bogolyubov&Shirkov}
or~\cite[\S~2a]{Pauli-RelativisticTheories}), the Lagrangian of a free vector
field is represented as a sum of the Lagrangians, corresponding to
$\tU_0,\dots,\tU_3$ considered as independent scalar fields, and a
`correction' term(s) ensuring the energy positivity (and, in fact, defining
$\tU$ as spin~1 quantum field).  As pointed in the discussion
in~\cite[\S~4.1 and \S~5.3]{Bogolyubov&Shirkov}, the Lagrangian of a free
vector field (and, possibly, conditions additional to the Lagrangian
formalism) can be chosen in different ways, which lead to identical theories,
\ie to coinciding field equations and dynamical variables.%
\footnote{~%
This may not be the case when \emph{interacting} fields are considered.%
}

	Between a number of possibilities for describing a \emph{massive}
vector field of mass $m\not=0$, we choose the Lagrangian
as~\cite{Bjorken&Drell-2}
	 \begin{equation}	\label{3.1}
\tope{L}
=
\frac{m^2c^4}{1+\tau(\tU)} \tU_\mu^\dag \circ \tU^\mu
-
\frac{c^2\hbar^2}{1+\tau(\tU)}	(\pd_\mu\tU_\nu^\dag) \circ
				(\pd^\mu\tU^\nu)
+
\frac{c^2\hbar^2}{1+\tau(\tU)}	(\pd_\mu\tU^{\mu\dag}) \circ
				(\pd_\nu\tU^\nu) ,
	\end{equation}
where the function $\tau$ takes care of is the field neutral (Hermitian) or
charged (non\ndash Hermitian) according to
	\begin{equation}	\label{3.2}
\tau(\tU) :=
	\begin{cases}
1	&\text{for $\tU_\mu^\dag=\tU_\mu$ (Hermitian (neutral) field)}
\\
0	&\text{for $\tU_\mu^\dag\not=\tU_\mu$ (non-Hermitian (charged) field)}
	\end{cases}\ .
	\end{equation}
Since
\[
 (\pd_\mu\tU^{\mu\dag}) \circ (\pd_\nu\tU^\nu)
-
 (\pd_\mu\tU_\nu^{\dag}) \circ (\pd^\nu\tU^\mu)
=
\pd_\mu
\{  \tU^{\mu\dag} \circ (\pd_\nu\tU^\nu)
  - \tU_\nu^{\dag} \circ (\pd^\nu\tU^\mu) \} ,
\]
the theory arising from the Lagrangian~\eref{3.1} is equivalent to the one
build from~\cite{Itzykson&Zuber,Ryder-QFT}
	 \begin{equation}	\label{3.3}
	\begin{split}
\tope{L}
& =
\frac{m^2c^4}{1+\tau(\tU)} \tU_\mu^\dag \circ \tU^\mu
-
\frac{c^2\hbar^2}{1+\tau(\tU)}	(\pd_\mu\tU_\nu^\dag) \circ
				(\pd^\mu\tU^\nu)
+
\frac{c^2\hbar^2}{1+\tau(\tU)}	(\pd_\mu\tU_\nu^{\dag}) \circ
				(\pd^\nu\tU^\mu)
\\
& =
\frac{m^2c^4}{1+\tau(\tU)} \tU_\mu^\dag \circ \tU^\mu
- \frac{c^2\hbar^2}{2(1+\tau(\tU))}
  \tope{F}_{\mu\nu}^\dag \circ \tope{F}^{\mu\nu} ,
	\end{split}
	\end{equation}
where
	\begin{equation}	\label{3.4}
\tope{F}_{\mu\nu} := \pd_\mu \tU_\nu - \pd_\nu \tU_\mu .
	\end{equation}
The first two terms in~\eref{3.1} (or in the first row in~\eref{3.3})
correspond to a sum of four independent Lagrangians for the components
$\tU_0,\dots,\tU_3$, considered as free scalar
fields~\cite{Bjorken&Drell-2,Bogolyubov&Shirkov,bp-QFTinMP-scalars}. The
remaining terms in~\eref{3.1} or~\eref{3.3} represent the afore\ndash
mentioned `correction' which reduces the independent components (degrees of
freedom) of a vector field from 4 to 3 and ensures the positivity of the
field's energy~\cite{Bjorken&Drell-2,Bogolyubov&Shirkov,Itzykson&Zuber}.

	Before proceeding with the description in momentum picture, we notice
that the Euler\ndash Lagrange equations for the Lagrangians~\eref{3.1}
and~\eref{3.3} coincide and are
	\begin{subequations}	\label{3.5}
	\begin{align}	\label{3.5a}
0 &= m^2c^2 \tU_\mu + \hbar^2 \tope{\square}(\tU_\mu)
		   - \hbar^2 \pd_\mu(\pd^\lambda\tU_\lambda)
  = m^2c^2 \tU_\mu + \hbar^2 \pd^\lambda \tope{F}_{\lambda\mu}
\\			\label{3.5b}
0 &= m^2c^2 \tU_\mu^\dag + \hbar^2 \tope{\square}(\tU_\mu^\dag)
		   - \hbar^2 \pd_\mu(\pd^\lambda\tU_\lambda^\dag)
  = m^2c^2 \tU_\mu^\dag + \hbar^2 \pd^\lambda \tope{F}_{\lambda\mu}^\dag ,
	\end{align}
	\end{subequations}
where $\tope{\square}:=\pd_\lambda\pd^\lambda$ is the D'Alembert operator (in
Heisenberg picture). For $m\not=0$, these equations, known as the
\emph{Proca equations} for $\tU_\mu$ and $\tU_\mu^\dag$, can be written
equivalently as
	\begin{subequations}	\label{3.6}
	\begin{alignat}{2}	\label{3.6a}
(m^2c^2 + \hbar^2 \tope{\square}) \tU_\mu &= 0
&\qquad
(m^2c^2 + \hbar^2 \tope{\square}) \tU_\mu^\dag &= 0
\\			\label{3.6b}
\pd^\mu\tU_\mu &= 0
&
\pd^\mu\tU_\mu^\dag &= 0 ,
	\end{alignat}
	\end{subequations}
due to $\pd^\mu\pd^\nu\tope{F}_{\mu\nu}\equiv0$ and $m\not=0$, and show that
there is a bijective correspondence between $\tU_\mu$ and $\tope{F}_{\mu\nu}$
(in the case $m\not=0$)~\cite[\S~2]{Pauli-RelativisticTheories}. Therefore
the field operators $\tU_\mu$ and $\tU_\mu^\dag$ are solutions of the
\emph{Klein\ndash Gordon equations}~\eref{3.6a} with mass $m$~($\not=0$) and
satisfy the conditions~\eref{3.6b}, known as the \emph{Lorenz conditions}.
We shall say that a vector field satisfies the Lorenz condition, if the
equations~\eref{3.6b} hold for it.

	According to~\eref{2.5}, the Lagrangians~\eref{3.1} and~\eref{3.3} in
momentum picture are
	\begin{align}	\label{3.7}
\ope{L}
& =
\frac{m^2c^4}{1+\tau(\U)} \U_\mu^\dag\circ\U^\mu
+
\frac{c^2}{1+\tau(\U)}
\bigl\{
[ \U_\nu^\dag , \ope{P}_\mu ]_{\_} \circ [ \U^\nu , \ope{P}^\mu ]_{\_}
-
[ \U^{\mu\dag} , \ope{P}_\mu ]_{\_} \circ [ \U^\nu , \ope{P}_\nu ]_{\_}
\bigr\}
\\			\label{3.8}
	\begin{split}
\ope{L}
& =
\frac{m^2c^4}{1+\tau(\U)} \U_\mu^\dag\circ\U^\mu
+
\frac{c^2}{1+\tau(\U)}
\bigl\{
[ \U_\nu^\dag , \ope{P}_\mu ]_{\_} \circ [ \U^\nu , \ope{P}^\mu ]_{\_}
-
[ \U_\nu^\dag , \ope{P}_\mu ]_{\_} \circ [ \U^\mu , \ope{P}^\nu ]_{\_}
\bigr\}
\\
&=
\frac{m^2c^4}{1+\tau(\U)} \U_\mu^\dag\circ\U^\mu
-
\frac{c^2\hbar^2}{2(1+\tau(\U))} \ope{F}_{\mu\nu}^\dag\circ \ope{F}^{\mu\nu} ,
	\end{split}
	\end{align}
respectively, where
	\begin{gather}	\label{3.10-1}
\U_\mu(x) := \ope{U}(x,x_0)\circ \tU_\mu(x) \circ \ope{U}^{-1}(x,x_0)
\quad
\U_\mu^\dag(x) := \ope{U}(x,x_0)\circ \tU_\mu^\dag(x) \circ \ope{U}^{-1}(x,x_0)
\\			\label{3.9}
\tau(\U) :=
	\begin{cases}
1	&\text{for $\U_\mu^\dag=\U_\mu$ (Hermitian (neutral) field)}
\\
0	&\text{for $\U_\mu^\dag\not=\U_\mu$ (non-Hermitian (charged) field)}
	\end{cases}
= \tau(\tU)
\\			\label{3.10}
\ope{F}_{\mu\nu}(x)
= \ope{U}(x,x_0)\circ (\tope{F}_{\mu\nu}(x)) \circ \ope{U}^{-1}(x,x_0)
= -\iih \{ [\U_\mu,\ope{P}_\nu]_{\_} - [\U_\nu,\ope{P}_\mu]_{\_} \} .
	\end{gather}

	Regarding $\U_\mu$ and $\U_\mu^\dag$ as independent variables,
from~\eref{3.7}, we get%
\footnote{~%
The derivatives in~\eref{3.11} are calculating according to the classical
rules of commuting variables, which requires additional rules for ordering
the operators in the expressions for dynamical variables; for details,
see~\cite{bp-QFT-action-principle}. The Lagrangian~\eref{3.8} has different
derivatives, but leads to the same field equations and dynamical variables
and, for this reason, will not be considered further in this work.%
}
	\begin{equation}	\label{3.11}
	\begin{split}
\frac{\pd\ope{L}}{\pd\U^\mu} = \frac{\pd\tope{L}}{\pd\U^\mu}
& = m^2c^4 \U_\mu^\dag
\qquad
\pi_{\mu\lambda}
:=  \frac{\pd\ope{L}}{\pd y^{\mu\lambda}}
=   \ih c^2 [\U_\mu^\dag,\ope{P}_\lambda]_{\_}
  - \ih c^2 \eta_{\mu\lambda} [\U^{\nu\dag},\ope{P}_\nu]_{\_}
\\
\frac{\pd\ope{L}}{\pd\U^{\mu\dag}} = \frac{\pd\tope{L}}{\pd\U^{\mu\dag}}
& = m^2c^4 \U_\mu
\qquad
\pi_{\mu\lambda}^\dag
:=  \frac{\pd\ope{L}}{\pd y^{\mu\lambda\dag}}
=   \ih c^2 [\U_\mu,\ope{P}_\lambda]_{\_}
  - \ih c^2 \eta_{\mu\lambda} [\U^{\nu},\ope{P}_\nu]_{\_}
	\end{split}
	\end{equation}
with
	\begin{equation*}
y_{\mu\lambda} := \iih [\U_\mu,\ope{P}_\lambda]_{\_}
\quad
y_{\mu\lambda}^\dag := \iih [\U_\mu^\dag,\ope{P}_\lambda]_{\_} .
	\end{equation*}
Therefore the field equations~\eref{12.129} now read
	\begin{subequations}	\label{3.12}
	\begin{align}	\label{3.12a}
m^2c^2 \U_\mu
- [ [ \U_\mu , \ope{P}_\lambda]_{\_} , \ope{P}^\lambda ]_{\_}
+ [ [ \U_\nu , \ope{P}^\nu]_{\_} , \ope{P}^\mu]_{\_}
& = 0
\\			\label{3.12b}
m^2c^2 \U_\mu^\dag
- [ [ \U_\mu^\dag , \ope{P}_\lambda]_{\_} , \ope{P}^\lambda ]_{\_}
+ [ [ \U_\nu^\dag , \ope{P}^\nu]_{\_} , \ope{P}^\mu]_{\_}
& = 0
	\end{align}
	\end{subequations}
or, using the notation~\eref{3.4},
	\begin{equation}	\label{3.13}
m^2c^2 \U_\mu - \ih [ \ope{F}_{\lambda\mu} , \ope{P}^\lambda]_{\_} = 0
\quad
m^2c^2 \U_\mu^\dag - \ih [ \ope{F}_{\lambda\mu}^\dag , \ope{P}^\lambda]_{\_}
= 0 .
	\end{equation}
These are the systems of the \emph{Proca equations in momentum picture} for a
massive free spin~1 (vector) fields.

	Since the equality $\pd_\mu\pd_\nu=\pd_\nu\pd_\mu$ (valid when
applied on $C^2$ functions or operators) in momentum picture takes the form
(see~\eref{12.114})
	\begin{equation}	\label{3.14}
[[\cdot,\ope{P}_\mu]_{\_},\ope{P}_\nu]_{\_}
=
[[\cdot,\ope{P}_\nu]_{\_},\ope{P}_\mu]_{\_},
	\end{equation}
from~\eref{3.13},~\eref{3.14} and $\ope{F}_{\mu\nu}=-\ope{F}_{\nu\mu}$ follow
the equalities
	\begin{equation}	\label{3.14-1}
m^2 [\U_\mu,\ope{P}^\mu]_{\_} = 0
\quad
m^2 [\U_\mu^\dag,\ope{P}^\mu]_{\_} = 0.
	\end{equation}
Consequently, in the massive case, \ie $m\not=0$, the system of Proca
equations~\eref{3.12} splits into the system of Klein\ndash Gordon equations
in momentum picture~\cite{bp-QFTinMP-scalars}
	\begin{gather}	\label{3.15}
m^2c^2\U_\mu - [ [ \U_\mu , \ope{P}_\nu]_{\_} , \ope{P}^\nu]_{\_} = 0
\quad
m^2c^2\U_\mu^\dag - [ [ \U_\mu^\dag , \ope{P}_\nu]_{\_} , \ope{P}^\nu]_{\_} =0
\\\intertext{and the system of \emph{Lorenz conditions}}
			\label{3.16}
[ \U_\mu , \ope{P}^\mu]_{\_} = 0
\quad
[ \U_\mu^\dag , \ope{P}^\mu]_{\_} = 0
	\end{gather}
for the field operators $\U_\mu$ and $\U_\mu^\dag$. This result is a momentum
picture analogue of~\eref{3.6}. From technical point of view, it is quite
important as it allows a partial application of most of the results obtained
for free scalar fields, satisfying (systems of) Klein\ndash Gordon equation(s),
to the case of massive vector fields. Evidently, for the solutions
of~\eref{3.15}--\eref{3.16}, the Lagrangian~\eref{3.7} reduces to%
\footnote{~%
The same result holds, up to a full divergence, for the Lagrangian~\eref{3.8}
too.%
}
	\begin{equation}	\label{3.17}
\ope{L}
=
\frac{m^2c^4}{1+\tau(\U)} \U_\mu^\dag\circ\U^\mu
+
\frac{c^2}{1+\tau(\U)}
[ \U_\nu^\dag , \ope{P}_\mu ]_{\_} \circ [ \U^\nu , \ope{P}^\mu ]_{\_} ,
	\end{equation}
which equals to a sum of four Lagrangians corresponding to
$\U_0,\dots,\U_3$ considered as free scalar fields.

	The above consideration show that the Lagrangian theory of massive
free vector field can be constructed equivalently from the
Lagrangian~\eref{3.17} under the additional conditions~\eref{3.16}. This
procedure is realized in Heisenberg picture in~\cite{Bogolyubov&Shirkov}.

	Consider now the above theory in the massless case, \ie for $m=0$. It
is easily seen, all of the above conclusions remain valid in the massless
case too with one very important exception. Namely, in it the
equations~\eref{3.14-1} are identically valid and, consequently, in this case
the massless Proca equations, i.e.~\eref{3.12} with $m=0$, do \emph{not}
imply the Klein\ndash Gordon equations~\eref{3.15} and the Lorenz
conditions~\eref{3.16}.%
\footnote{~%
As it is well know~\cite{Bogolyubov&Shirkov,Bjorken&Drell,Itzykson&Zuber}, in
this important case a gauge symmetry arises, \ie an invariance of the
theory under the gauge transformations
 $\tU_\mu\mapsto \tU_\mu + \pd_\mu \tope{K}$,
 $\tope{K}$ being a $C^2$ operator, in Heisenberg picture or
$\U_\mu\mapsto \tU_\mu + \iih [K,\ope{P}_\mu]$ in momentum picture.%
}
However, one can verify, \eg in momentum representation in Heisenberg
picture, that the Lorenz conditions~\eref{3.16} are compatible with the
massless Proca equations~\eref{3.12} with $m=0$; said differently, the system
of Klein\ndash Gordon equations~\eref{3.15} with $m=0$ and Lorenz
conditions~\eref{3.16}, \ie
	\begin{alignat}{2}	\label{3.15-0}
[ [ \U_\mu , \ope{P}_\nu]_{\_} , \ope{P}^\nu]_{\_} &= 0
&\quad
[ [ \U_\mu^\dag , \ope{P}_\nu]_{\_} , \ope{P}^\nu]_{\_} &= 0
\\			\label{3.16-0}
[ \U_\mu , \ope{P}^\mu]_{\_} & = 0
&
[ \U_\mu^\dag , \ope{P}^\mu]_{\_} &= 0 ,
	\end{alignat}
does not contain contradictions and possesses non\ndash trivial solutions.
Moreover, one can consider this system of equations as the one describing a
free electromagnetic field in Lorenz gauge \emph{before} second
quantization, \ie before imposing a suitable commutation relations between
the field's components.

	For these reasons, in the present investigation, with an exception of
Sect.~\ref{Sect13}, we shall consider a
\emph{quantum field theory build according to the Lagrangian formalism
arising from the Lagrangian~\eref{3.7} to which, in the massless case, are
added the Lorenz conditions~\eref{3.16}}
as additional requirements. In other words, with an exception of
Sect.~\ref{Sect13}, vector fields satisfying the Lorenz conditions will be
explored in this work.

	A free vector field possesses energy-momentum, (possibly vanishing)
charge, and angular momentum. The corresponding to them density operators,
the energy\ndash momentum tensor $\tope{T}_{\mu\nu}$, current density
$\ope{J}_\mu$ and (total) angular momentum density
$\tope{M}_{\mu\nu}^{\lambda}$, in Heisenberg picture for the
Lagrangian~\eref{3.7} are as follows:%
\footnote{~%
As a consequence of~\eref{3.16}, the expressions~\eref{3.18}--\eref{3.21} are
sums of the ones corresponding to $\tU_0,\dots,\tU_3$ considered as free
scalar fields~\cite{Bogolyubov&Shirkov,Bjorken&Drell-2,Itzykson&Zuber}. For
a rigorous derivation of~\eref{3.18}--\eref{3.22}, see the general rules
described in~\cite{bp-QFT-action-principle}.%
}
	\begin{align}
			\label{3.18}
	\begin{split}
\tope{T}_{\mu\nu}
& =
\frac{1}{1+\tau(\tU)}
\bigl\{
\tope{\pi}_{\lambda\mu}\circ(\pd_\nu\tU^\lambda) +
(\pd_\nu\tU^{\lambda\dag})\circ\tope{\pi}_{\lambda\mu}^\dag
\bigr\}
- \eta_{\mu\nu}\tope{L}
\\ & =
- \frac{c^2\hbar^2}{1+\tau(\tU)}
\bigl\{
(\pd_\mu\tU_\lambda^\dag)\circ(\pd_\nu\tU^\lambda) +
(\pd_\nu\tU_\lambda^\dag)\circ(\pd_\mu\tU^\lambda)
\bigr\}
- \eta_{\mu\nu}\tope{L}
= \tope{T}_{\nu\mu}
	\end{split}
\displaybreak[1]\\			\label{3.19}
	\begin{split}
\tope{J}_{\mu}
& =
\frac{q}{\ih}
\bigl\{
\tope{\pi}_{\lambda\mu}\circ\tU^\lambda -
\tU^{\lambda\dag}\circ\tope{\pi}_{\lambda\mu}^\dag
\bigr\}
=
\ih q c^2
\bigl\{
(\pd_\mu\tU_\lambda^\dag)\circ\tU^\lambda -
\tU_\lambda^\dag\circ(\pd_\mu\tU^\lambda)
\bigr\}
	\end{split}
\displaybreak[1]\\			\label{3.20}
  \tope{M}_{\mu\nu}^{\lambda}
& = \tope{L}_{\mu\nu}^{\lambda} + \tope{S}_{\mu\nu}^{\lambda} \, ,
\\\intertext{where the Lorenz conditions were taken into account, $q$ is the
charge of the field (of field's particles), and}
			\label{3.21}
\tope{L}_{\mu\nu}^{\lambda}
& :=
x_\mu\Sprindex[\tope{T}]{\nu}{\lambda} -
x_\nu\Sprindex[\tope{T}]{\mu}{\lambda}
\displaybreak[2]\\			\label{3.22}
	\begin{split}
\tope{S}_{\mu\nu}^{\lambda}
& : =
\frac{1}{1+\tau(\tU)}
\bigl\{
\tope{\pi}^{\rho\lambda} \circ ( I_{\rho\mu\nu}^{\sigma}\tU_\sigma ) +
( I_{\rho\mu\nu}^{\dag\, \sigma}\tU_\sigma^\dag )
				\circ \tope{\pi}^{\rho\lambda\dag}
\bigr\}
\\ & =
\frac{\hbar^2c^2}{1+\tau(\tU)}
\bigl\{
(\pd^\lambda \tU_\mu^\dag) \circ \tU_\nu -
(\pd^\lambda \tU_\nu^\dag) \circ \tU_\mu -
\tU_\mu^\dag \circ (\pd^\lambda\tU_\nu) +
\tU_\nu^\dag \circ (\pd^\lambda\tU_\mu)
\bigr\}
	\end{split}
\displaybreak[3]\\\intertext{with the numbers}	\label{3.23}
I_{\rho\mu\nu}^{\sigma}
= & I_{\rho\mu\nu}^{\dag\, \sigma}
=
\delta_\mu^\sigma \eta_{\nu\rho} - \delta_\nu^\sigma \eta_{\mu\rho}
	\end{align}
being characteristics of a vector field under
4-rotations~\cite[eq.~(0-43)]{Roman-QFT}. It should be noticed, since the
energy\ndash momentum operator~\eref{3.18} is \emph{symmetric},
$\tope{T}_{\mu\nu}=\tope{T}_{\nu\mu}$, the spin and orbital angular momentum
density operators satisfy the continuity equations
	\begin{equation}	\label{3.24}
\pd_\lambda\tope{S}_{\mu\nu}^{\lambda} = 0 \quad
\pd_\lambda\tope{L}_{\mu\nu}^{\lambda} = 0
	\end{equation}
and, consequently, the spin and orbital angular momentum operators of a free
vector field are conserved ones, \ie
	\begin{equation}	\label{3.25}
\frac{\od}{\od x^0} \tope{S}_{\mu\nu}^{\lambda} = 0 \quad
\frac{\od}{\od x^0} \tope{L}_{\mu\nu}^{\lambda} = 0  .
	\end{equation}

	According to~\eref{12.114},~\eref{3.10-1},~\eref{3.11},
and~\eref{3.16}, the densities of the dynamical characteristics of a free
vector field in momentum picture are:
	\begin{align}
			\label{3.26}
	\begin{split}
\ope{T}_{\mu\nu}
& =
\frac{c^2}{1+\tau(\U)}
\bigl\{
[\U_\lambda^\dag,\ope{P}_\mu]_{\_} \circ [\U^\lambda,\ope{P}_\nu]_{\_} +
[\U_\lambda^\dag,\ope{P}_\nu]_{\_} \circ [\U^\lambda,\ope{P}_\mu]_{\_}
\bigr\}
\\ & \hphantom{=} -
\frac{\eta_{\mu\nu}c^2}{1+\tau(\U)}
\bigl\{
m^2c^2 \U_\lambda^\dag \circ \U^\lambda +
[\U_\varkappa^\dag,\ope{P}_\lambda]_{\_} \circ
[\U^\varkappa,\ope{P}^\lambda]_{\_}
\bigr\}
	\end{split}
\displaybreak[2]\\	\label{3.27}
\ope{J}_\mu
& = q c^2
\bigl\{
[\U_\lambda^\dag,\ope{P}_\mu]_{\_} \circ \U^\lambda -
\U_\lambda^\dag \circ [\U^\lambda,\ope{P}_\mu]_{\_}
\bigr\}
\displaybreak[2]\\	\label{3.28}
\ope{L}_{\mu\nu}^{\lambda}
& =
x_\mu\Sprindex[\ope{T}]{\nu}{\lambda} -
x_\nu\Sprindex[\ope{T}]{\mu}{\lambda}
\displaybreak[1]\\	\label{3.29}
	\begin{split}
\ope{S}_{\mu\nu}^\lambda
& =
- \frac{\ih c^2}{1+\tau(\U)}
\bigl\{
  [\U_\mu^\dag,\ope{P}^\lambda]_{\_} \circ \U_\nu
- [\U_\nu^\dag,\ope{P}^\lambda]_{\_} \circ \U_\mu
- \U_\mu^\dag \circ [\U_\nu,\ope{P}^\lambda]_{\_}
+ \U_\nu^\dag \circ [\U_\mu,\ope{P}^\lambda]_{\_}
\bigr\} .
	\end{split}
	\end{align}

	Comparing~\eref{3.26} and~\eref{3.27} with the corresponding
expressions for $\U_0,\dots,\U_3$, considered as free scalar
fields~\cite{bp-QFTinMP-scalars}, we see that the terms originating from
$\U_1$, $\U_2$ and $\U_3$ enter in~\eref{3.26} and~\eref{3.27} with right
signs if $\U_1$, $\U_2$ and $\U_3$ were free scalar fields. But the terms, in
which $\U_0$ enters, are with signs opposite to the ones if $\U_0$ was a free
scalar field. In particular, this means that the contribution of $\U_0$ in
the field's energy is negative. All this points to the known fact that $\U_0$
is a carrier of an unphysical degree of freedom, which must be eliminated
(via the Lorenz conditions~\eref{3.16}). The second new moment, with respect
to the scalar field case, is the existence of a, generally, non\ndash
vanishing spin angular momentum density operator~\eref{3.29} to which a
special attention will be paid (see Sect.~\ref{Sect6}).

	Since $\U_\mu$ are solutions of the Klein-Gordon
equations~\eref{3.15}, the operator
\(
\frac{1}{c^2} [ [ \cdot , \ope{P}_\lambda ]_{\_} , \ope{P}^\lambda ]_{\_}
\)
has a meaning of a square-of-mass operator of the vector field under
consideration. At the same time, the operator
$\frac{1}{c^2}\ope{P}_\lambda\circ\ope{P}^\lambda$ has a meaning of
square-of-mass operator of the field's states (state vectors).

	We shall specify the relation~\eref{2.26} for a vector field by
putting $\varepsilon(\U_\mu)=+1$ and $\varepsilon(\U_\mu^\dag)=-1$.
Therefore the relations~\eref{2.26} and~\eref{2.27} take the form
	\begin{equation}	\label{3.30}
[\U_\mu, \ope{Q}]_{\_} =  q \U_\mu
\quad
[\U_\mu^\dag, \ope{Q}]_{\_} = - q \U_\mu^\dag
	\end{equation}
\vspace{-4.4ex}
	\begin{subequations}	\label{3.31}
	\begin{align}
			\label{3.31a}
& [\U_\lambda, \ope{M}_{\mu\nu}(x,x_0)]_{\_}
=
x_\mu [\U_\lambda ,\ope{P}_\nu]_{\_} - x_\nu [\U_\lambda ,\ope{P}_\mu]_{\_}
+ \ih ( \U_\mu \eta_{\nu\lambda} - \U_\nu \eta_{\mu\lambda} )
\\			\label{3.31b}
& [\U_\lambda^\dag, \ope{M}_{\mu\nu}(x,x_0)]_{\_}
=
  x_\mu [\U_\lambda^\dag ,\ope{P}_\nu]_{\_}
- x_\nu [\U_\lambda^\dag ,\ope{P}_\mu]_{\_}
+ \ih ( \U_\mu^\dag \eta_{\nu\lambda} - \U_\nu^\dag \eta_{\mu\lambda} ) ,
	\end{align}
	\end{subequations}
where~\eref{3.23} was used. It is clear, the last terms in~\eref{3.31} are
due to the spin angular momentum, while the other ones originate from the
orbital angular momentum.


\section {Analysis of the field equations}
\label{Sect4}

	The analysis of the Dirac equations in~\cite{bp-QFTinMP-spinors} can
\emph{mutatis mutandis} be applied to the case of vector fields satisfying
the Lorenz condition. This can be done as follows

	At first, we distinguish the `degenerate' solutions
	\begin{gather}
			\label{4.1}
[\U,\ope{P}_\mu]_{\_} = 0
\quad
[\U^\dag,\ope{P}_\mu]_{\_} = 0
\qquad\text{for $m=0$}
\\\intertext{of the Klein-Gordon equations~\eref{3.15}, which solutions, in
view of~\eref{2.28}, in Heisenberg picture read}
			\label{4.2}
\tU_\mu(x) = \tU_\mu(x_0) = \tU_\mu \ (=\const_\mu)
\quad
\tU_\mu^\dag(x) = \tU_\mu^\dag(x_0) = \tU_\mu^\dag \ (=\const_\mu^\dag)
\quad\text{for $m=0$} .
	\end{gather}

	According to equations~\eref{3.26}--\eref{3.29}, the energy\ndash
momentum, charge and  angular momentum density operators for the
solutions~\eref{4.1} respectively are:
	\begin{align}
				\label{4.2-1}
& \ope{T}_{\mu\nu} = 0 \quad
  \ope{J}_\mu = 0 \quad
  \ope{L}_{\mu\nu}^{\lambda} =
  \ope{S}_{\mu\nu}^{\lambda} =
  \ope{M}_{\mu\nu}^{\lambda} =0
\\\intertext{Since~\eref{4.2-1} and~\eref{2.6}--\eref{2.15} imply}
				\label{4.2-2}
& \ope{P}_\mu = 0 \quad
  \ope{Q}_\mu = 0 \quad
  \ope{L}_{\mu\nu} =
  \ope{S}_{\mu\nu} =
  \ope{M}_{\mu\nu} =0 ,
	\end{align}
the solutions~\eref{4.1} (or~\eref{4.2} in Heisenberg picture) describe a
massless vector field with vanishing dynamical characteristics. Such a field
cannot lead to any predictable observable results and, in this sense is
unphysical.%
\footnote{~%
This case is similar to the one of free scalar fields describe
in~\cite{bp-QFTinMP-scalars}. Note, the so\ndash arising situation is
completely different from a similar one, when free spinor Dirac fields are
concerned as in it solutions, like~\eref{4.1}, are in principle observable
--- see~\cite{bp-QFTinMP-spinors}.
}%

	The further analysis of the field equations will be done similarly to
the one of free spinor fields in~\cite{bp-QFTinMP-spinors}. For the purpose,
one should replace the Dirac equations with the Lorenz
conditions~\eref{3.16} and take into account that now the field equations
are~\eref{3.15}--\eref{3.16}, not only~\eref{3.16}.%
\footnote{~%
It is interesting to be noted, the Dirac equation
 $\ih\gamma^\mu\pd_\mu\tope{\psi} - mc\tope{\psi}=0$,
 $\gamma^\mu$ being the $\gamma$\ndash matrices and $\tope{\psi}$ a 4\ndash
spinor, in the massless case, $m=0$, takes the form of a Lorenz condition,
\viz $\pd_\mu\tU^\mu=0$ with $\tU^\mu=\gamma^\mu\tope{\psi}$.%
}

	Taking into account the above facts, we can describe the structure of
the solutions of the field equations~\eref{3.15}--\eref{3.16} as follows.

	\begin{Prop}	\label{Prop4.1}
The solutions of the equations~\eref{3.15}--\eref{3.16} and~\eref{2.28} can
be written as (do not sum over $\mu$!)
	\begin{subequations}	\label{4.3}
	\begin{gather}
			\label{4.3a}
\U_\mu
=
\int\Id^3\bk
	\bigl\{
	  f_{\mu,+}(\bk) \U_\mu(k)\big|_{k_0=+\sqrt{m^2c^2+{\bk}^2} }
	+ f_{\mu,-}(\bk) \U_\mu(k)\big|_{k_0=-\sqrt{m^2c^2+{\bk}^2} }
	\bigr\}
\\			\label{4.3b}
\U_\mu^\dag
=
\int\Id^3\bk
	\bigl\{
  f^\dag_{\mu,+}(\bk) \U_\mu^\dag(k)\big|_{k_0=+\sqrt{m^2c^2+{\bk}^2} }
+ f^\dag_{\mu,-}(\bk) \U_\mu^\dag(k)\big|_{k_0=-\sqrt{m^2c^2+{\bk}^2} }
	\bigr\}
	\end{gather}
	\end{subequations}
or, equivalently as
	\begin{gather}
			\label{4.4}
\U_\mu = \int\Id^4k \delta(k^2-m^2c^2) f_\mu(k) \U_\mu(k)
\quad
\U_\mu^\dag = \int\Id^4k \delta(k^2-m^2c^2) f_\mu^\dag(k) \U_\mu^\dag(k) .
	\end{gather}
Here: $k=(k^0,k^1,k^2,k^3)$ is a 4-vector with dimension of 4-momentum,
$k^2=k_\mu k^\mu=k_0^2 -k_1^2 -k_2 -k_3^2=k_0^2 -{\bk}^2$ with $k_\mu$ being
the components of $k$ and $\bk:=(k^1,k^2,k^3)=-(k_1,k_2,k_3)$ being the
3\ndash dimensional part of $k$,
$\delta(\cdot)$ is the (1\ndash dimensional) Dirac delta function,
the operators $\U_\mu(k),\U_\mu^\dag(k)\colon\Hil\to\Hil$ are solutions of
the equations
	\begin{subequations}	\label{4.5}
	\begin{align}	\label{4.5a}
& [ \U_\mu(k), \ope{P}_\nu ]_{\_} = - k_\nu \U_\mu(k)
&&
[ \U_\mu^\dag(k), \ope{P}_\nu ]_{\_} = - k_\nu \U_\mu^\dag(k)
\\	   		\label{4.5b}
& \{ k^\mu \U_\mu(k) \} \big|_{k^2=m^2c^2} = 0
&&
\{ k^\mu \U_\mu^\dag(k)  \} \big|_{k^2=m^2c^2} = 0 ,
	\end{align}
	\end{subequations}
$f_{\mu,\pm}(\bk)$ and $f^\dag_{\mu,\pm}(\bk)$ are complex\ndash valued
functions (resp.\ distributions (generalized functions)) of $\bk$ for
solutions different from~\eref{4.1} (resp.\ for the solutions~\eref{4.1}),
and $f_\mu$ and $f_\mu^\dag$ are complex\ndash valued functions (resp.\
distribution) of $k$ for solutions different from~\eref{4.1} (resp.\ for the
solutions~\eref{4.1}). Besides, we have the relations
\(
f_\mu(k)|_{k_0=\pm\sqrt{m^2c^2+{\bs k}^2} }
	= 2\sqrt{m^2c^2+{\bs k}^2} f_{\mu,\pm}(\bs k)
\)
and
\(
f_\mu^\dag(k)|_{k_0=\pm\sqrt{m^2c^2+{\bs k}^2} }
	= 2\sqrt{m^2c^2+{\bs k}^2} f^\dag_{\mu,\pm}(\bs k)
\)
for solutions different from~\eref{4.1}.
	\end{Prop}

	\begin{Rem}	\label{Rem4.1}
	Evidently, in~\eref{4.3} and~\eref{4.4} enter only the solutions
of~\eref{4.5} for which
	\begin{align}	\label{4.6}
k^2 := k_\mu k^\mu =k_0^2 - {\bk}^2 = m^2c^2 .
	\end{align}
This circumstance is a consequence of the fact that $\U_\mu$ the solutions
of the Klein\ndash Gordon equations~\eref{3.15}.
	\end{Rem}

	\begin{Rem}	\label{Rem4.2}
	Obviously, to the solutions~\eref{4.1} corresponds~\eref{4.5a} with
$\ope{P}_\mu=0$. Hence
	\begin{gather}	\label{4.6-1}
\tope{\U_\mu}(x,0) = \U_\mu(0) = \const
\quad
\tope{\U_\mu^\dag}(x,0) = \U_\mu^\dag(0) = \const
\quad
\ope{P}_\mu = \tope{P}_\mu = 0
\intertext{with (see~\eref{12.114})}
			\label{4.6-2}
	\begin{split}
\tope{\U_\mu}(x,k) := \ope{U}^{-1}(x,x_0)\circ \U_\mu(k)\circ \ope{U}(x,x_0)
\quad
\tope{\U_\mu^\dag}(x,k):= \ope{U}^{-1}(x,x_0)\circ \U_\mu^\dag(k)\circ \ope{U}(x,x_0) .
	\end{split}
	\end{gather}
These solutions, in terms of~\eref{4.3} or~\eref{4.4}, are described by  $m=0$
and, for example,
$f_{\mu,\pm}(\bk)=f^\dag_{\mu,\pm}(\bk)=(\frac{1}{2}\pm a)\delta^3(\bk)$ for
some $a\in\field[C]$ or $f_\mu(k)=f_\mu^\dag(k)$ such that $
f_\mu(k)|_{k_0=\pm|\bs k|} = (1\pm 2a) |\bs k| \delta^3(\bs k)$,
respectively.  (Here $\delta^3(\bk):=\delta(k^1)\delta(k^2)\delta(k^3)$ is
the 3\ndash dimensional Dirac delta\ndash function. Note the equality
$\delta(y^2-b^2)=\frac{1}{b}(\delta(y+b)+\delta(y-b))$ for $b>0$.)
	\end{Rem}

	\begin{Rem}	\label{Rem4.3}
	Since $\U_\mu^\dag:=(\U_\mu)^\dag$, from~\eref{4.3}
(resp.~\eref{4.4}) is clear that there should exist some connection between
$f_{\mu,\pm}(\bk) \U_\mu(k)$ and
$f^\dag_{\mu,\pm}(\bk) \U_\mu^\dag(k)$ with
$k_0=+\sqrt{m^2c^2+{\bk}^2}$ (resp.\ between $f_\mu(k) \U_\mu(k)$ and
$f_\mu^\dag(k) \U_\mu^\dag(k)$).
A simple examination of ~\eref{4.3} (resp.~\eref{4.4}) reveals that the
Hermitian conjugation can either transform these expressions into each other
or `change' the signs plus and minus in them according to:
	\begin{subequations}	\label{4.7}
	\begin{align}
			\label{4.7a}
&
\bigl( f_{\mu\pm}(\bk)\U_\mu(k) \big|_{k_0=\pm\sqrt{m^2c^2+{\bs k}^2} }
\bigr)^\dag
=
 - f^\dag_{\mu,\mp}(-\bk)\U_\mu^\dag(-k)
	\big|_{k_0=\mp\sqrt{m^2c^2+{\bs k}^2} } .
\\			\label{4.7b}
&
\bigl( f^\dag_{\mu,\pm}(\bk)\U_\mu^\dag(k)
		\big|_{k_0=\pm\sqrt{m^2c^2+{\bs k}^2} }
\bigr)^\dag
=
 - f_{\mu,\mp}(-\bk)\U_\mu(-k) \big|_{k_0=\mp\sqrt{m^2c^2+{\bs k}^2} }
	\end{align}
	\end{subequations}
\vspace{-4ex}
	\begin{subequations}	\label{4.8}
	\begin{align}
			\label{4.8a}
&  \bigl( f_\mu(k)\U_\mu(k) \bigr)^\dag
=
  f_\mu^\dag(-k)\U_\mu^\dag(-k)
\\			\label{4.8b}
& \bigl( f_\mu^\dag(k)\U_\mu^\dag(k) \bigr)^\dag
=
  {f}(-k)\U_\mu(-k) .
	\end{align}
	\end{subequations}
 From the below presented proof of proposition~\ref{Prop4.1} and the comments
after it, it will be clear that ~\eref{4.7} and~\eref{4.8} should be
accepted. Notice, the above equations mean that $\U_\mu^\dag(k)$ is \emph{not} the
Hermitian conjugate of $\U_\mu(k)$.
	\end{Rem}

	\begin{Proof}
	The proposition was proved for the solutions~\eref{4.1} in
remark~\ref{Rem4.2}. So, below we suppose that $(k,m)\not=(0,0)$.

	The equivalence of~\eref{4.3} and~\eref{4.4} follows from
$\delta(y^2-b^2)=\frac{1}{b}(\delta(y+b)+\delta(y-b))$ for $b>0$.

	Since $\U_\mu$ and $\U_\mu^\dag$ are solutions of the Klein\ndash
Gordon equations~\eref{3.15}, the representations~\eref{4.3} and the
equalities~\eref{4.7} and~\eref{4.8}, with  $\U_\mu(k)$ and $\U_\mu^\dag(k)$
satisfying~\eref{4.5a}, follow from the proved in~\cite{bp-QFTinMP-scalars}
similar proposition~4.1\typeout{!!!!!!!!!!!!!!!!! Bozho, check this citation
before a publication!!!!!!!!!!!!!!!!}
describing the structure of the solutions of the Klein\ndash Gordon equation
in momentum picture.%
\footnote{~%
One can prove the representations~\eref{4.3}, under the conditions~\eref{4.5},
by repeating \emph{mutatis mutandis} the proof
of~\cite[proposition~4.1]{bp-QFTinMP-scalars}\typeout{!!!!!!!!!!!!!!!!!
Bozho, check this citation before a publication!!!!!!!!!!!!!!!!}. From it the
equalities~\eref{4.7} and~\eref{4.8} rigorously follow too.%
}

	At the end, inserting~\eref{4.3} or~\eref{4.4} into~\eref{3.16}, we
obtain the equations~\eref{4.5b}, due to~\eref{4.5a}.
	\end{Proof}

	From the proof of proposition~\ref{Prop4.1}, as well as from the one
of~\cite[proposition~4.1]{bp-QFTinMP-scalars}\typeout{!!!!!!!!!!!!!!!!!
Bozho, check this citation before a publication!!!!!!!!!!!!!!!!}
the next two conclusions can be made. On one hand, the conditions~\eref{4.5a}
ensure that~\eref{4.3} and~\eref{4.4} are solutions of~\eref{2.28} and the
Klein\ndash Gordon equations~\eref{3.15}, while~\eref{4.5b} single out
between them the ones satisfying the Lorenz conditions~\eref{3.16}. On other
hand, since up to a phase factor and, possibly, normalization constant, the
expressions $f_\mu(k)\U_\mu(k)$ and $f^\dag_\mu(k)\U_\mu(k)$ coincide with
the Fourier images of respectively $\tU_\mu(x)$ and $\tU_\mu^\dag(x)$ in
Heisenberg picture, we can write
	\begin{equation}	\label{4.9}
\U_\mu = \int\delta(k^2-m^2c^2) \underline{\U_\mu}(k) \Id^4k
\quad
\tU_\mu (x)
=
\int\delta(k^2-m^2c^2) \underline{\U_\mu}(k)
\e^{\iu\frac{1}{\hbar}(x^\mu-x_0^\mu)k_\mu} \Id^4k
	\end{equation}
and similarly for $\U_\mu^\dag$ (with
\(
\bigl( \underline{\U_\mu^\dag}(k)\bigr)^\dag
= \underline{\U_\mu^\dag}(-k)
\)),
where $\underline{\U_\mu}(k)$ are suitably normalized solutions of~\eref{4.5}.
Therefore, up to normalization factor, the Fourier images of $\tope{\U_\mu}(x)$
and $\U_\mu^\dag(x)$ are
	\begin{equation}	\label{4.10}
\underline{\tope{\U_\mu}} (k)
=
\e^{\iih x_0^\mu k_\mu} \underline{\U_\mu} (k)
\qquad
\underline{\tope{\U_\mu^\dag}} (k)
=
\e^{\iih x_0^\mu k_\mu} \underline{\U_\mu^\dag} (k)
	\end{equation}
where $x_0$ is a fixed point (see Sect.~\ref{Sect2}). So, the momentum
representation of free vector field (satisfying the Lorenz condition) in
Heisenberg picture is an appropriately chosen operator base for the solutions
of the equations~\eref{3.15}--\eref{3.16} and~\eref{2.28} in momentum
picture. This conclusion allows us freely to apply in momentum picture the
existing results concerning that basis in Heisenberg picture.

	As anyone of the equations~\eref{4.5b} is a linear homogeneous
equation with respect to $\U_\mu$ and $\U_\mu^\dag$, each of these equations
has exactly \emph{three linearly independent} solutions, which will be
labeled by indices $s,s',t,\dots$ taking the values~1,~2 and~3,
$s,s',t=1,2,3$.%
\footnote{~%
These indices, which will be referred as the \emph{polarization} or
\emph{spin} indices, have nothing common with the spacial indices
$a,b,\dots=1,2,3$ labeling the spacial components of 4\ndash vectors or
tensors.%
}
Define the operator\ndash valued vectors $\U_{s,(\pm)}^{\mu}(k)$ and
$\U_{s,(\pm)}^{\mu\dag}(k)$ , where $s=1,2,3$ and the index $(\pm)$ indicates
the sign of $k_0=\pm\sqrt{m^2c^2+\bk^2}$ in~\eref{4.5b}, as linearly
independent solutions of the equations
	\begin{align}	\label{4.11}
& k_\mu\big|_{k_0=\pm\sqrt{m^2c^2+\bk^2}} \U_{s,(\pm)}^{\mu}(k) = 0
&& k_\mu\big|_{k_0=\pm\sqrt{m^2c^2+\bk^2}} \U_{s,(\pm)}^{\mu\dag}(k)  = 0.
\intertext{As a consequence of~\eref{4.5a}, they also satisfy the relations}
			\label{4.12}
& [ \U_{s,(\pm)}^{\mu}(k) , \ope{P}_\mu]_{\_} = - k_\nu \U_{s,(\pm)}^{\mu}(k)
&& [ \U_{s,(\pm)}^{\mu\dag}(k) , \ope{P}_\mu]_{\_}
= - k_\nu \U_{s,(\pm)}^{\mu\dag}(k) .
	\end{align}
Since any solution of the first (resp.\ second) equation in~\eref{4.5b} can
be represented as a linear combination of $\U_{s,(\pm)}^{\mu}(k)$ and
$\U_{s,(\pm)}^{\mu\dag}(k)$, $s=1,2,3$, we can rewrite~\eref{4.3} as
(do not sum over $\mu$!)
	\begin{subequations}	\label{4.13}
	\begin{gather}
			\label{4.13a}
\U_\mu
=
\sum_{s} \int\Id^3\bk
	\bigl\{
     f_{\mu,s,+}(\bk) \U_{\mu,s,+}(k)
   + f_{\mu,s,-}(\bk) \U_{\mu,s,-}(k)
	\bigr\}\big|_{k_0=+\sqrt{m^2c^2+{\bk}^2} }
\\			\label{4.13b}
\U_{\mu}^\dag
=
\sum_{s} \int\Id^3\bk
	\bigl\{
  f^\dag_{\mu,s,+}(\bk) \U_{\mu,s,+}^{\dag}(k)
+ f^\dag_{\mu,s,-}(\bk) \U_{\mu,s,-}^{\dag}(k)
	\bigr\} \big|_{k_0=+\sqrt{m^2c^2+{\bk}^2} } ,
	\end{gather}
	\end{subequations}
where $f_{\mu,s,\pm}(\bk)$ and $f^\dag_{\mu,s,\pm}(\bk)$ are some
complex-valued (generalized) functions of $\bk$ such that
	\begin{equation}	\label{4.13-1}
	\begin{split}
\sum_{s}
f_{\mu,s,\pm}(\bk) \U_{\mu,s,\pm}(k)\big|_{k_0=+\sqrt{m^2c^2+{\bk}^2} }
& =
f_{\mu,\pm}(\bk) \U_{\mu}(k)\big|_{k_0=\pm\sqrt{m^2c^2+{\bk}^2} }
\\
\sum_{s}
f^\dag_{\mu, s,\pm}(\bk) \U_{\mu,s,\pm}^{\dag}(k)
	\big|_{k_0=+\sqrt{m^2c^2+{\bk}^2} }
& =
f^\dag_{\mu,\pm}(\bk) \U_\mu^{\dag}(k)\big|_{k_0=\pm\sqrt{m^2c^2+{\bk}^2} } .
	\end{split}
	\end{equation}

	In what follows, we shall need a system of \emph{classical}, not
operator\ndash valued, suitably normalized solutions of the
equations~\eref{4.5b}, which equations reflect the Lorenz
conditions~\eref{3.16}, The idea of their introduction lies in the separation
of the frame\ndash independent properties of a free vector field from the
particular representation of that field in a particular frame of reference.
It will be realized below in Sect.~\ref{Sect5}.

	Consider the equation
	\begin{equation}	\label{4.14}
k^\mu \big|_{k_0=+\sqrt{m^2c^2+{\bk}^2} } v_\mu(\bk) = 0
	\end{equation}
where $v_\mu(\bk)$ is a \emph{classical} 4-vector field (over the
$\bk$-space). This is a \emph{single} linear and homogeneous equation with
respect to \emph{four} functions  $v_\mu(\bk)$, $\mu=0,1,2,3$.
Therefore~\eref{4.14} admits \emph{three} linearly independent solutions.
Define $v_{\mu}^{s}(\bk)$, with $s=1,2,3$ and $\mu=0,1,2,3$, as linearly
independent solutions of
	\begin{equation}	\label{4.15}
k^\mu \big|_{k_0=+\sqrt{m^2c^2+{\bk}^2} } v_\mu^s(\bk) = 0
	\end{equation}
satisfying the conditions
	\begin{equation}	\label{4.16}
v_\mu^s(\bk) v^{\mu,s'}(\bk)
= - \delta^{ss'} (1-\delta_{0m}\delta_{s3})
= - \delta^{ss'}
\times
	\begin{cases}
1			&\text{for } m\not=0 \\
\delta^{1s}+\delta^{2s}	&\text{for } m=0
	\end{cases}
	\end{equation}
where $v^{\mu,s}(\bk):=\eta^{\mu\nu}v_\nu^{s}(\bk)$. In more
details, the relations~\eref{4.16} read
	\begin{subequations}	\label{4.18}
	\begin{align}	\label{4.18a}
& v_\mu^s(\bk) v^{\mu,s'}(\bk) = - \delta^{ss'} \qquad\text{for } m\not=0
\\			\label{4.18b}
& \bigl\{ v_\mu^s(\bk) v^{\mu,s'}(\bk) \bigr\} \big|_{m=0}
 =
	\begin{cases}
- \delta^{ss'} 	&\text{if } (s,s')\not=(3,3) \\
0	 	&\text{if } (s,s')=(3,3)
	\end{cases}
 =
	\begin{cases}
- \delta^{ss'} 	&\text{if } s,s'=1,2 \\
0	 	&\text{otherwise}
	\end{cases}
	\end{align}
	\end{subequations}
The reader may verify by a direct calculation, an explicit solution
of~\eref{4.15}--\eref{4.16}, for $\bk\not=\bs0$, is provided by
	\begin{align}	\label{4.19}
	\begin{split}
& v_a^1(\bk) =e_a^1(\bk) \quad v_a^2(\bk) =e_a^2(\bk)
\\
& v_a^3(\bk)
=
\frac{k_a}{\sqrt{\bk^2}}
\Bigl( \frac{m^2c^2+\bk^2} {m^2c^2+\bk^2\delta_{0m}} \Bigr)^{1/2}
=
\frac{k_a}{\sqrt{\bk^2}} \sqrt{m^2c^2+\bk^2}
\times
	\begin{cases}
\frac{1}{mc}  		&\text{for } m\not=0 \\
\frac{1}{\sqrt{{\pmb{\bk}}^2}}	 	&\text{for } m=0
	\end{cases}
\\
&
v_0^s(\bk)
= - \frac{1}{\sqrt{m^2c^2+\bk^2}} \sum_{a=1}^{3} k^a v_a^s(\bk)
= \Bigl( \frac{\bk^2} {m^2c^2+\bk^2\delta_{0m}} \Bigr)^{1/2} \delta^{3s}
	\end{split}
	\end{align}
where the vectors $e_a^1(\bk)$ and $e_a^2(\bk)$ are such that
	\begin{equation}	\label{4.20}
\bs{e}^{s}(\bk) \cdot \bs{e}^{s'}(\bk) =
\sum_{a}e_a^s(\bk) e_a^{s'}(\bk) = \delta^{ss'}
\quad
\bs{e}^{s}(\bk) \cdot \bs{v}^{3}(\bk) =
\sum_{a}e_a^s(\bk) v_a^{3}(\bk) = 0
\quad \text{for } s,s'=1,2 ,
	\end{equation}
\ie the 3-vectors $\bs{v}^1(\bk)$, $\bs{v}^2(\bk)$ and $\bs{v}^3(\bk)$ form
an orthogonal (orthonormal for $m=0$) basis in the $\field[R]^3$
$\bk$\ndash space with $\bs{v}^3(\bk)$ being proportional to (having the
direction of) $\bk$. Here $\delta_{0m}:=0$ for $m\not=0$ and $\delta_{0m}:=1$
for $m=0$. If $\bk=\bs0$, one can put
	\begin{equation}	\label{4.21}
	\begin{split}
& v_0^s(\bs0)= 0 \quad
v_a^1(\bs0)= e_a^1(\bs0) \quad
v_a^2(\bs0)= e_a^2(\bs0) \quad
\\ &
v_a^3(\bs0)
=
 - (1-\delta_{0m})  \sum_{bc} \varepsilon^{abc} e_b^1(\bs0) e_c^2(\bs0)
=
	\begin{cases}
- \varepsilon^{abc} e_b^1(\bs0) e_c^2(\bs0) 	&\text{for } m\not=0 \\
0	 					&\text{for } m=0
	\end{cases}
	\end{split}
	\end{equation}
with $e_a^1(\bs0)$ and $e_a^2(\bs0)$ satisfying~\eref{4.20} with $\bk=0$; in
particular, one can put $e_a^1(\bs0)=-\delta_a^1$ and
$e_a^2(\bs0)=-\delta_a^2$.%
\footnote{~%
The expression for $v_a^3(\bk)$ in~\eref{4.19} is not defined for $\bk=\bs0$.
Indeed, the limit of $k_a/\sqrt{\bk^2}$, when $\bk\to\bs0$, depends on how
$\bk$ approaches to the zero vector $\bs0$; one can force $k_a/\sqrt{\bk^2}$
to tends to any real number by an appropriate choice of the limiting process
$\bk\to\bs0$. For instance
\(
\lim_{\alpha\to0} \frac{k_a}{\sqrt{\bk^2}}
=
- \frac{ \delta_{1a}+\delta_{2a}\beta }{\sqrt{1+\beta^2}}
\)
if $\bk=\alpha(1,\beta,0)$ for some $\beta\in\field[R]$.%
}

	The solutions~\eref{4.19} and~\eref{4.21}
of~\eref{4.15}--\eref{4.16} satisfy the following relations for summation with
respect to the polarization index ($k_0=\sqrt{m^2c^2+\bk^2}$):
	\begin{subequations}	\label{4.22}
	\begin{align}	\label{4.22a}
\sum_{s=1}^{3} \big\{ v_\mu^s(\bk) v_\nu^s(\bk) \bigr\} \big|_{\bk\not=\bs0}
& =
- \eta_{\mu\nu} + \frac{k_\mu k_\nu}{m^2c^2+\bk^2 \delta_{0m}}
\times
	\begin{cases}
1+\delta_{0m}	&\text{for } \mu=\nu=0		\\
1-\delta_{0m}	&\text{for } \mu=\nu=1,2,3	\\
1	&\text{otherwise}
	\end{cases}
\\			\label{4.22b}
\sum_{s=1}^{3} v_\mu^s(\bs0) v_\nu^s(\bs0)
& =
	\begin{cases}
\delta_{\mu\nu}	&\text{if $m\not=0$ and $\mu,\nu=1,2,3$
		    or if $m=0$ and $\mu,\nu=1,2$}	\\
 0	&\text{otherwise}
	\end{cases}
	\end{align}
	\end{subequations}
\vspace{-4ex}
%
	\begin{align}	\label{4.23}
\sum_{s=1,2}  v_\mu^s(\bk) v_\nu^s(\bk)
& =
	\begin{cases}
\delta_{\mu\nu}	&\text{for } \mu,\nu=1,2	\\
 0	&\text{otherwise}
	\end{cases}
\ \ .
	\end{align}
These equations are simple corollaries of~\eref{4.10}--\eref{4.21} in a
3-frame specified by the unit vectors defined by $v_a^s(\bk)$, $s=1,2,3$. If
we introduce the vectors $\eta^\mu:=(1,0,0,0)$ and
\(
\hat{k}^\mu
:= \bigl( 0,\frac{\bk}{\sqrt{\bk^2}} \bigr)
 = \frac{k^\mu-(k\eta)\eta^\mu}{\sqrt{(k\eta)-k^2}}
\)
with $(k\eta):=k_\mu\eta^\mu$ and $\hat{k}^\mu\big|_{\bk=\bs0}:=(0,0,0,1)$,
then the covariant form of~\eref{4.23} is
	\begin{equation}	\label{4.24}
\sum_{s=1,2}  v_\mu^s(\bk) v_\nu^s(\bk)
=
- \eta_{\mu\nu} + \eta_\mu\eta_\nu - \hat{k}_\mu\hat{k}_\nu ,
	\end{equation}
which, for $m=0$, coincides with~\cite[eq.~(14.53)]{Bjorken&Drell-2}, where
$v_\mu^s(\bk)|_{m=0}$ with $s=1,2$ is denoted by $\varepsilon_\mu(k,s)$, but
it is supposed that $\varepsilon_0(k,s):=0$ and the value $s=3$ is excluded
by definition. Notice, the last multiplier in~\eref{4.22a} can be written in
a covariant form as $(1+\eta_{\mu\nu}\delta_{0m})$.
The easiest way for proving these equalities for $\mu,\nu=1,2,3$ is in a
frame in which $k^1=k^2=0$. The rest of the equations are consequences
of the ones with $\mu,\nu=1,2,3$, \eref{4.19} and~\eref{4.21}.


\section
[Frequency decompositions and creation and annihilation operators]
{Frequency decompositions and \\creation and annihilation operators}
\label{Sect5}

	The frequency decompositions of a free vector field, satisfying the
Lorenz condition, can be introduced similarly to the ones of a free scalar
field~\cite{bp-QFTinMP-scalars}, if~\eref{4.3} is used, or of a free spinor
field~\cite{bp-QFTinMP-spinors}, if~\eref{4.13} is used, \ie if the spin of
the field is taken into account. Respectively, we put:
	\begin{gather}
			\label{5.1}
	\begin{split}
\U_\mu^\pm(k)  :=
	\begin{cases}
f_{\mu,\pm}(\pm \bs k) \U_\mu(\pm k)	& \text{for $k_0\ge0$} \\
0			   			& \text{for $k_0<0$}
	\end{cases}
\quad
\U_{\mu}^{\dag\,\pm}(k) :=
	\begin{cases}
f_{\mu,\pm}^\dag(\pm \bs k) \U_\mu^\dag(\pm k)
	& \text{for $k_0\ge0$} \\
0	& \text{for $k_0<0$}
	\end{cases}
	\end{split}
\\
			\label{5.2}
	\begin{split}
& \U_{\mu,s}^\pm(k)  :=
	\begin{cases}
f_{\mu,s,\pm}(\pm \bs k) \U_{\mu,s,(\pm)}(\pm k) & \text{for $k_0\ge0$} \\
0			   			 & \text{for $k_0<0$}
	\end{cases}
\\
& \U_{\mu,s}^{\dag\,\pm}(k) :=
	\begin{cases}
f_{\mu,s,\pm}^\dag(\pm \bs k) \U_{\mu,s,(\pm)}^\dag(\pm k)
	& \text{for $k_0\ge0$} \\
0	& \text{for $k_0<0$}
	\end{cases}
\ ,
	\end{split}
	\end{gather}
where $k^2=m^2c^2$ and $s=1,2,3$. As a consequence of~\eref{4.13-1}, we have
	\begin{equation}	\label{5.3}
\U_{\mu}^{\pm}(k) 	= \sum_{s=1}^{3} \U_{\mu,s}^{\pm}(k) \quad
\U_{\mu}^{\dag\,\pm}(k)	= \sum_{s=1}^{3} \U_{\mu,s}^{\dag\,\pm}(k) .
	\end{equation}
These operators satisfy the equations
	\begin{equation}	\label{5.4}
\bigl( \U_{\mu}^{\pm}(k) \bigr)^\dag = \U_{\mu}^{\dag\,\mp}(k)  \quad
\bigl( \U_{\mu}^{\dag\,\pm}(k)\bigr)^\dag = \U_{\mu}^{\mp}(k)
	\end{equation}
due to~\eref{4.7}.

	It will be convenient for the following the definitions~\eref{5.1}
and~\eref{5.2} to be specified when $k_0=+\sqrt{m^2c^2+\bk^2}\ge0$:
	\begin{equation}	\label{5.4-1}
	\begin{split}
\U_{\mu}^{\pm}(\bk)
:= \U_{\mu}^{\pm}(k) \big|_{k_0=+\sqrt{m^2c^2+\bk^2}}
= \sum_{s=1}^{3} \U_{\mu,s}^{\pm}(\bk)
\quad
\U_{\mu,s}^{\pm}(\bk) := \U_{\mu,s}^{\pm}(k) \big|_{k_0=+\sqrt{m^2c^2+\bk^2}}
\\
\U_{\mu}^{\dag\,\pm}(\bk)
:= \U_{\mu}^{\dag\,\pm}(k) \big|_{k_0=+\sqrt{m^2c^2+\bk^2}}
= \sum_{s=1}^{3} \U_{\mu,s}^{\dag\,\pm}(\bk)
\quad
\U_{\mu,s}^{\dag\,\pm}(\bk)
	:= \U_{\mu,s}^{\dag\,\pm}(k) \big|_{k_0=+\sqrt{m^2c^2+\bk^2}} .
	\end{split}
	\end{equation}

	Combining~\eref{5.1}--\eref{5.3},~\eref{4.3},~\eref{4.5}
and~\eref{4.13}, we get
	\begin{gather}
			\label{5.5}
\U_\mu = \U_\mu^+ + \U_\mu^-
\qquad
\U_\mu^\dag = \U_\mu^{\dag\,+} + \U_\mu^{\dag\,-}
\\			\label{5.6}
	\begin{split}
\U_\mu^\pm
:= \sum_{s}\int\Id^3\bs k \U_{\mu,s}^\pm (\bk)
 = \int\Id^3\bs k \U_\mu^\pm (\bk)
\quad
\U_\mu^{\dag\,\pm}
:= \sum_{s}\int\Id^3\bs k \U_{\mu,s}^{\dag\,\pm} (\bk)
 = \int\Id^3\bs k \U_\mu^\pm (\bk)
	\end{split}
	\end{gather}
\vspace{-3ex}
	\begin{subequations}	\label{5.7}
	\begin{align}
			\label{5.7a}
 [\U_\mu^\pm(\bk),\ope{P}_\nu]_{\_}
& = \mp k_\nu\big|_{k_0=\sqrt{m^2c^2+{\bs k}^2}} \U_\mu^\pm(\bk)
\\			\label{5.7b}
 [\U_\mu^{\dag\,\pm}(\bk),\ope{P}_\nu]_{\_}
& = \mp k_\nu\big|_{k_0=\sqrt{m^2c^2+{\bs k}^2}} \U_\mu^{\dag\,\pm}(\bk) .
	\end{align}
	\end{subequations}
The equations~\eref{4.5b} are incorporated in the above equalities
via~\eref{5.2}, \eref{5.3} and~\eref{4.13}.

	The physical meaning of the above-introduced operators is a
consequence of~\eref{5.7} and the equations
	\begin{equation}	\label{5.8}
[\U_\mu^\pm(\bk), \ope{Q}]_{\_} =  q \U_\mu^\pm(\bk)
\qquad
[\U_\mu^{\dag\,\pm}(\bk), \ope{Q}]_{\_} = - q \U_\mu^{\dag\,\pm}(\bk)
	\end{equation}
\vspace{-3.4ex}
	\begin{subequations}	\label{5.9}
	\begin{align}
			\label{5.9a}
& [\U_\lambda^\pm(\bk), \ope{M}_{\mu\nu}(x)]_{\_}
=
\{
\mp (x_\mu k_\nu - x_\nu k_\mu)
			\big|_{k_0=\sqrt{m^2c^2+\bk^2}}	\delta_\lambda^\sigma
+ \ih( \delta_\mu^\sigma \eta_{\nu\lambda}
-      \delta_\nu^\sigma \eta_{\mu\lambda} )
\} \U_\sigma^\pm(\bk)
\\			\label{5.9b}
& [\U_\lambda^{\dag\,\pm}(\bk), \ope{M}_{\mu\nu}(x)]_{\_}
=
\{
\mp (x_\mu k_\nu - x_\nu k_\mu)
			\big|_{k_0=\sqrt{m^2c^2+\bk^2}}	\delta_\lambda^\sigma
+ \ih( \delta_\mu^\sigma \eta_{\nu\lambda}
-      \delta_\nu^\sigma \eta_{\mu\lambda} )
\} \U_\sigma^{\dag\,\pm}(\bk) ,
	\end{align}
	\end{subequations}
which follow from~\eref{3.30} and~\eref{3.31}. Recall, these equations are
external to the Lagrangian formalism, but, in general, they agree with the
particle interpretation of the theory.
	Therefore the below\ndash presented results, in particular the
physical interpretation of the creation and annihilation operators, should be
accepted with some reserve. However, after the establishment of the particle
interpretation of the theory (see Sect~\ref{Sect10}), the results of this
section will be confirmed.

	Let  $\ope{X}_p$, $\ope{X}_e$ and $\ope{X}_m$ denote state vectors of
a vector field with fixed respectively 4\ndash momentum $p_\mu$, (total)
charge $e$ and (total) angular momentum $m_{\mu\nu}(x)$, \ie
	\begin{subequations}	\label{5.10}
	\begin{align}
			\label{5.10a}
& \ope{P}_\mu(\ope{X}_p) = p_\mu \ope{X}_p
\\			\label{5.10b}
& \ope{Q}(\ope{X}_e) = e \ope{X}_e
\\			\label{5.10c}
& \ope{M}_{\mu\nu}(x)(\ope{X}_m) = m_{\mu\nu}(x) \ope{X}_m.
	\end{align}
	\end{subequations}
Combining these equations
with~\eref{5.5}--\eref{5.9}, we obtain
	\begin{subequations}	\label{5.11}
	\begin{align}
			\label{5.11a}
& \ope{P}_\mu\bigl( \U_\mu^\pm(\bk) (\ope{X}_p) \bigr)
  = (p_\mu \pm k_\mu) \U_\mu^\pm(\bk) (\ope{X}_p)
	\qquad k_0=\sqrt{m^2c^2+\bk^2}
\\			\label{5.11b}
& \ope{P}_\mu\bigl( \U_\mu^{\dag\,\pm}(\bk) (\ope{X}_p) \bigr)
  = (p_\mu \pm k_\mu) \U_\mu^{\dag\,\pm}(\bk) (\ope{X}_p)
	\qquad k_0=\sqrt{m^2c^2+\bk^2}
	\end{align}
	\end{subequations}
\vspace{-4ex}
	\begin{subequations}	\label{5.12}
	\begin{align}
			\label{5.12a}
	\begin{split}
 \ope{Q}\bigl( \U_\mu (\ope{X}_e) \bigr)  = (e-q) \U_\mu (\ope{X}_e)
\quad
& \ope{Q}\bigl( \U_\mu^\dag (\ope{X}_e) \bigr) = (e+q) \U_\mu^\dag (\ope{X}_e)
	\end{split} 
\displaybreak[1]\\			\label{5.12b}
	\begin{split}
 \ope{Q}\bigl( \U_\mu^\pm (\ope{X}_e) \bigr)
  = (e-q) \U_\mu^\pm (\ope{X}_e)
\quad
& \ope{Q}\bigl( \U_\mu^{\dag\,\pm} (\ope{X}_e) \bigr)
  = (e+q) \U_\mu^{\dag\,\pm} (\ope{X}_e)
	\end{split} 
\displaybreak[1]\\			\label{5.12c}
	\begin{split}
 \ope{Q}\bigl( \U_\mu^\pm(\bk) (\ope{X}_e) \bigr)
  = (e-q) \U_\mu^\pm(\bk) (\ope{X}_e)
\quad
& \ope{Q}\bigl( \U_\mu^{\dag\,\pm}(\bk) (\ope{X}_e) \bigr)
  = (e+q) \U_\mu^{\dag\,\pm}(\bk) (\ope{X}_e)
	\end{split} 
	\end{align}
	\end{subequations}
\vspace{-4ex}
	\begin{subequations}	\label{5.13}
	\begin{align}
			\label{5.13a}
	\begin{split}
\ope{M}_{\mu\nu}(x)
  \bigl( \U_\lambda^\pm(\bk) (\ope{X}_m) \bigr)
=
\{ ( m_{\mu\nu}(x)
&  \pm (x_\mu k_\nu - x_\nu k_\mu) \big|_{k_0=\sqrt{m^2c^2+\bk^2}}
   ) \delta_\lambda^\sigma
\\
& - \ih ( \delta_\mu^\sigma \eta_{\nu\lambda}
      - \delta_\nu^\sigma \eta_{\mu\lambda} )
\}  \U_\sigma^\pm(\bk) (\ope{X}_m)
	\end{split}
\\			\label{5.13b}
	\begin{split}
\ope{M}_{\mu\nu}(x)
 \bigl( \U_\lambda^{\dag\,\pm}(\bk) (\ope{X}_m) \bigr)
=
\{ ( m_{\mu\nu}(x)
& \pm (x_\mu k_\nu - x_\nu k_\mu) \big|_{k_0=\sqrt{m^2c^2+\bk^2}}
   ) \delta_\lambda^\sigma
\\
& - \ih ( \delta_\mu^\sigma \eta_{\nu\lambda}
      - \delta_\nu^\sigma \eta_{\mu\lambda} )
\}
\U_\sigma^{\dag\,\pm}(\bk)(\ope{X}_m) .
	\end{split}
	\end{align}
	\end{subequations}

	The equations~\eref{5.11} (resp.~\eref{5.12}) show that the
eigenvectors of the momentum (resp.\ charge) operator are mapped into such
vectors by the operators $\U_\mu^{\pm}(\bk)$ and $\U_\mu^{\dag\,\pm}(\bk)$
(resp.\
$\U_\mu$, $\U_\mu^{\pm}$, $\U_\mu^{\pm}(\bk)$,
$\U_\mu^\dag$, $\U_\mu^{\dag\,\pm}$, and $\U_\mu^{\dag\,\pm}(\bk)$).
However, by virtue of the equalities~\eref{5.13}, no one of the operators
$\U_\mu$, $\U_\mu^{\pm}$, $\U_\mu^{\pm}(\bk)$,
$\U_\mu^\dag$, $\U_\mu^{\dag\,\pm}$, and $\U_\mu^{\dag\,\pm}(\bk)$
maps an eigenvector of the angular momentum operator into such a vector. The
cause for this fact are the matrices
\(
I_{\mu\nu} :=
[ \delta_\mu^\sigma \eta_{\nu\lambda}
- \delta_\nu^\sigma \eta_{\mu\lambda}
]_{\lambda,\sigma=0}^{3}
\)
appearing in~\eref{5.13}, which generally are  non\ndash diagonal and,
consequently mix the components of the vectors
$\U_\mu(\ope{X}_m)$, $\U_\mu^{\pm}(\ope{X}_m)$,
 $\U_\mu^{\pm}(\bk)(\ope{X}_m)$,
$\U_\mu^\dag(\ope{X}_m)$, $\U_\mu^{\dag\,\pm}(\ope{X}_m)$, and
$\U_\mu^{\dag\,\pm}(\bk)(\ope{X}_m)$ in~\eref{5.13}. Since the matrices
$\pm\ih I_{\mu\nu}$ have a dimension of angular momentum and, obviously,
originate from the `pure spin' properties of vector fields, we shall refer to
them as \emph{spin\ndash mixing angular momentum matrices} or simply as
\emph{spin\ndash mixing matrices}; by definition, the spin\ndash mixing
matrix of the field $\U_\mu$ and its Hermitian conjugate $\U_\mu^\dag$ is
$-\ih I_{\mu\nu}$. More generally, if $\ope{X}$ is a state vector and
\(
 \ope{M}_{\mu\nu}(x)  \bigl( \U_\lambda^\pm(\bk) (\ope{X}) \bigr)
=
\{ l_{\mu\nu}(x) \delta_\lambda^\sigma + s_{\lambda\mu\nu}^\sigma \}
     \U_\sigma^\pm(\bk) (\ope{X})
\)
or
\(
 \ope{M}_{\mu\nu}(x)  \bigl( \U_\lambda^{\dag\,\pm}(\bk) (\ope{X}) \bigr)
=
\{ l_{\mu\nu}^\dag(x) \delta_\lambda^\sigma
 + s_{\lambda\mu\nu}^{\dag\,\sigma} \}
     \U_\sigma^{\dag\,\pm}(\bk) (\ope{X})
\)
where $l_{\mu\nu}$ and $l_{\mu\nu}^\dag$ are some operators
and
$s_{\mu\nu}:=[s_{\lambda\mu\nu}^\sigma]_{\lambda,\sigma=0}^{3}$ and
$s_{\mu\nu}^\dag:=[s_{\lambda\mu\nu}^{\dag\,\sigma}]_{\lambda,\sigma=0}^{3}$
are matrices, not proportional
to the unit matrix $\openone_4$, with operator entries, then we shall say
that the operators $\U_\mu^\pm(\bk)$ or $\U_\mu^{\dag\,\pm}(\bk)$ have,
respectively, spin\ndash mixing (angular momentum) matrices $s_{\mu\nu}$ and
$s_{\mu\nu}^\dag$ relative to the state vector $\ope{X}$; we shall
abbreviate this by saying that the states
$\U_\mu^\pm(\bk) (\ope{X})$ and $\U_\mu^{\dag\,\pm}(\bk) (\ope{X})$
have spin\ndash mixing matrices $s_{\mu\nu}$ and $s_{\mu\nu}^\dag$,
respectively.

	The other additional terms in equations~\eref{5.13} are
$\pm (x_\mu k_\nu - x_\nu k_\mu) \big|_{k_0=\sqrt{m^2c^2+\bk^2}} \openone_4$.
They do not mix the components of
$\U_\mu^{\pm}(\bk)(\ope{X}_m)$ and $\U_\mu^{\dag\,\pm}(\bk)(\ope{X}_m)$. These
terms may be associated with the orbital angular momentum of the state
vectors
$\U_\mu^{\pm}(\bk)(\ope{X}_m)$ and
$\U_\mu^{\dag\,\pm}(\bk)(\ope{X}_m)$.

	Thus, from~\eref{5.11}--\eref{5.13}, the following conclusions can be
made:
\renewcommand{\theenumi}{\roman{enumi}}
	\begin{enumerate}
\item
	The operators $\U_\mu^+(\bk)$ and $\U_\mu^{\dag\,+}(\bk)$
(respectively $\U_\mu^-(\bk)$ and $\U_\mu^{\dag\,-}(\bk)$) increase
(respectively decrease) the state's 4\ndash momentum by the quantity
 $(\sqrt{m^2c^2+\bk^2},\bk)$.

\item
	The operators $\U_\mu$, $\U_\mu^\pm$ and $\U_\mu^\pm(\bk)$
(respectively $\U_\mu^\dag$, $\U_\mu^{\dag\,\pm}$ and
 $\U_\mu^{\dag\,\pm}(\bk)$) decrease (respectively increase) the states'
 charge by $q$.

\item
	The operators $\U_\mu^+(\bk)$ and $\U_\mu^{\dag\,+}(\bk)$
(respectively $\U_\mu^-(\bk)$ and $\U_\mu^{\dag\,-}(\bk)$) increase
 (respectively decrease) the state's orbital angular momentum by
$(x_\mu k_\nu -x_\nu k_\mu)\bigr|_{k_0=\sqrt{m^2c^2+\bk^2}}$.

\item
	The operators $\U_\mu^\pm(\bk)$ and $\U_\mu^{\dag\,\pm}(\bk)$
possess spin\ndash mixing angular momentum matrices $-\ih I_{\mu\nu}$
relative to states with fixed total angular momentum.
       	\end{enumerate}

	In this way, the operators $\U_\mu^\pm(\bk)$ and
$\U_\mu^{\dag\,\pm}(\bk)$ obtain an interpretation of creation and
annihilation operators of particles (quanta) of a vector field, \viz
\\\indent
	(a) the operator $\U_\mu^+(\bk)$ (respectively $\U_\mu^-(\bk)$)
creates (respectively annihilates) a particle with 4\ndash momentum
$(\sqrt{m^2c^2+\bk^2},\bk)$, charge $(-q)$ (resp.\ $(+q)$), orbital angular
momentum $(x_\mu k_\nu -x_\nu k_\mu)\bigr|_{k_0=\sqrt{m^2c^2+\bk^2}}$, and
spin\ndash mixing angular momentum matrix $\bigl(-\ih I_{\mu\nu}\bigr)$ and
\\\indent
	(b) the operator $\U_\mu^{\dag\,+}(\bk)$ (respectively
$\U_\mu^{\dag\,-}(\bk)$) creates (respectively annihilates) a particle with
4\ndash momentum $(\sqrt{m^2c^2+\bk^2},\bk)$, charge $(+q)$ (resp.\ $(-q)$),
orbital angular momentum
$(x_\mu k_\nu -x_\nu k_\mu)\bigr|_{k_0=\sqrt{m^2c^2+\bk^2}}$, and spin\ndash
mixing angular momentum matrix $\bigl(-\ih I_{\mu\nu}\bigr)$.

	Since $\U_{s,(\pm)}^{\mu}(k)$ and $\U_{s,(\pm)}^{\mu,\dag}(k)$ are, by
definition, arbitrary linearly independent solutions of~\eref{4.11}, the
operators $\U_{\mu,s}^{\pm}(\bk)$ and $\U_{\mu,s}^{\dag\,\pm}(\bk)$ are
linearly independent solutions of the operator equations (see~\eref{5.1}
and~\eref{5.5})
	\begin{align}	\label{5.19}
k^\mu\big|_{k_0=\sqrt{m^2c^2+\bk^2}} \U_{\mu,s}^{\pm}(\bk) &= 0
&&
k^\mu\big|_{k_0=\sqrt{m^2c^2+\bk^2}}  \U_{\mu,s}^{\dag\,\pm}(\bk) = 0,
\\\intertext{which, by virtue of~\eref{5.3}, imply}
			\label{5.19-1}
k^\mu\big|_{k_0=\sqrt{m^2c^2+\bk^2}} \U_{\mu}^{\pm}(\bk) &= 0
&&
k^\mu\big|_{k_0=\sqrt{m^2c^2+\bk^2}}  \U_{\mu}^{\dag\,\pm}(\bk) = 0.
	\end{align}
The fact that the classical vector fields $v_\mu^s(\bk)$ are linearly
independent solutions of the same equations (see~\eref{4.15}) gives us the
possibility to separate the invariant, frame\ndash independent,
operator part of the field operators and the frame\ndash dependent their
properties by writing (do not sum over $s$!)
	\begin{gather}
			\label{5.20}
	\begin{split}
\U_{\mu,s}^{\pm} (\bs k)
& :=
\bigl\{ 2 c (2\pi\hbar)^3 \sqrt{m^2c^2+\bk^2} \bigr\}^{-1/2}
					a_s^{\pm} (\bk)  v_\mu^s(\bk)
\\
\U_{\mu,s}^{\dag\,\pm} (\bs k)
& :=
\bigl\{ 2 c (2\pi\hbar)^3 \sqrt{m^2c^2+\bk^2} \bigr\}^{-1/2}
					a_s^{\dag\,\pm} (\bk)  v_\mu^s(\bk) ,
	\end{split}
\\\intertext{which is equivalent to expend $\U_\mu^{\pm}$ and
$\U_\mu^{\dag\,\pm}$ as}
			\label{5.20new}
	\begin{split}
\U_{\mu}^{\pm} (\bs k)
& :=
\bigl\{ 2 c (2\pi\hbar)^3 \sqrt{m^2c^2+\bk^2} \bigr\}^{-1/2}
       \sum_{s=1}^{3} a_s^{\pm} (\bk)  v_\mu^s(\bk)
\\
\U_{\mu}^{\dag\,\pm} (\bs k)
& :=
\bigl\{ 2 c (2\pi\hbar)^3 \sqrt{m^2c^2+\bk^2} \bigr\}^{-1/2}
       \sum_{s=1}^{3}  a_s^{\dag\,\pm} (\bk)  v_\mu^s(\bk) ,
	\end{split}
\\\intertext{where $a_s^{\pm}(\bk), a_s^{\dag\,\pm}(\bk)\colon\Hil\to\Hil$ are
some operators such that}
			\label{5.21}
(a_s^{\pm}(\bk))^\dag = a_s^{\dag\,\mp}(\bk) \quad
(a_s^{\dag\,\pm}(\bk))^\dag = a_s^{\mp}(\bk),
	\end{gather}
due to~\eref{5.4}. The normalization constant
$\bigl\{ 2 c (2\pi\hbar)^3 \sqrt{m^2c^2+\bk^2} \bigr\}^{-1/2}$
is introduced in~\eref{5.20} and~\eref{5.20new} for future convenience (see
Sect.~\ref{Sect6}). The operators $a_s^{+}(\bk)$ and $a_s^{\dag\,+}(\bk)$
(resp.\ $a_s^{-}(\bk)$ and $a_s^{\dag\,-}(\bk)$) will be referred as the
\emph{creation} (resp.\ \emph{annihilation}) \emph{operators} (of the field).

	The physical meaning of the creation and annihilation operators is
similar to the one of $\U_{\mu}^{\pm}$ and $\U_{\mu}^{\dag\,\pm}$. To
demonstrate this, we insert~\eref{5.20new} into~\eref{5.11}--\eref{5.13}
and, using~\eref{4.18}, we get:
	\begin{subequations}	\label{5.22}
	\begin{gather}
			\label{5.22a}
	\begin{split}
 \ope{P}_\mu\bigl( a_s^\pm(\bk) (\ope{X}_p) \bigr)
& = (p_\mu \pm k_\mu) a_s^\pm(\bk) (\ope{X}_p)
\\
 \ope{P}_\mu\bigl( a_s^{\dag\,\pm}(\bk) (\ope{X}_p) \bigr)
& = (p_\mu \pm k_\mu) a_s^{\dag\,\pm}(\bk) (\ope{X}_p)
	\end{split} \Bigg\} \quad
s = \begin{cases}
    1,2,3	&\text{for } m\not=0	\\
    1,2	&\text{for } m=0
	\end{cases}
\displaybreak[1]\\			\label{5.22b}
	\begin{split}
 \ope{Q}\bigl( a_s^\pm(\bk) (\ope{X}_e) \bigr)
& = (e-q) a_s^\pm(\bk) (\ope{X}_e)
\\
 \ope{Q}\bigl( a_s^{\dag\,\pm}(\bk) (\ope{X}_e) \bigr)
& = (e+q) a_s^{\dag\,\pm}(\bk) (\ope{X}_e)
	\end{split} \Bigg\} \quad
s = \begin{cases}
    1,2,3	&\text{for } m\not=0	\\
    1,2	&\text{for } m=0
	\end{cases}
\displaybreak[1]\\			\label{5.22c}
\left.	\begin{split}
\ope{M}_{\mu\nu}(x)  \bigl( a_s^\pm(\bk) (\ope{X}_m) \bigr)
 =  \{ m_{\mu\nu}(x)
& \pm (x_\mu k_\nu - x_\nu k_\mu) \} a_s^\pm(\bk) (\ope{X}_m)
\\
&	+ \ih \sum_{t=1}^3 \sigma_{\mu\nu}^{st}(\bk)
 	a_t^\pm(\bk) (\ope{X}_m)
\\
\ope{M}_{\mu\nu}(x)   \bigl( a_s^{\dag\,\pm}(\bk) (\ope{X}_m) \bigr)
=  \{ m_{\mu\nu}(x)
& \pm (x_\mu k_\nu - x_\nu k_\mu) \} a_s^{\dag\,\pm}(\bk) (\ope{X}_m)
\\
&	+ \ih \sum_{t=1}^3 \sigma_{\mu\nu}^{st}(\bk)
	a_t^{\dag\,\pm}(\bk) (\ope{X}_m)
\\
\sum_{t=1}^{2} \sigma_{\mu\nu}^{3t} a_t^\pm(\bk)  = 0 \quad
\sum_{t=1}^{2} \sigma_{\mu\nu}^{3t} a_t^{\dag\,\pm}(\bk)  &= 0
\qquad\text{if } m=0
	\end{split}\right\}
s = \begin{cases}
    1,2,3	&\text{for } m\not=0	\\
    1,2	&\text{for } m=0
	\end{cases}
	\end{gather}
	\end{subequations}
where
	\begin{equation}	\label{5.23}
	\begin{split}
\sigma_{\mu\nu}^{st}(\bk)
:& = - v^{\lambda,s}(\bk) I_{\lambda\mu\nu}^\varkappa v_\varkappa^{t}(\bk)
= - v^{\lambda,s}(\bk)
(\delta_\mu^\varkappa\eta_{\nu\lambda} -\delta_\nu^\varkappa\eta_{\mu\lambda})
    v_\varkappa^{t}(\bk)
\\
& =  v_\mu^s(\bk) v_\nu^t(\bk) - v_\nu^s(\bk) v_\mu^t(\bk)
= - \sigma_{\nu\mu}^{st}(\bk)
= - \sigma_{\mu\nu}^{ts}(\bk)
	\end{split}
	\end{equation}
with $s,t=1,2,3$.%
\footnote{~%
Notice, the last equations in~\eref{5.22c}, valid only in the massless case,
impose, generally, 6~conditions on any one of the pairs of operators
$a_1^+(\bk)$ and $a_2^+(\bk)$  and
$a_1^-(\bk)$ and $a_2^-(\bk)$.
However, one should not be
worried about that as these conditions originate from the \emph{external} to
the Lagrangian formalism equations~\eref{2.27} and, consequently, they may
not hold in particular theory based on the Lagrangian formalism; see the
paragraph containing equations~\eref{6.30} below.%
}

	As a consequence of~\eref{5.22}, the interpretation of
$a_s^\pm(\bk)$ and $a_s^{\dag\,\pm}(\bk)$ is almost the same as the one of
$\U_\mu^\pm(\bk)$ and $\U_\mu^{\dag\,\pm}(\bk)$, respectively, with an only
change regarding the angular momentum in the massless case.
Equations~\eref{5.22} do not say anything about the dynamical
characteristics of the states $a_3^{+}(\ope{X}_m)$ and
$a_3^{\dag\,+}(\ope{X}_m)$ for a vanishing mass. All this indicates
possible problems with the degree of freedom arising from the value $s=3$ of
the polarization variable in the massless case. Indeed, as we shall see
below in Sect.~\ref{Sect11}, this is an `unphysical' variable; this agrees
with the known fact that a massless vector field possesses only two, not
three, independent components.


\section
[The dynamical variables in terms of creation and annihilation operators]
{The dynamical variables in\\ terms of creation and annihilation operators}
\label{Sect6}

	The Lagrangian~\eref{3.7} (under the Lorenz
conditions~\eref{3.16}), energy-momentum operator~\eref{3.26}, current
operator~\eref{3.27}, and orbital angular momentum operator~\eref{3.28} are
sums of similar ones corresponding to the components  $\U_0$, $\U_1$, $\U_2$
and $\U_3$ of a vector field, considered as independent free scalar fields
(see~\cite{bp-QFTinMP-scalars}). Besides, the operators $\U_\mu^{\pm}(\bk)$
and $\U_\mu^{\dag\,\pm}(\bk)$, defined via~\eref{5.4-1}, \eref{5.1}
and~\eref{5.2}, are up to the normalization constant
$\bigl\{ 2 c (2\pi\hbar)^3 \sqrt{m^2c^2+\bk^2} \bigr\}^{1/2}$ equal to the
creation/annihilation operators for $\U_0$, $\U_1$, $\U_2$ and $\U_3$
(considered as independent scalar fields~\cite{bp-QFTinMP-scalars}).
Consequently, we can automatically write the expressions for the momentum,
charge and angular momentum operators in terms of
 $\U_\mu^{\pm}(\bk)$ and $\U_\mu^{\dag\,\pm}(\bk)$
by applying the results obtained in~\cite{bp-QFTinMP-scalars} for arbitrary
free scalar fields. In this way, we get the following representations for the
momentum operator $\ope{P}_\mu$, charge operator $\tope{Q}$ and orbital
angular momentum operator $\tope{L}_{\mu\nu}$:%
\footnote{~%
The choice of the Lagrangian~\eref{3.7} corresponds to the Lagrangian
$\lindex[\mspace{-5mu}\tope{L}]{}{\prime}$ and energy\ndash momentum operator
$\tope{T}_{\mu\nu}^{(3)}$ in~\cite{bp-QFTinMP-scalars}. So, the below\ndash
presented operators are consequences of the expressions for
 ${ \lindex[\mspace{-3mu}\ope{P}]{}{\prime} }_{\mu}^{(3)}$,
 ${ \lindex[\mspace{-3mu}\ope{Q}]{}{\prime} }_{\mu}^{(3)}$, and
 ${ \lindex[\mspace{-3mu}\ope{L}]{}{\prime} }_{\mu\nu}^{(3)}$,
in \emph{loc.\ cit.}%
}
	\begin{gather}
			\label{6.1}
	\begin{split}
\tope{P}_\mu
=
- \frac{1}{1+\tau(\U)}\int
&  \bigl\{ k_\mu   \bigl( 2 c (2\pi\hbar)^3 k_0 \bigr)
   \bigr\} \big|_{ k_0=\sqrt{m^2c^2+{\bs k}^2} }
\\
& \times \{
\U_\lambda^{\dag\,+}(\bs k)\circ \U^{\lambda,-}(\bs k) +
\U_\lambda^{\dag\,-}(\bs k)\circ \U^{\lambda,+}(\bs k)
  \}
\Id^3\bs k
	\end{split}
\\\displaybreak[1]		\label{6.2}
\tope{Q}
=
- q  \int  \Id^3\bs k  \bigl( 2 c (2\pi\hbar)^3 \sqrt{m^2c^2+\bk^2} \bigr)
\bigl\{
\U_\lambda^{\dag\,+}(\bs k)\circ \U^{\lambda,-}(\bs k) -
\U_\lambda^{\dag\,-}(\bs k)\circ \U^{\lambda,+}(\bs k)
\bigr\}
	\end{gather}
\vspace{-4ex}
	\begin{multline}	\label{6.3}
\tope{L}_{\mu\nu}
=
x_{0\,\mu} \ope{P} - x_{0\,\nu} \ope{P}
-
\frac{\ih}{2(1+\tau(\U))}   \int\Id^3\bk
		\bigl( 2 c (2\pi\hbar)^3 \sqrt{m^2c^2+\bk^2} \bigr)
\\\times
\Bigl\{
\U_\lambda^{\dag\,+}(\bk)
\Bigl( \xlrarrow{ k_\mu \frac{\pd}{\pd k^\nu} }
     - \xlrarrow{ k_\nu \frac{\pd}{\pd k^\mu} } \Bigr)
\circ \U^{\lambda,-}(\bk)
\\  -
\U_\lambda^{\dag\,-}(\bk)
\Bigl( \xlrarrow{ k_\mu \frac{\pd}{\pd k^\nu} }
     - \xlrarrow{ k_\nu \frac{\pd}{\pd k^\mu} } \Bigr)
\circ \U^{\lambda,+}(\bk)
\Bigr\}
\Big|_{	k_0=\sqrt{m^2c^2+\bk^2} } ,
	\end{multline}
where
	\begin{multline}	\label{6.4}
A(\bk) \xlrarrow{ k_\mu\frac{\pd}{\pd k^\nu} } \circ B(\bk)
:=
-
\Bigl( k_\mu\frac{\pd A(\bk)}{\pd k^\nu} \Bigr) \circ B(\bk)
+
\Bigl( A(\bk) \circ k_\mu\frac{\pd B(\bk)}{\pd k^\nu} \Bigr)
\\ =
k_\mu \Bigl(  A(\bk) \xlrarrow{ \frac{\pd}{\pd k^\nu} } \circ B(\bk) \Bigr)
	\end{multline}
for operators $A(\bk)$ and $B(\bk)$ having $C^1$ dependence on $\bk$ (and
common domains).~%
\footnote{~%
More generally, if $\omega\colon\{\Hil\to\Hil\}\to\{\Hil\to\Hil\}$ is a
mapping on the operator space over the system's Hilbert space, we put
 $A\xlrarrow{\omega}\circ B := -\omega(A)\circ B + A\circ \omega(B)$
for any $A,B\colon\Hil\to\Hil$. Usually~\cite{Bjorken&Drell,Itzykson&Zuber},
this notation is used for $\omega=\pd_\mu$.%
}

	Now we shall express these operators in terms of the creation and
annihilation operators $a_s^{\pm}(\bk)$ and $a_s^{\dag\,\pm}(\bk)$,
introduced in Sect.~\ref{Sect5}. For the purpose, one should
substitute~\eref{5.20new} into~\eref{6.1}--\eref{6.3} and to take into
account the normalization conditions~\eref{4.18} for the vectors
$v_\mu^s(\bk)$, $s=1,2,3$. The result of this procedure reads:
	\begin{align}	\label{6.5}
\ope{P}_\mu
& =
\frac{1}{1+\tau(\U)} \sum_{s=1}^{3-\delta_{0m}} \int
  k_\mu |_{ k_0=\sqrt{m^2c^2+{\bs k}^2} }
\{
a_s^{\dag\,+}(\bk)\circ a_s^-(\bk) +
a_s^{\dag\,-}(\bk)\circ a_s^+(\bk)
\}
\Id^3\bk
\displaybreak[1]\\	\label{6.6}
\tope{Q}
& =
q \sum_{s=1}^{3-\delta_{0m}}\int
\{
a_s^{\dag\,+}(\bk)\circ a_s^-(\bk) -
a_s^{\dag\,-}(\bk)\circ a_s^+(\bk)
\} \Id^3\bk
	\end{align}
\vspace{-4ex}
	\begin{multline}	\label{6.7}
\tope{L}_{\mu\nu}
=
\frac{1}{1+\tau(\U)} \sum_{s=1}^{3-\delta_{0m}} \!\! \int\! \Id^3\bk
 ( x_{0\,\mu}k_\nu - x_{0\,\nu}k_\mu ) |_{ k_0=\sqrt{m^2c^2+{\bs k}^2} }
\{
a_s^{\dag\,+}(\bk)\circ a_s^-(\bk) +
a_s^{\dag\,-}(\bk)\circ a_s^+(\bk)
\}
\\
+
\frac{\ih}{1+\tau(\U)}  \sum_{s,s'=1}^{3} \int \Id^3\bk \,
l_{\mu\nu}^{ss'}(\bk)
\bigl\{
 a_s^{\dag\,+}(\bk) \circ a_{s'}^-(\bk) -
 a_s^{\dag\,-}(\bk) \circ a_{s'}^+(\bk)
\bigr\}
\displaybreak[2]\\
+
\frac{\ih}{2(1+\tau(\U))}  \sum_{s=1}^{3-\delta_{0m}} \int \Id^3\bk
\Bigl\{
a_s^{\dag\,+}(\bk)
\Bigl( \xlrarrow{ k_\mu \frac{\pd}{\pd k^\nu} }
     - \xlrarrow{ k_\nu \frac{\pd}{\pd k^\mu} } \Bigr)
\circ a_s^-(\bk)
\\ -
a_s^{\dag\,-}(\bk)
\Bigl( \xlrarrow{ k_\mu \frac{\pd}{\pd k^\nu} }
     - \xlrarrow{ k_\nu \frac{\pd}{\pd k^\mu} } \Bigr)
\circ a_s^+(\bk)
\Bigr\} \Big|_{ k_0=\sqrt{m^2c^2+{\bs k}^2} } \ ,
	\end{multline}
where
	\begin{equation}	\label{6.8}
	\begin{split}
l_{\mu\nu}^{ss'}(\bk)
: & =
\frac{1}{2} v_\lambda^{s}(\bk)
\Bigl( \xlrarrow{ k_\mu \frac{\pd}{\pd k^\nu} }
     - \xlrarrow{ k_\nu \frac{\pd}{\pd k^\mu} } \Bigr)
v^{\lambda,s'}(\bk)
= - l_{\nu\mu}^{ss'}(\bk)
= - l_{\mu\nu}^{s's}(\bk)
\\
& =
- \Bigl( k_\mu \frac{\pd v_\lambda^{s}(\bk)}{\pd k^\nu}
        - k_\nu \frac{\pd v_\lambda^{s}(\bk)}{\pd k^\mu}
  \Bigr) v^{\lambda,s'}(\bk)
\\
& =
+ v_\lambda^{s}(\bk) \Bigl(
  k_\mu \frac{\pd v^{\lambda,s'}(\bk)}{\pd k^\nu}
- k_\nu \frac{\pd v^{\lambda,s'}(\bk)}{\pd k^\mu} \Bigr) .
 	\end{split}
	\end{equation}
with the restriction $k_0=\sqrt{m^2c^2+{\bs k}^2}$ done after the
differentiation (so that the derivatives with respect to $k_0$ vanish).
The last two equalities in~\eref{6.8} are consequences of
(see~\eref{4.18})
	\begin{gather}	\label{6.9}
\frac{\pd v_\mu^{s}(\bk)}{\pd k^\lambda} v^{\mu,s'}(\bk)
+
v_\mu^{s}(\bk) \frac{\pd v^{\mu,s'}(\bk)}{\pd k^\lambda}
= 0 ,
\intertext{so that}		\label{6.10}
v_\lambda^{s}(\bk)
\xlrarrow{ k_\mu \frac{\pd}{\pd k^\nu} }
v^{\lambda,s'}(\bk)
=-2k_\mu \frac{\pd v_\lambda^{s}(\bk)}{\pd k^\nu} v^{\lambda,s'}(\bk)
= 2k_\mu v_\lambda^{s}(\bk) \frac{\pd v^{\lambda,s'}(\bk)}{\pd k^\nu} .
	\end{gather}
Since, $v_\mu^{s}(\bk)$ are real (see~\eref{4.14}--\eref{4.21}), the
definition~\eref{6.8} implies
	\begin{equation}	\label{6.11}
\bigl( l_{\mu\nu}^{ss'}(\bk) \bigr)^\ast
=   l_{\mu\nu}^{ss'}(\bk)
= - l_{\mu\nu}^{s's}(\bk),
	\end{equation}
where the asterisk $\ast$ denotes complex conjugation. So,
$l_{\mu\nu}^{ss'}(\bk)$ are real and, by virtue of~\eref{5.21}, the sums of
the first/second terms in the last integrand in~\eref{6.7} are Hermitian.

	A peculiarity of~\eref{6.5}--\eref{6.7} is the presence in them of
the Kronecker symbol $\delta_{0m}$, which equals to zero in the massive case,
$m\not=0$, and to one in the massless case, $m=0$. Thus, in the massless case,
the modes with polarization $s=3$ do \emph{not} contribute to the momentum
and charge operators, but they \emph{do} contribute to the orbital angular
momentum operator only via the numbers~\eref{6.8} in the second term
in~\eref{6.7}. Notice, in this way the arbitrariness in the
definition~\eref{4.15}--\eref{4.16} of the vectors $v_\mu^s(\bk)$ enters in
the orbital angular momentum operator. This is more or less an expected
conclusion as the last operator is generally a frame\ndash dependent object.

	Let us turn now our attention to the spin angular momentum
operator~\eref{2.15} with density operator $\ope{S}_{\mu\nu}^{\lambda}$
given by~\eref{3.29}. Substituting~\eref{5.5}--\eref{5.6} into~\eref{3.29},
we get the following representation of the spin angular momentum density
operator:
	\begin{multline}	\label{6.24}
\ope{S}_{\mu\nu}^{\lambda}
=
- \frac{\ih c^2}{1+\tau(\U)} \int \Id^3\bk\Id^3\bk'
\bigl\{
k^\lambda \big|_{k_0=\sqrt{m^2c^2+\bk^2}}
\bigl( - \U_\mu^{\dag\,+}(\bk) + \U_\mu^{\dag\,-}(\bk) \bigr) \circ
\bigl(   \U_\nu^{+}(\bk') + \U_\nu^{-}(\bk') \bigr)
\\ -
k^\lambda \big|_{k_0=\sqrt{m^2c^2+\bk^2}}
\bigl( - \U_\nu^{\dag\,+}(\bk) + \U_\nu^{\dag\,-}(\bk) \bigr) \circ
\bigl(   \U_\mu^{+}(\bk')      + \U_\mu^{-}(\bk') \bigr)
\\ -
k^{\prime\lambda} \big|_{k'_0=\sqrt{m^2c^2+{\bk'}^2}}
\bigl(   \U_\mu^{\dag\,+}(\bk) + \U_\mu^{\dag\,-}(\bk) \bigr) \circ
\bigl( - \U_\nu^{+}(\bk') + \U_\nu^{-}(\bk') \bigr)
\\ +
k^{\lambda} \big|_{k_0=\sqrt{m^2c^2+\bk^2}}
\bigl(   \U_\nu^{\dag\,+}(\bk) + \U_\nu^{\dag\,-}(\bk) \bigr) \circ
\bigl( - \U_\mu^{+}(\bk')      + \U_\mu^{-}(\bk') \bigr) .
	\end{multline}

	We shall calculate the spin angular momentum operator in Heisenberg
picture by inserting~\eref{6.24}, with $\lambda=0$, into~\eref{2.15}. Then
one should `move' the operators $\U_\mu^{\pm}$ and $\U_\mu^{\dag\,\pm}$ to
the right of $\ope{U}(x,x_0)
$ according to the relation
	\begin{gather}
			\label{6.25}
\varphi^\varepsilon(\bk)\circ \varphi^{\varepsilon'}(\bk')\circ\ope{U}(x,x_0)
=
\e^{-\iih(x^\mu-x_0^\mu)(\varepsilon k_\mu + \varepsilon' k'_\mu)}
 \ope{U}(x,x_0) \varphi^\varepsilon(\bk) \circ \varphi^{\varepsilon'}(\bk')
	\end{gather}
where $\varepsilon,\varepsilon'=+,-$,
$k_0=\sqrt{m^2c^2+\bk^2}$, $k'_0=\sqrt{m^2c^2+(\bk')^2}$,
\(
\varphi^\varepsilon(\bk)
= \U_\mu^{\varepsilon}(\bk),\U_\mu^{\varepsilon}(\bk),
\)
and $\ope{U}(x,x_0)$ being the operator~\eref{12.112} by means of which the
transition from Heisenberg to momentum picture is performed. This relation
is valid for any $\varphi^\varepsilon(\bk)$ such that
\(
[ \varphi^\varepsilon(\bk) , \ope{P}_\nu ]_{\_}
=
- k_\nu \varphi^\varepsilon(\bk)
\)
--
see~\eref{5.7} and~\cite[eq.~(6.4)]{bp-QFTinMP-scalars}.\typeout{Bozho,
check this reference, please do this Bozho!!!!!!!!!!!!!!!}
At last, performing the integration over $\bs x$, which results in the terms
$(2\pi\hbar)^3\delta^3(\bk\pm\bk')$, and the trivial integration over $\bk'$
by means of the $\delta$\ndash functions $\delta^3(\bk\pm\bk')$, we find:
	\begin{multline}	\label{6.26}
\tope{S}_{\mu\nu}
=
- \frac{\ih }{1+\tau(\U)} \int \Id^3\bk
   \bigl( 2c(2\pi\hbar)^3\sqrt{m^2c^2+\bk^2} \bigr)
\bigl\{
  - \U_\mu^{\dag\,+}(\bk) \circ \U_\nu^{-}(\bk)
  + \U_\mu^{\dag\,-}(\bk) \circ \U_\nu^{+}(\bk)
\\
  + \U_\nu^{\dag\,+}(\bk) \circ \U_\mu^{-}(\bk)
  - \U_\nu^{\dag\,-}(\bk) \circ \U_\mu^{+}(\bk)
\bigr\} .
	\end{multline}
To express $\tope{S}_{\mu\nu}$ via the creation and annihilation operators,
we substitute~\eref{5.20new} into~\eref{6.26} and get
	\begin{equation}	\label{6.27}
\tope{S}_{\mu\nu}
=
  \frac{\ih }{1+\tau(\U)} \sum_{s,s'=1}^{3} \int \Id^3\bk
\sigma_{\mu\nu}^{ss'}(\bk)
\bigl\{
  a_s^{\dag\,+}(\bk) \circ a_{s'}^{-}(\bk)
- a_s^{\dag\,-}(\bk) \circ a_{s'}^{+}(\bk)
\bigr\} ,
	\end{equation}
where the functions
	\begin{equation}	\label{6.28}
\sigma_{\mu\nu}^{st}(\bk)
:= v_\mu^s(\bk) v_\nu^t(\bk) - v_\nu^s(\bk) v_\mu^t(\bk)
 = - \sigma_{\nu\mu}^{st}(\bk)
 = - \sigma_{\mu\nu}^{ts}(\bk)
 = + \sigma_{\nu\mu}^{ts}(\bk)
	\end{equation}
were introduced earlier by~\eref{5.23}.

	From~\eref{6.27}, we observe that $\tope{S}_{\mu\nu}$, generally,
depends on the mode with polarization index $s=3$ even in the massless case.
Evidently, a necessary and sufficient condition for the independence of the
spin angular momentum form this mode is
	\begin{multline}	\label{6.29}
\sum_{s=1,2} \int\Id^3\bk \bigl\{
\bigl( \sigma_{\mu\nu}^{s3}(\bk) a_s^{\dag\,+}(\bk) \bigr) \circ a_3^{-}(\bk)
-
\bigl( \sigma_{\mu\nu}^{s3}(\bk) a_s^{\dag\,-}(\bk) \bigr) \circ a_3^{+}(\bk)
\\ +
a_3^{\dag\,+}(\bk) \circ \bigl( \sigma_{\mu\nu}^{3s}(\bk) a_s^{-}(\bk) \bigr)
-
a_3^{\dag\,-}(\bk) \circ \bigl( \sigma_{\mu\nu}^{3s}(\bk) a_s^{+}(\bk)
\bigr)
\bigr\}
= 0 .
	\end{multline}
In particular,~\eref{6.29} is fulfilled in the massless case if
	\begin{equation}	\label{6.30}
\sum_{s=1}^{2} \sigma_{\mu\nu}^{3s} a_s^\pm(\bk)  = 0 \quad
\sum_{s=1}^{2} \sigma_{\mu\nu}^{3s} a_s^{\dag\,\pm}(\bk)  = 0
\qquad \text{for } m=0 ,
	\end{equation}
which are exactly the conditions appearing in~\eref{5.22c}.%
\footnote{~%
Recall, the equation~\eref{6.30} was derived in Sect.~\ref{Sect5} on the
base of the relations~\eref{3.31}, which are external to the Lagrangian
formalism.%
}
Thus, a strange situation arises: the massless modes with $s=3$ do \emph{not}
contribute to the momentum and charge operators, but they \emph{do}
contribute to the spin and orbital angular momentum operators unless
additional conditions, like~\eref{6.30}, are valid. However, one can prove
that~\eref{6.30} is either equivalent to
$a_s^{\pm}(\bk)=a_s^{\dag\,\pm}(\bk)=0$, if $s=1,2$ and
$(\bk,m)\not=(\bs0,0)$, or it is identically valid, if $(\bk,m)=(\bs0,0)$.%
\footnote{~
Let $\bk\not=\bs0$. Substituting~\eref{6.28}, with $t=3$, into the first
equation in~\eref{6.30} and, then, using~\eref{4.19}, we find
	\begin{equation}	\tag{*}	\label{*}
(k_a e_b^1(\bk) - k_b e_a^1(\bk)) a_1^{\pm}(\bk) +
(k_a e_b^2(\bk) - k_b e_a^2(\bk)) a_2^{\pm}(\bk) = 0
\quad a,b=1,2,3 .
	\end{equation}
Multiplying this equality with $e_a^1(\bk)$ or $e_a^2(\bk)$ and summing over
$a=1,2,3$, we, in view of~\eref{4.20}, get
$k_b a_1^{\pm}(\bk)=k_b a_2^{\pm}(\bk)=0$ for any $b=1,2,3$. Therefore
$a_1^{\pm}(\bk)=a_2^{\pm}(\bk)=0$, as we supposed $\bk\not=0$. If $\bk=0$,
repeating the above method by using~\eref{4.21} for~\eref{4.19}, we get
	\begin{equation}	\tag{**}	\label{**}
(1-\delta_{0m}) \{
(\delta_a^3 e_b^1(\bs0) - \delta_b^3 e_a^1(\bs0))  a_1^{\pm}(\bs0) +
(\delta_a^3 e_b^2(\bs0) - \delta_b^3 e_a^2(\bs0))  a_2^{\pm}(\bs0)
\} = 0
	\end{equation}
instead of~\eref{*}. This equation is identically valid for $m=0$, but for
$m\not=0$ it, by virtue of~\eref{4.20}, implies
$a_1^{\pm}(\bk)=a_2^{\pm}(\bk)=0$. The assertion for the operators
$a_1^{\dag\,\pm}(\bk)$ and $a_2^{\dag\,\pm}(\bk)$ can be proved similarly;
alternatively, it follow from the just proved results and~\eref{5.21}.%
} 
In view of~\eref{6.5}--\eref{6.8} and~\eref{6.27}, this means
that~\eref{6.30} may be valid only for free vector fields with vanishing
momentum, charge and spin and angular momentum operators, which fields are
completely unphysical as they cannot lead to any physically
observable/predictable results. Thus, only for such unphysical massless free
vector fields the Heisenberg relations~\eref{3.31} (for the
Lagrangian~\eref{3.7}) may be valid. For these reasons, the
relations~\eref{3.31} and~\eref{6.30} will not be considered further in the
present work.

	However, the above conclusions do not exclude the validity of the more
general than~\eref{6.30} equation~\eref{6.29}. We shall return on this
problem in Sect.~\ref{Sect11}.

	To get a concrete idea of the spin angular momentum
operator~\eref{6.27}, we shall calculate the quantities~\eref{6.28} for
particular choices of the vectors $e_\mu^1(\bk)$ and $e_\mu^2(\bk)$
in~\eref{4.19}--\eref{4.20}.

	To begin with, we notice that, as a result of the antisymmetry of
$\sigma_{\mu\nu}^{st}(\bk)$ in the superscripts and subscripts, only
$\frac{3(3-1)}{2}\times\frac{4(4-1)}{2}=18$ of all of the $3^2\times4^2=144$
of these quantities are independent. As such we shall choose the ones with
 $(s,t)=(1,2),(2,3),(3,1)$ and
 $(\mu,\nu)=(0,1),(0,2),(0,3),(1,2),(2,3),(3,1)$.

	For $\bk\not=\bs0$, we choose a frame such that
	\begin{equation}	\label{6.31}
\bk=(0,0,0,k^3)\big|_{k^3=\sqrt{\bk^2}\ge0} \quad
e_a^1(\bk) = - \delta_a^1 \quad
e_a^2(\bk) = - \delta_a^2.
	\end{equation}
Let
\(
\Lambda
:= \frac{k^3}{\sqrt{\bk^2}}
\Bigl( \frac{m^2c^2+\bk^2}{m^2c^2+\bk^2\delta_{0m}} \Bigr)^{1/2}
= v^{3,3}(\bk) = - v_3^{3}(\bk) .
\)
The results of a straightforward calculation, by means of~\eref{6.28}, of the
chosen independent quantities $\sigma_{\mu\nu}^{st}(\bk)$ are presented in
table~\ref{Table:SpinCoefficients1}. Similar results for $\bk=\bs0$ in a frame
in which
	\begin{equation}	\label{6.32}
v_0^s(\bs0) = 0 \quad
e_a^1(\bs0) = - \delta_a^1 \quad
e_a^2(\bs0) = - \delta_a^2 \quad
v_a^3(\bs0) = - \delta_a^3 (1-\delta_{0m})
	\end{equation}
are given in table~\ref{Table:SpinCoefficients2}.

	\begin{table}[ht!] 
\caption{\small The quantities~\eref{6.28} for $\bk\not=\bs0$ in the
basis~\eref{6.31}. \vspace{1.2ex} }
\label{Table:SpinCoefficients1}
	\begin{tabular}{c||ccc|ccc}
\hline
\((s,t)\downarrow  \mspace{-4mu}
\overset{(\mu,\nu)}{\xrightarrow{\hphantom{(\mu,\nu)}}}
\)
	&(0,1) &(0,2) &(0,3)
	&(1,2) &(2,3) &(3,1)
\\ \hline
(1,2) 	& 0 & 0 &0
	& 1 & 0 & 0
\\
(2,3)	& 0	& $ \frac{k^3}{\sqrt{m^2c^2+\bk^2}}\Lambda $	& 0
	& 0	& $ \Lambda $					& 0
\\
(3,1)	& $ - \frac{k^3}{\sqrt{m^2c^2+\bk^2}}\Lambda $	& 0	& 0
	& 0	& 0						& $ \Lambda $
\\ \hline
	\end{tabular}
	\end{table}

	\begin{table}[ht!] 
\caption{\small The quantities~\eref{6.28} for $\bk=\bs0$ in the
basis~\eref{6.32}. \vspace{1.2ex} }
\label{Table:SpinCoefficients2}
	\begin{tabular}{c||ccc|ccc}
\hline
\((s,t)\downarrow  \mspace{-4mu}
\overset{(\mu,\nu)}{\xrightarrow{\hphantom{(\mu,\nu)}}}
\)
	&(0,1) &(0,2) &(0,3)
	&(1,2) &(2,3) &(3,1)
\\ \hline
(1,2) 	& 0 & 0 &0
	& 1 & 0 & 0
\\
(2,3)	& 0	& 0  			& 0
	& 0	& $1-\delta_{0m}$ 	& 0
\\
(3,1)	& 0 & 0	& 0
	& 0 & 0	& $1-\delta_{0m}$
\\ \hline
	\end{tabular}
	\end{table}

	Consider now the so-called spin vector(s). Since $\tope{S}_{\mu\nu}$
is antisymmetric in $\mu$ and $\nu$, $\tope{S}_{\mu\nu}=-\tope{S}_{\nu\mu}$,
the spin angular momentum operator has~6 independent components, from which
can be formed two 3\ndash dimensional vectors, \viz
	\begin{equation}	\label{6.33}
\tope{R}_a := \tope{S}_{0a}
\qquad
\tope{S}^{a} := \frac{1}{2} \varepsilon^{abc} \tope{S}^{bc}
	      = \sum_{b,c=1}^{3} \frac{1}{2} \varepsilon^{abc} \tope{S}^{bc}
	\end{equation}
where $\varepsilon^{abc}$ is the 3-dimensional Levi-Civita's symbol (which
equals to +1 (resp.\ -1) if $(a,b,c)$ is an even (resp.\ odd) permutation of
$(1,2,3)$ and to zero otherwise). Defining the cross (vector) product of
3\ndash vectors $\bs A$ and $\bs B$ in Cartesian coordinates by
\(
(\bs A \times \bs B)^a := \varepsilon^{abc} A_b B_c ,
\)
where $A_b=-A^b$ are the covariant Cartesian components of
$\bs A=(A^1,A^2,A^3)$, from~\eref{6.26} we find
	\begin{gather}	\label{6.34}
	\begin{split}
\bs{\tope{R}}
=
- \frac{\ih}{1+\tau(\U)} \int\Id^3\bk
  \bigl( 2c(2\pi\hbar\sqrt{m^2c^2+\bk^2})^3 \bigr)
\bigl\{
& - \U_0^{\dag\,+}(\bk) \circ \bs{\U}^{-}(\bk)
  + \U_0^{\dag\,-}(\bk) \circ \bs{\U}^{+}(\bk)
\\
& + \bs{\U}^{\dag\,+}(\bk) \circ  \U_0^{-}(\bk)
- \bs{\U}^{\dag\,-}(\bk) \circ  \U_0^{+}(\bk)
\bigr\}
	\end{split}
\\
	\begin{split}
\bs{\tope{S}}
=
 \frac{\ih}{1+\tau(\U)} \int\Id^3\bk
  \bigl( 2c(2\pi\hbar\sqrt{m^2c^2+\bk^2})^3 \bigr)
\bigl\{
  \bs{\U}^{\dag\,+}(\bk) \ocross \bs{\U}^{-}(\bk)
- \bs{\U}^{\dag\,-}(\bk) \ocross \bs{\U}^{+}(\bk)
\bigr\} ,
	\end{split}
	\end{gather}
where
\(
\bs{\U}^\pm(\bk)
:=  ( \U^{1,\pm}(\bk),\U^{2,\pm}(\bk),\U^{3,\pm}(\bk) )
:= -( \U_1^{\pm}(\bk),\U_2^{\pm}(\bk),\U_3^{\pm}(\bk) )
\),
\(
\bs\U^{\dag\,\pm}(\bk)
:= -( \U_1^{\dag\,\pm}(\bk),\U_2^{\dag\,\pm}(\bk),\U_3^{\dag\,\pm}(\bk) )
\),
and $\ocross$ means a cross product combined with operator composition, \eg
 $(\bs A\ocross\bs B)^1=A_2\circ B_3-A_3\circ B_2$ for operator\ndash valued
vectors $\bs A$ and $\bs B$.

	To write the spin vectors $\bs{\tope{R}}$ and $\bs{\tope{S}}$ in
terms of creation and annihilation operators, we notice that the vectors
$\bs{e}^s(\bk) : =-(e_1^s(\bk),e_2^s(\bk),e_3^s(\bk))$, $s=1,2,3$, with
$\bs{e}^3(\bk):=\bs{e}^1(\bk)\times\bs{e}^2(\bk)$ form an orthonormal basis
of the $\field[R]^3$ $\bk$\ndash space, such that
(see~\eref{4.19}--\eref{4.21} and do not sum over $s$)
	\begin{equation}	\label{6.35new}
	\begin{split}
& \bs{v}^s(\bk)
:=-(v_1^s(\bk),v_2^s(\bk),v_3^s(\bk))
 = \omega^s(\bk) \bs{e}^s(\bk)
\\
& \omega^1(\bk)=\omega^2(\bk):=1
\quad
\omega^3(\bk)\big|_{\bk\not=\bs0}
:= \frac{m^2c^2+\bk^2}{m^2c^2+\bk^2\delta_{0m}}
\quad
\omega^3(\bs0) :=
	\begin{cases}
1	&\text{for } m\not=0	\\
0	&\text{for } m=0
	\end{cases}
~ .
	\end{split}
	\end{equation}
Then, we have
	\begin{equation*}
	\begin{split}
\frac{1}{2}\varepsilon^{abc} \sigma_{bc}^{st}(\bk)
& = \varepsilon^{abc} v_b^s(\bk) v_c^t(\bk)
= \omega^s(\bk)\omega^t (\bk)  \varepsilon^{abc} e_b^s(\bk) e_c^t(\bk)
= \omega^s(\bk)\omega^t (\bk)
	\bigl( \bs{e}^s(\bk)\times \bs{e}^t(\bk) \bigr)^a
\\
\sigma_{0a}^{st}(\bk) \big|_{\bk\not=\bs0}
& = v_0^s(\bk)v_a^t(\bk) - v_a^s(\bk)v_0^t(\bk)
= - \frac{k^b}{\sqrt{m^2c^2+\bk^2}} \sigma_{ba}^{st}(\bk)
\\ &
= - \frac{k^b}{\sqrt{m^2c^2+\bk^2}}
    \frac{1}{2} \varepsilon_{bac} \varepsilon^{cdf} \sigma_{df}^{st}(\bk)
= + \frac{k^b}{\sqrt{m^2c^2+\bk^2}}
    \varepsilon_{abc} (\bs{v}^s(\bk) \times \bs{v}^t(\bk) )^c
\\
\sigma_{0a}^{st}(\bs0) & = 0 ,
	\end{split}
	\end{equation*}
where~\eref{4.19},~\eref{4.21}, and the equality
\(
\varepsilon_{abc}\varepsilon^{efc}
 = \delta_a^e\delta_b^f- \delta_b^e\delta_a^f
\)
were applied. Therefore, from~\eref{6.33} and~\eref{6.27}, we get
	\begin{align}	\label{6.36}
\tope{R}_a
& =
\frac{\ih}{1+\tau(\U)} \varepsilon_{abc} \int \Id^3\bk
r^b (\bk)
\bigl\{
\bs a^{\dag\,+}(\bk) \ocross \bs a^{-}(\bk) -
\bs a^{\dag\,-}(\bk) \ocross \bs a^{+}(\bk)
\bigr\}^c
\displaybreak[1]\\			\label{6.37}
\bs{\tope{S}}
& =
\frac{\ih}{1+\tau(\U)}  \int \Id^3\bk
\bigl\{
\bs a^{\dag\,+}(\bk) \ocross \bs a^{-}(\bk) -
\bs a^{\dag\,-}(\bk) \ocross \bs a^{+}(\bk)
\bigr\} ,
	\end{align}
where the operator-valued vectors
	\begin{gather}	\label{6.38}
\bs a^\pm(\bk) := \sum_{s=1}^{3} \bs v^s(\bk) a_s^\pm(\bk)
\quad
\bs a^{\dag\,\pm}(\bk) := \sum_{s=1}^{3} \bs v^s(\bk) a_s^{\dag\,\pm}(\bk)
\\\intertext{were introduced, the index $c$ in $\{\cdots\}^c$ means the
$c^{\text{th}}$ component of $\{\cdots\}$, and the function}
			\label{6.39}
r^b(\bk) :=
	\begin{cases}
\frac{k^b}{\sqrt{m^2c^2+\bk^2}}	&\text{for } (\bk,m)\not=(\bs0,0) 	\\
0				&\text{for } (\bk,m)=(\bs0,0)
	\end{cases}
	\end{gather}
takes care of the above-obtained expressions for $\sigma_{0a}^{st}(\bk)$.

	In connection with the particle interpretation of the creation and
annihilation operators, the component $\tope{S}^3$ is of particular interest
as, for $\bk\not=\bs0$, its integrand describes the spin projection on the
direction of the 4\ndash momentum $\bk$. From~\eref{6.37}, we obtain
	\begin{equation}	\label{6.40}
	\begin{split}
\tope{S}^3
 =
\frac{\ih}{1+\tau(\U)}  \int \Id^3\bk
\bigl\{
& a_1^{\dag\,+}(\bk) \circ a_2^{-}(\bk) -
  a_2^{\dag\,+}(\bk) \circ a_1^{-}(\bk)
\\ -
&  a_1^{\dag\,-}(\bk) \circ a_2^{+}(\bk) +
  a_2^{\dag\,-}(\bk) \circ a_1^{+}(\bk)
\bigr\} ,
	\end{split}
	\end{equation}
Since in~\eref{6.40} enter only `mixed' products, like
$a_1^{\dag\,\pm}(\bk) \circ a_2^{\mp}(\bk) $, the states/(anti)particles
created/annihilated by $a_s^{\pm}(\bk)$ and $a_s^{\dag\,\pm}(\bk)$ for
$s=1,2$ do not have definite projection on $\bk$ (for $\bk\not=\bs0$). This
situation can be improved by introducing operators
$b_s^{\pm}(\bk)$ and $b_s^{\dag\,\pm}(\bk)$ such
that~\cite[eq.~(4.28)]{Bogolyubov&Shirkov}
	\begin{equation}	\label{6.41}
	\begin{split}
a_1^{\pm}(\bk) = 	\frac{b_1^{\pm}(\bk) + b_2^{\pm}(\bk)} {\sqrt{2}}
\quad
a_2^{\pm}(\bk) = -\iu\, \frac{b_1^{\pm}(\bk) - b_2^{\pm}(\bk)} {\sqrt{2}}
\quad
a_3^{\pm}(\bk) = b_3^{\pm}(\bk)
\\
a_1^{\dag\,\pm}(\bk)
	=       \frac{b_1^{\dag\,\pm}(\bk) + b_2^{\dag\,\pm}(\bk)} {\sqrt{2}}
\quad
a_2^{\dag\,\pm}(\bk)
	= +\iu\, \frac{b_1^{\dag\,\pm}(\bk) - b_2^{\dag\,\pm}(\bk)} {\sqrt{2}}
\quad
a_3^{\dag\,\pm}(\bk) = b_3^{\dag\,\pm}(\bk) .
	\end{split}
	\end{equation}
In terms of the operators $b_s^{\pm}(\bk)$ and $b_s^{\dag\,\pm}(\bk)$, the
momentum~\eref{6.5}, charge~\eref{6.6} and third spin vector
projection~\eref{6.40} operators take respectively the forms:
	\begin{align}	\label{6.42}
\ope{P}_\mu
& =
\frac{1}{1+\tau(\U)} \sum_{s=1}^{3-\delta_{0m}} \int
  k_\mu |_{ k_0=\sqrt{m^2c^2+{\bs k}^2} }
\{
b_s^{\dag\,+}(\bk)\circ b_s^-(\bk) +
b_s^{\dag\,-}(\bk)\circ b_s^+(\bk)
\}
\Id^3\bk
\displaybreak[1]\\	\label{6.43}
\tope{Q}
& =
q \sum_{s=1}^{3-\delta_{0m}}\int
\{
b_s^{\dag\,+}(\bk)\circ b_s^-(\bk) -
b_s^{\dag\,-}(\bk)\circ b_s^+(\bk)
\} \Id^3\bk
\displaybreak[1]\\	\label{6.44}
\tope{S}^3
& =
\frac{\hbar}{1+\tau(\U)} \sum_{s=1}^{2} \int
(-1)^{s+1}
\bigl\{
b_s^{\dag\,+}(\bk)\circ b_s^-(\bk) -
b_s^{\dag\,-}(\bk)\circ b_s^+(\bk)
\bigr\}
 \Id^3\bk .
	\end{align}
From these formulae is clear that the states (particles) created/annihilated
by $b_s^{\pm}(\bk)$ and $b_s^{\dag\,\pm}(\bk)$ have 4\ndash momentum
$(\sqrt{m^2c^2+\bk^2},\bk)$, charge $\pm q$, and spin projection on the
direction of movement equal to $\pm\hbar\times(1+\tau(\U))^{-1}$ for $s=1,2$
or equal to zero if $s=3$.%
\footnote{~%
One can get rid of the factor $(1+\tau(\U))^{-1}$ by rescaling the operators
$b_s^{\pm}(\bk)$ and $b_s^{\dag\,\pm}(\bk)$ by the factor
$(1+\tau(\U))^{1/2}$.%
}

	We would like now to make a comparison with the expressions for the
dynamical variables in terms of the creation/annihilation operators
$\tilde{a}_{s}^{\pm}(\bk)$ and $\tilde{a}_{s}^{\dag\,\pm}(\bk)$ in (the
momentum representation of) Heisenberg picture of
motion~\cite{Bogolyubov&Shirkov,Bjorken&Drell,Itzykson&Zuber,Roman-QFT}.
As a consequence of~\eref{4.10}, the analogues of the creation/annihilation
operators, defined in terms of the vector field frequency operators
via~\eref{5.1} and~\eref{5.2}, are
	\begin{equation}	\label{6.46}
	\begin{split}
\tU_{\mu,s}^\pm(\bk)
& =
\e^{\pm\frac{1}{\ih} x_0^\mu k_\mu} \U_{\mu,s}^\pm(\bk)
\quad
\tU_{\mu,s}^{\dag\,\pm}(\bk)
=
\e^{\pm\frac{1}{\ih} x_0^\mu k_\mu} \U_{\mu,s}^{\dag\,\pm}(\bk)
\quad
\bigl( k_0=\sqrt{m^2c^2+\bk^2} \bigr)
\\
\tU_{\mu}^\pm(\bk)
& =
\e^{\pm\frac{1}{\ih} x_0^\mu k_\mu} \U_{\mu}^\pm(\bk)
\quad
\tU_{\mu}^{\dag\,\pm}(\bk)
=
\e^{\pm\frac{1}{\ih} x_0^\mu k_\mu} \U_{\mu}^{\dag\,\pm}(\bk)
\quad
( k_0=\sqrt{m^2c^2+\bk^2} )
	\end{split}
	\end{equation}
in Heisenberg picture. Therefore, defining (cf.~\eref{5.20})
	\begin{equation}	\label{6.47}
	\begin{split}
\tU_{\mu,s}^\pm(\bk)
& =:
\bigl\{ 2 c (2\pi\hbar)^3 \sqrt{m^2c^2+\bk^2} \bigr\}^{-1/2}
			\tilde{a}_s^\pm(\bk) v_\mu^{s}(\bk)
\\
\tU_{\mu,s}^{\dag\,\pm}(\bk)
&=:
\bigl\{ 2 c (2\pi\hbar)^3 \sqrt{m^2c^2+\bk^2} \bigr\}^{-1/2}
			\tilde{a}_s^{\dag\,\pm}(\bk) v_\mu^{s}(\bk) ,
	\end{split}
	\end{equation}
we get the creation/annihilation operators in Heisenberg picture as
	\begin{gather}	\label{6.48}
\tilde{a}_s^\pm(\bk)
 = \e^{ \pm\frac{1}{\ih} x_0^\mu k_\mu } a_s^\pm(\bk)
\quad
\tilde{a}_s^{\dag\,\pm}(\bk)
 = e^{ \pm\frac{1}{\ih} x_0^\mu k_\mu } a_s^{\dag\,\pm}(\bk)
\quad
	\bigl( k_0=\sqrt{m^2c^2+\bk^2} \bigr) .
\intertext{Evidently, these operators satisfy the equations}
			\label{6.49}
\bigl( \tilde{a}_s^\pm(\bk) \bigr)^\dag = \tilde{a}_s^{\dag\,\mp}(\bk)
\quad
\bigl( \tilde{a}_s^{\dag\,\pm}(\bk) \bigr)^\dag = \tilde{a}_s^\mp(\bk) ,
	\end{gather}
due to~\eref{5.21}, and have all other properties of their momentum picture
counterparts described in Sect.~\ref{Sect5}.

	The connection~\eref{12.114} is not applicable to the
creation/annihilation operators, as well as to operators in momentum
representation (of momentum picture), \ie to ones depending on the momentum
variable $\bk$. In particular, the reader may verify, by using the results of
sections~\ref{Sect4} and~\ref{Sect5}, the formulae
	\begin{equation}	\label{6.49-1}
	\begin{split}
a_s^\pm(\bk)
& =
\e^{\mp\iih x^\mu k_\mu}
\ope{U}(x,x_0)\circ \tilde{a}_s^\pm(k) \circ \ope{U}^{-1}(x,x_0)
\\
a_s^{\dag\,\pm}(\bk)
& =
\e^{\mp\iih x^\mu k_\mu}
\ope{U}(x,x_0)\circ \tilde{a}_s^{\dag\,\pm}(\bk) \circ \ope{U}^{-1}(x,x_0)
	\end{split}
\ \ \Bigg\} \quad
k_0=\sqrt{m^2c^2+{\bs k}^2} ,
	\end{equation}
from which equations~\eref{6.48} follow for $x=x_0$. (Notice, the right hand
sides of the equations~\eref{6.49-1} are independent of $x$, due to the
Heisenberg relations~\eref{2.28}.)

	From~\eref{6.5},~\eref{6.6},~\eref{6.27}
and~\eref{6.46}--\eref{6.48}, it is clear that all of the obtained
expressions for the momentum, charge and spin angular momentum operators in
terms of the (invariant) creation/annihilation operators remain unchanged in
Heisenberg picture; to obtain a Heisenberg version of these equations, one
has formally to add a tilde over the creation/annihilation operators in
momentum picture. However, this is not the case with the orbital
operator~\eref{6.7} because of the presents of derivatives in the integrands
in~\eref{6.7}. We leave to the reader to prove as exercise that, in terms of
the operators~\eref{6.48}, in~\eref{6.7} the term
$x_{0\,\mu}\ope{P}_\nu - x_{0\,\nu}\ope{P}_\mu$, \ie the first sum in it,
should be deleted and tildes over the creation/annihilation operators must be
added. Correspondingly, equation~\eref{6.7} will read
	\begin{multline}	\label{6.50}
\tope{L}_{\mu\nu}
=
\frac{\ih}{1+\tau(\U)}  \sum_{s,s'=1}^{3} \int \Id^3\bk \,
l_{\mu\nu}^{ss'}(\bk)
\bigl\{
 \tilde{a}_s^{\dag\,+}(\bk) \circ \tilde{a}_{s'}^-(\bk) -
 \tilde{a}_s^{\dag\,-}(\bk) \circ \tilde{a}_{s'}^+(\bk)
\bigr\}
\displaybreak[2]\\
+
\frac{\ih}{2(1+\tau(\U))}  \sum_{s=1}^{3-\delta_{0m}} \int \Id^3\bk
\Bigl\{
\tilde{a}_s^{\dag\,+}(\bk)
\Bigl( \xlrarrow{ k_\mu \frac{\pd}{\pd k^\nu} }
     - \xlrarrow{ k_\nu \frac{\pd}{\pd k^\mu} } \Bigr)
\circ \tilde{a}_s^-(\bk)
\\ -
\tilde{a}_s^{\dag\,-}(\bk)
\Bigl( \xlrarrow{ k_\mu \frac{\pd}{\pd k^\nu} }
     - \xlrarrow{ k_\nu \frac{\pd}{\pd k^\mu} } \Bigr)
\circ \tilde{a}_s^+(\bk)
\Bigr\} \Big|_{ k_0=\sqrt{m^2c^2+{\bs k}^2} } \ .
	\end{multline}


\section
[The field equations in terms of creation and annihilation operators]
{The field equations in terms of creation and\\ annihilation operators}
\label{Sect7}

	For a free vector field (satisfying the Lorenz condition), the
equalities~\eref{3.15}, \eref{3.16}, \eref{2.16} and~\eref{3.26} form a
closed algebraic\ndash functional system of equations for determination of
the field operators $\tU_\mu$ and $\tU_\mu^\dag$. As these operators and the
field's dynamical variables are expressible in terms of the creation and
annihilation operators, it is clear that the mentioned system of equations
can equivalently be represented in terms of creation and annihilation
operators. The derivation of the so\ndash arising system of equations and
some its consequences are the main contents of this section.

	As a result of~\eref{5.5},~\eref{5.6} and~\eref{5.20}
(or~\eref{5.20new}), we have the decompositions
	\begin{equation}	\label{7.1}
	\begin{split}
\U_\mu
& =
\sum_{s=1}^3 \int \Id^3\bk
\bigl( \U_{\mu,s}^{+}(\bk) + \U_{\mu,s}^{-}(\bk) \bigr)
\\&  =
\sum_{s=1}^3 \int \Id^3\bk
\bigl\{ 2 c (2\pi\hbar)^3 \sqrt{m^2c^2+\bk^2} \bigr\}^{-1/2}
\bigl\{ a_{s}^{+}(\bk) + a_{s}^{-}(\bk) \bigr\} v_\mu^s(\bk)
\\
\U_\mu^\dag
& =
\sum_{s=1}^3 \int \Id^3\bk
\bigl( \U_{\mu,s}^{\dag\,+}(\bk) + \U_{\mu,s}^{\dag\,-}(\bk) \bigr)
\\& =
\sum_{s=1}^3 \int \Id^3\bk
\bigl\{ 2 c (2\pi\hbar)^3 \sqrt{m^2c^2+\bk^2} \bigr\}^{-1/2}
\bigl\{ a_{s}^{\dag\,+}(\bk) + a_{s}^{\dag\,-}(\bk) \bigr\} v_\mu^s(\bk) .
	\end{split}
	\end{equation}
The definitions of the quantities entering in these equations ensure the
fulfillment of the equations~\eref{4.5b}. Therefore the
conditions~\eref{4.5a}, which are equivalent to~\eref{5.7}, are the only
restrictions on the operators $a_{s}^{\pm}(\bk)$ and $a_{s}^{\dag\,\pm}(\bk)$.
Inserting~\eref{5.20new} into~\eref{5.7}, multiplying the result by
$v^{\mu,s'}(\bk)$, summing over $\mu$, and applying the normalization
conditions~\eref{4.18}, we obtain:
	\begin{subequations}	\label{7.2}
	\begin{align}
			\label{7.2a}
  [a_s^\pm(\bk),\ope{P}_\mu]_{\_}
& = \mp k_\mu a_s^\pm(\bk)
  = \mp \sum_{t=1}^{3-\delta_{0m}} \int
	q_\mu a_s^\pm(\bk) \delta_{st} \delta^3(\bk-\bs q) \Id^3\bs q
\\			\label{7.2b}
  [a_s^{\dag\,\pm}(\bk),\ope{P}_\mu]_{\_}
& = \mp k_\mu a_s^{\dag\,\pm}(\bk)
  = \mp \sum_{t=1}^{3-\delta_{0m}} \int
	q_\mu a_s^{\dag\,\pm}(\bk) \delta_{st} \delta^3(\bk-\bs q) \Id^3\bs q
\\			\label{7.2c}
s & =
	\begin{cases}
1,2,3	&\text{for } m\not=0	\\
1,2	&\text{for } m=0
	\end{cases}
\qquad
k_0 =\sqrt{m^2c^2+{\bs k}^2}
\qquad
q_0=\sqrt{m^2c^2+{\bs q}^2} .
	\end{align}
	\end{subequations}
Substituting in~\eref{7.2} the equation~\eref{6.5}, with integration
variable $\bs q$ for $\bk$ and summation index $t$ for $s$, we get (do not
sum over $s$!)
	\begin{subequations}	\label{7.3}
	\begin{align}
			\label{7.3a}
	\begin{split}
\sum_{t=1}^{3-\delta_{0m}} \int q_\mu\big|_{q_0=\sqrt{m^2c^2+{\bs q}^2}}
\bigl\{
\bigl[ a_s^{\pm}(\bk)
& ,
a_t^{\dag\,+}(\bs q) \circ a_t^{-}(\bs q)
+
a_t^{\dag\,-}(\bs q) \circ a_t^{+}(\bs q)
\bigr]_{\_}
\\
& \pm (1+\tau(\U)) a_s^{\pm}(\bk) \delta_{st} \delta^3(\bk-\bs q)
\bigr\} \Id^3\bs q = 0
	\end{split}
\\			\label{7.3b}
	\begin{split}
\sum_{t=1}^{3-\delta_{0m}} \int q_\mu\big|_{q_0=\sqrt{m^2c^2+{\bs q}^2}}
\bigl\{
\bigl[ a_s^{\dag\,\pm}(\bk)
& ,
a_t^{\dag\,+}(\bs q) \circ a_t^{-}(\bs q)
+
a_t^{\dag\,-}(\bs q) \circ a_t^{+}(\bs q)
\bigr]_{\_}
\\
& \pm (1+\tau(\U)) a_s^{\dag\,\pm}(\bk) \delta_{st} \delta^3(\bk-\bs q)
\bigr\} \Id^3\bs q = 0 .
	\end{split}
\\			\label{7.3c}
s =
	\begin{cases}
1,2,3	&\text{for } m\not=0	\\
1,2	&\text{for } m=0
	\end{cases}
\ .
	\end{align}
	\end{subequations}
Consequently, the operators $a_s^{\pm}(\bk)$ and $a_s^{\dag\,\pm}(\bk)$ must
be solutions of
	\begin{subequations}	\label{7.4}
	\begin{align}
			\label{7.4a}
	\begin{split}
& \bigl[ a_s^{\pm}(\bk) ,
a_t^{\dag\,+}(\bs q) \circ a_t^{-}(\bs q)
 +
a_t^{\dag\,-}(\bs q) \circ a_t^{+}(\bs q)
\bigr]_{\_}
 \pm (1+\tau(\U)) a_s^{\pm}(\bk) \delta_{st} \delta^3(\bk-\bs q)
= f_{st}^\pm(\bk,\bs q)
	\end{split}
\\			\label{7.4b}
	\begin{split}
& \bigl[ a_s^{\dag\,\pm}(\bk) ,
a_t^{\dag\,+}(\bs q) \circ a_t^{-}(\bs q)
 +
a_t^{\dag\,-}(\bs q) \circ a_t^{+}(\bs q)
\bigr]_{\_}
\pm (1+\tau(\U))  a_s^{\dag\,\pm}(\bk) \delta_{st} \delta^3(\bk-\bs q)
= f_{st}^{\dag\,\pm}(\bk,\bs q)
	\end{split}
\\			\label{7.4c}
& s,t =
	\begin{cases}
1,2,3	&\text{for } m\not=0	\\
1,2	&\text{for } m=0
	\end{cases}
\ ,
	\end{align}
where $f_{st}^{\pm}(\bk,\bs q)$ and $f_{st}^{\dag\,\pm}(\bk,\bs q)$ are
(generalized) functions such that
	\begin{equation}	\label{7.4d}
	\begin{split}
\sum_{t=1}^{3-\delta_{0m}} \int q_\mu\big|_{q_0=\sqrt{m^2c^2+{\bs q}^2}}
f_{st}^{\pm}(\bk,\bs q) \Id^3\bs{q}
= 0
\qquad
\sum_{t=1}^{3-\delta_{0m}} \int q_\mu\big|_{q_0=\sqrt{m^2c^2+{\bs q}^2}}
f_{st}^{\dag\,\pm}(\bk,\bs q) \Id^3\bs{q}
& = 0 .
	\end{split}
	\end{equation}
	\end{subequations}

	Since any solution of the field equations~\eref{3.15}--\eref{3.16}
can be written in the form~\eref{7.1} with $a_s^{\pm}(\bk)$ and
$a_s^{\dag\,\pm}(\bk)$ being solutions of~\eref{7.4} or, equivalently,
of~\eref{7.3}, the system of equations~\eref{7.4} or~\eref{7.3} is equivalent
to the initial system of field equations, consisting of the Klein\ndash
Gordon equations~\eref{3.15} and the Lorenz conditions~\eref{3.16}.%
\footnote{~%
Recall, the Heisenberg relations~\eref{2.28} are incorporated in the
constancy of the field operators and are reflected in~\eref{4.5a}
or~\eref{7.2a}.%
}
In this sense,~\eref{7.4} or~\eref{7.3} is the \emph{system of field
equations (of a vector field satisfying the Lorenz condition) in terms of
creation and annihilation operators in momentum picture}. If we neglect the
polarization indices, this system of equations is identical with the one for
an arbitrary free scalar field, obtained in~\cite{bp-QFTinMP-scalars}. The
reader may also wish to compare~\eref{7.4} with a similar system of field
equations for a free spinor field, found in~\cite{bp-QFTinMP-spinors}.

	It is important to be mentioned, in the massless case, $m\not=0$, the
field equations~\eref{7.4} contain only the polarization modes with $s=1,2$
and, consequently, they \emph{do not impose any restrictions on the operators
 $a_3^{\pm}(\bk)$ and $a_3^{\dag\,\pm}(\bk)$}. We shall comment on this
phenomenon in Sect.~\ref{Sect11}.

	The commutators of the dynamical variables with the momentum operator
can easily be found by means of the field equations~\eref{7.4}. Indeed,
from~\eref{6.5}, \eref{6.6}, and~\eref{6.27} is evident that for the momentum,
charge and spin operators these commutators are expressible as integrals
whose integrands are linear combinations of the terms
	\begin{equation}	\label{7.5}
B_{ss'}^{\mu,\mp}(\bs q)
:=
\sum_{t=1}^{3-\delta_{0m}} \! \int\Id^3\bk
k^\mu \big|_{k_0=\sqrt{m^2c^2+\bk^2}}
[
 a_{s}^{\dag\,\pm}(\bs q)\circ a_{s'}^{\mp}(\bs q)
,
 a_{t}^{\dag\,+}(\bk)\circ a_{t}^{-}(\bk) +
 a_{t}^{\dag\,-}(\bk)\circ a_{t}^{+}(\bk)
]_{\_} ,
	\end{equation}
where
\(
s,s'= \big\{
	\begin{smallmatrix}
1,2,3		&\text{if } m\not=0	\\
1,2\hphantom{,3}&\text{if } m=0
	\end{smallmatrix}
\)
for the momentum and charge operators and $s,s,=1,2,3$ for the spin angular
momentum operator. Applying the identity
 $[A\circ B,C]_{\_}\equiv[A,C]_{\_}\circ B + A\circ[B,C]_{\_}$,
for operators $A$, $B$ and $C$, and~\eref{7.3} (which is equivalent
to~\eref{7.4}), we get (do not sum over $s$ and $s'$!)
	\begin{multline*}
B_{ss'}^{\mu,\mp}(\bs q)
:=
(1+\tau(\U)) \sum_{t=1}^{3-\delta_{0m}} \int\Id^3\bk
k^\mu \big|_{k_0=\sqrt{m^2c^2+\bk^2}}
(\mp \delta_{st} \pm \delta_{s't}) \delta^3(\bk-\bs q)
 a_{s}^{\dag\,\pm}(\bs q)\circ a_{s'}^{\mp}(\bs q)
\\
=
(1+\tau(\U)) q^\mu \big|_{q_0=\sqrt{m^2c^2+{\bs q}^2}}
\sum_{t=1}^{3-\delta_{0m}}
(\mp \delta_{st} \pm \delta_{s't})
 a_{s}^{\dag\,\pm}(\bs q)\circ a_{s'}^{\mp}(\bs q)
	\end{multline*}
for
\(
s,s'= \big\{
	\begin{smallmatrix}
1,2,3		&\text{if } m\not=0	\\
1,2\hphantom{,3}&\text{if } m=0
	\end{smallmatrix}
\).
Obviously, the summation over $t$ in the last expression results in the
multiplier $\pm\delta_{ss} \pm\delta_{ss'}=\mp1 \pm1 \equiv 0$, and hence
	\begin{subequations}	\label{7.6}
	\begin{equation}	\label{7.6a}
B_{ss'}^{\mu,\mp}(\bs q) = 0
\quad\text{for }
s,s'=
	\begin{cases}
1,2,3		&\text{if } m\not=0	\\
1,2\hphantom{,3}&\text{if } m=0
	\end{cases}
\ .
	\end{equation}
Similarly, when $m=0$ and $s$ or $s'$ is equal to 3, we find
	\begin{multline}	\label{7.6b}
B_{ss'}^{\mu,\mp}(\bs q) \big|_{m=0}
=
\pm (1+\tau(\U)) q^\mu \big|_{q_0=\sqrt{m^2c^2+{\bs q}^2}}
\times
	\begin{cases}
  a_{3}^{\dag\,\pm}(\bs q)\circ a_{s'}^{\mp}(\bs q)
					&\text{if $s=3$ and $s'=1,2$}
\\
- a_{s}^{\dag\,\pm}(\bs q)\circ a_{3}^{\mp}(\bs q)
					&\text{if $s=1,2$ and $s'=3$}
	\end{cases}
\\ +
(1+\tau(\U))
\times
	\begin{cases}
[ a_{3}^{\dag\,\pm}(\bs q) , \ope{P}^\mu ]_{\_} \circ a_{s'}^{\mp}(\bs q)
					&\text{if $s=3$ and $s'=1,2$}
\\
  a_{s}^{\dag\,\pm}(\bs q)\circ [ a_{3}^{\mp}(\bs q), \ope{P}^\mu ]_{\_}
					&\text{if $s=1,2$ and $s'=3$}
	\end{cases}
\ .
	\end{multline}
	\end{subequations}
At last, the case $s=s'=3$ is insignificant for us as the quantities
$B_{33}^{\mu,-}(\bs q)$ have a vanishing contribution in~\eref{6.27}
and~\eref{6.7}, due to $\sigma_{33}^{ss'}(\bk)=l_{33}^{ss'}(\bk)=0$
(see~\eref{6.28} and~\eref{6.8}). Now, it is trivial to be seen
that~\eref{6.5}, \eref{6.6}, \eref{6.27}, and~\eref{6.8}, on one hand,
and~\eref{7.6}, on another hand, imply the commutation relations
	\begin{align}
			\label{7.8}
[ \ope{P}_\mu , \ope{P}_\nu ]_{\_} & = 0
\\			\label{7.9}
[ \tope{Q} , \ope{P}_\mu ]_{\_} & = 0
\\			\label{7.10}
[ \tope{S}_{\mu\nu} , \ope{P}_\lambda ]_{\_}
& =
\delta_{0m} \lindex[C]{}{S} {}_{\mu\nu}^{\lambda}
\\			\label{7.11}
[ \tope{L}_{\mu\nu} , \ope{P}_\lambda ]_{\_}
& =
- \ih \{ \eta_{\lambda\mu}\ope{P}_\nu - \eta_{\lambda\nu}\ope{P}_\mu \}
- \delta_{0m}  \lindex[C]{}{L} {}_{\mu\nu}^{\lambda}
	\end{align}
where
	\begin{align}	\label{7.10-1}
	\begin{split}
\lindex[C]{}{S} {}_{\mu\nu}^{\lambda}
& :=
\frac{\ih}{(1+\tau(\U))^2} \sum_{s=1,2} \int \Id^3\bk
\sigma_{\mu\nu}^{3s}(\bk)
\bigl\{
  B_{3s}^{\lambda,-}(\bk) - B_{3s}^{\lambda,+}(\bk)
- B_{s3}^{\lambda,-}(\bk) + B_{s3}^{\lambda,+}(\bk)
\bigr\} \big|_{m=0}
\\
\lindex[C]{}{L} {}_{\mu\nu}^{\lambda}
& :=
\frac{2\ih}{(1+\tau(\U))^2} \sum_{s=1,2} \int \Id^3\bk
l_{\mu\nu}^{3s}(\bk)
\bigl\{
  B_{3s}^{\lambda,-}(\bk) - B_{3s}^{\lambda,+}(\bk)
- B_{s3}^{\lambda,-}(\bk) + B_{s3}^{\lambda,+}(\bk)
\bigr\} \big|_{m=0}
	\end{split}
	\end{align}
and we have also used~\eref{6.28},~\eref{6.8} and the equality
	\begin{multline}	\label{7.12}
\sum_{s=1}^{3-\delta_{0m}} \int \Id^3\bs q
\sum_{t=1}^{3-\delta_{0m}} \int \Id^3\bk
k_\lambda \big|_{k_0=\sqrt{m^2c^2+\bk^2}}
\\ \times
[
      a_{s}^{\dag\,\pm}(\bs q) \xlrarrow{q_\mu\frac{\pd}{\pd q^\mu}}
\circ a_{s}^{\mp}(\bs q)
,
a_{t}^{\dag\,+}(\bk) \circ a_{t}^{-}(\bk) +
a_{t}^{\dag\,-}(\bk) \circ a_{t}^{+}(\bk)
]_{\_}
\\ =
\pm 2 (1+\tau(\U)) \eta_{\lambda\nu}
\sum_{s=1}^{3-\delta_{0m}} \int \Id^3\bk
k_\mu \big|_{k_0=\sqrt{m^2c^2+\bk^2}}
a_{s}^{\dag\,\pm}(\bk) \circ a_{s}^{\mp}(\bk) ,
	\end{multline}
which can be proved analogously to~\eref{7.6a} and which is responsible for
the term proportional to $\ih$ in~\eref{7.11}.

	As in Sect~\ref{Sect6}, the polarization along the 3-momentum,
characterized by $s=3$, is a cause for the appearance of `abnormal' terms
in~\eref{7.10} and~\eref{7.11} in the massless case, $m=0$.%
\footnote{~%
The reader may wish to compare~\eref{7.10} and~\eref{7.11} with similar
relations for a free Dirac field in~\cite{bp-QFTinMP-spinors} or for a free
scalar field in~\cite{bp-QFTinMP-scalars} (with $\ope{S}_{\mu\nu}=0$ in the
last case). The mentioned abnormal terms destroy also the `ordinary'
commutation relation between the total angular momentum
$\ope{M}_{\mu\nu}=\ope{L}_{\mu\nu}+\ope{S}_{\mu\nu}$ and the momentum
operator $\ope{P}_\lambda$; see equation~\eref{7.13} below and,
e.g.,~\cite{Roman-QFT,Ryder-QFT,
Bogolyubov&et_al.-AxQFT,Bogolyubov&et_al.-QFT}.%
}
Combining~\eref{7.10}, \eref{7.11} and~\eref{2.11}, we get the commutator
between the total angular momentum $\ope{M}_{\mu\nu}$ and the momentum
operator $\ope{P}_\lambda$ as
	\begin{equation}	\label{7.13}
[ \ope{M}_{\mu\nu} , \ope{P}_\lambda ]_{\_}
=
- \ih \{ \eta_{\lambda\mu}\ope{P}_\nu - \eta_{\lambda\nu}\ope{P}_\mu \}
+ \delta_{0m}
\{ \lindex[C]{}{S} {}_{\mu\nu}^{\lambda}
 - \lindex[C]{}{L} {}_{\mu\nu}^{\lambda}
\} ,
	\end{equation}
in which the terms mentioned also change the `ordinary' commutation relation
in the massless case. We should also pay attention on the sign before the
constant $\ih$ in~\eref{7.13} which sign agrees with a similar one
in~\eref{2.21} and is opposite to the one, usually, accepted in the
literature~\cite{Roman-QFT,Ryder-QFT,
Bogolyubov&et_al.-AxQFT,Bogolyubov&et_al.-QFT}.

	The expressions for the dynamical variables in momentum picture can
be found from equations~\eref{7.8}--\eref{7.11} and the general
rule~\eref{12.114} with $\ope{U}(x,x_0)$ being the operator~\eref{12.112}.
However, the spin and orbital angular momentum operators in momentum picture
cannot be written, generally, in a closed form for $m=0$, due to the presents
of the terms proportional to $\delta_{0m}$ in~\eref{7.10} and~\eref{7.11}. To
simplify the situation, below it will be supposed that
	\begin{equation}	\label{7.14}
[ \lindex[C]{}{S} {}_{\mu\nu}^{\lambda} , \ope{P}_\varkappa ]_{\_} = 0
\quad
[ \lindex[C]{}{L} {}_{\mu\nu}^{\lambda} , \ope{P}_\varkappa ]_{\_} = 0 .
	\end{equation}
If required for some purpose, the reader may generalize, as an exercise, the
following results on the base of~\eref{12.112} and~\eref{12.114} in a case
if~\eref{7.14} do not hold.

	Since~\eref{7.9}--\eref{7.11},~\eref{7.14} and~\eref{12.112} entail
(see footnote~\ref{CommutativityWithMomentum})
	\begin{align}
			\label{7.15}
[ \tope{Q},\ope{U}(x,x_0) ]_{\_}
& = 0
\\			\label{7.16}
[ \tope{S}_{\mu\nu},\ope{U}(x,x_0) ]_{\_}
& =
\delta_{0m} \iih (x_\lambda- x_{0\,\lambda})
\lindex[C]{}{S} {}_{\mu\nu}^{\lambda} \circ \ope{U}(x,x_0)
\\			\label{7.17}
	\begin{split}
[ \tope{L}_{\mu\nu},\ope{U}(x,x_0) ]_{\_}
& =
- \{ (x_\mu - x_{0\,\mu}) \ope{P}_\nu -
     (x_\nu - x_{0\,\nu}) \ope{P}_\mu \} \circ \ope{U}(x,x_0)
\\ &\hphantom{=}
- \delta_{0m} \iih (x_\lambda - x_{0\,\lambda})
\lindex[C]{}{L} {}_{\mu\nu}^{\lambda} \circ \ope{U}(x,x_0) ,
	\end{split}
	\end{align}
by virtue of
\(
\ope{A}(x)
= \tope{A}(x) - [\tope{A}(x),\ope{U}(x,x_0)]_{\_} \circ \ope{U}^{-1}(x,x_0)
\)
(see~\eref{12.114}), it follows that the charge, spin and orbital angular
momentum operators in \emph{momentum picture} respectively are:
	\begin{align}
			\label{7.18}
\ope{Q} & = \tope{Q}
\\			\label{7.19}
\ope{S}_{\mu\nu}
&= \tope{S}_{\mu\nu}
- \delta_{0m} \iih (x_\lambda - x_{0\,\lambda})
  \lindex[C]{}{S} {}_{\mu\nu}^{\lambda}  .
\\			\label{7.20}
\ope{L}_{\mu\nu}
& =
\tope{L}_{\mu\nu}
+ (x_\mu - x_{0\,\mu}) \ope{P}_\nu -
  (x_\nu - x_{0\,\nu}) \ope{P}_\mu
+ \delta_{0m} \iih (x_\lambda - x_{0\,\lambda})
\lindex[C]{}{L} {}_{\mu\nu}^{\lambda}  .
	\end{align}
Explicitly, by virtue of~\eref{6.7}, the orbital angular momentum operator is
	\begin{multline}	\label{7.21}
\tope{L}_{\mu\nu}
=
\frac{1}{1+\tau(\U)} \sum_{s=1}^{3-\delta_{0m}} \!\! \int\! \Id^3\bk
 ( x_{\mu}k_\nu - x_{\nu}k_\mu ) |_{ k_0=\sqrt{m^2c^2+{\bs k}^2} }
\{
a_s^{\dag\,+}(\bk)\circ a_s^-(\bk) +
a_s^{\dag\,-}(\bk)\circ a_s^+(\bk)
\}
\\
+
\frac{\ih}{1+\tau(\U)}  \sum_{s,s'=1}^{3} \int \Id^3\bk \,
l_{\mu\nu}^{ss'}(\bk)
\bigl\{
 a_s^{\dag\,+}(\bk) \circ a_{s'}^-(\bk) -
 a_s^{\dag\,-}(\bk) \circ a_{s'}^+(\bk)
\bigr\}
\displaybreak[2]\\
+
\frac{\ih}{2(1+\tau(\U))}  \sum_{s=1}^{3-\delta_{0m}} \int \Id^3\bk
\Bigl\{
a_s^{\dag\,+}(\bk)
\Bigl( \xlrarrow{ k_\mu \frac{\pd}{\pd k^\nu} }
     - \xlrarrow{ k_\nu \frac{\pd}{\pd k^\mu} } \Bigr)
\circ a_s^-(\bk)
\\ -
a_s^{\dag\,-}(\bk)
\Bigl( \xlrarrow{ k_\mu \frac{\pd}{\pd k^\nu} }
     - \xlrarrow{ k_\nu \frac{\pd}{\pd k^\mu} } \Bigr)
\circ a_s^+(\bk)
\Bigr\} \Big|_{ k_0=\sqrt{m^2c^2+{\bs k}^2} } \ ,
	\end{multline}

	As we see again, these results differ in the massless case from the
ones, expected from the outcome
of~\cite{bp-QFTinMP-scalars,bp-QFTinMP-spinors}, by terms depending on the
operators $a_{3}^{\pm}(\bk)$ and $a_{3}^{\dag\,\pm}(\bk)$ with polarization
variable $s=3$.


\section{The commutation relations}
\label{Sect8}

	Comparing the field equations~\eref{7.4} with similar ones for an
arbitrary free scalar field, obtained in~\cite{bp-QFTinMP-scalars}, we see
that the only difference between them is that the creation and annihilation
operators depend on the polarization indices in the vector field case, which
indices are missing when scalar fields are concerned. It is a simple
observation, a polarization variable, say $s$, is coupled always to a
momentum variable, say $\bk$, and can be considered as its counterpart. This
allows $s$ and $\bk$ to be treated on equal footing in order that one takes
into account that $\bk\in\field[R]^3$ is a continuous variable, while $s$ is
a discrete one, taking the values $s=1,2,3$ for a massive vector field and
$s=1,2$ for a massless vector field satisfying the Lorenz condition.
Therefore the transformations
	\begin{equation}	\label{8.1}
	\begin{split}
& \bk \mapsto  (s,\bk)  \qquad
\varphi_{0}^{\pm}(\bk) \mapsto a_{s}^{\pm}(\bk) \qquad
\varphi_{0}^{\dag\,\pm}(\bk) \mapsto a_{s}^{\dag\,\pm}(\bk)
\\ &
\int\Id^3\bk \mapsto  \sum_{s=1}^{3-\delta_{0m}} \int\Id^3\bk   \qquad
\delta^3(\bk-\bs q) \mapsto  \delta_{st}\delta^3(\bk-\bs q) ,
	\end{split}
	\end{equation}
where $\varphi_{0}^{\pm}(\bk)$ and $\varphi_{0}^{\dag\,\pm}(\bk)$ are the
creation/annihilation operators for a free scalar field, allow us to transfer
automatically all results regarding the field equations of a free scalar
field to free vector field (satisfying the Lorenz condition). The same
conclusion is, evidently valid and with respect to results in which the
momentum and charge operators are involved.%
\footnote{~%
However, when the angular momentum operator is concerned, one should be quite
careful as the changes~\eref{8.1} will produce~\eref{6.7} with missing the
integral depending on $l_{\mu\nu}^{ss'}(\bk)$. Moreover, the rules~\eref{8.1}
cannot be applied at all to results in which the spin is involved; \eg
they will produce identically vanishing spin angular momentum operator of
free vector fields instead of the expression~\eref{6.27}.%
}
As a particular realization of these assertions, the commutation relations
for a free vector field (satisfying the Lorenz condition) will be considered
below; in other words, a second quantization of such a field will be
performed by their means. The reader can find a motivation for an
introduction of these relations in books
like~\cite{Bogolyubov&Shirkov,Bjorken&Drell,Itzykson&Zuber,Roman-QFT}.

	Before writing the commutation relations for a free vector field
satisfying the Lorenz condition, we would like to state explicitly the
\emph{additional to Lagrangian formalism conditions}, imposed on the field
operators, which \emph{reduce} the field equations~\eref{7.4} to these
relations. As a \emph{first} conditions, it is supposed the (anti)commutators
between all creation and/or annihilation operators to be $c$\ndash numbers,
\ie to be proportional to the identity mapping $\id_\Hil$ of the system's
(field's) Hilbert space $\Hil$ of states. This hypothesis reduces the field
equations~\eref{7.4} to a certain algebraic\ndash functional system of
equations which can be obtained from a similar one for a scalar field,
derived in~\cite{bp-QFTinMP-scalars}, by means of the rules~\eref{8.1}.
As a \emph{second} restriction, it is demanded the last system of equations to
be an identity with respect to the creation and annihilation operators. A
consequence of this restriction is that the (anti)commutators between
creation and/or annihilation operators are uniquely defined as operators
proportional to $\id_\Hil$.%
\footnote{~%
The mentioned system of equations does not give any information about the
operators $a_{s}^{\pm}(\bs0)|_{m=0}$ and $a_{s}^{\dag\,\pm}(\bs0)|_{m=0}$,
which describe massless particles with vanishing 4\ndash momentum and,
possibly, non\ndash vanishing charge and spin. By convention, we assume these
operators to satisfy the same (anti)commutation relations as the
creation/annihilation operators for $(\bk,m)\not=(\bs0,0)$.%
}
At this stage of the theory's development it remains undetermined whether a
vector field should be quantize via commutators or anticommutators; the
Lagrangian~\eref{3.7}, we started off, is insensitive with respect to that
choice. To be achieved a conformity with the experimental data, one should
choose, as a \emph{third} additional restriction, a quantization via
commutators, not via anticommutators.%
\footnote{~%
Equivalently, one may demand a charge symmetry of the theory, the validity of
the spin\ndash statistics theorem, etc.%
}%
$^{\text{,}}$~%
\footnote{~%
This condition can be incorporated in the Lagrangian formalism by a suitable
choice of a Lagrangian. It follows from the Lagrangian~\eref{3.7}, we started
off, if the field considered is neutral/Hermitian. For details, see
Sect.~\ref{Sect12}.%
}

	As a result of the described additional hypotheses, the field
equations~\eref{7.4} reduce to the following system of commutation relations,
which is obtainable from a similar one for a free scalar field, derived
in~\cite{bp-QFTinMP-scalars}, via the changes~\eref{8.1}:
	\begin{align}	\notag
&[a_{s}^{\pm}(\bs k), a_{t}^{\pm}(\bs q) ]_{\_}
	= 0
&&
[a_{s}^{\dag\,\pm}(\bs k), a_{t}^{\dag\,\pm}(\bs q) ]_{\_}
	= 0
\\	\notag
&[a_{s}^{\mp}(\bs k), a_{t}^{\pm}(\bs q) ]_{\_}
	= \pm \tau(\U) \delta_{st} \delta^3(\bs k-\bs q) \id_\Hil
&&
[a_{s}^{\dag\,\mp}(\bs k), a_{t}^{\dag\,\pm}(\bs q) ]_{\_}
	= \pm \tau(\U) \delta_{st} \delta^3(\bs k-\bs q) \id_\Hil
\\	\notag
&[a_{s}^{\pm}(\bs k), a_{t}^{\dag\,\pm}(\bs q) ]_{\_}
	= 0
&&
[a_{s}^{\dag\,\pm}(\bs k), a_{t}^{\pm}(\bs q) ]_{\_}
	= 0
\\	\label{8.2}
&[a_{s}^{\mp}(\bs k), a_{t}^{\dag\,\pm}(\bs q) ]_{\_}
	= \pm \delta_{st} \delta^3(\bs k-\bs q) \id_\Hil
&&
[a_{s}^{\dag\,\mp}(\bs k), a_{t}^{\pm}(\bs q) ]_{\_}
	= \pm \delta_{st} \delta^3(\bs k-\bs q) \id_\Hil
	\end{align}
where, as it was said above, the values of the polarization indices depend on
the mass parameter $m$ according to
	\begin{equation}	\label{8.2-1}
s,t =
	\begin{cases}
1,2,3	&\text{for } m\not=0	\\
1,2	&\text{for } m=0
	\end{cases}
\ ,
	\end{equation}
the zero operator of $\Hil$ is denoted by $0$, and $\tau(\U)$ takes
care of is the field neutral/Hermitian ($\U^\dag=\U$, $\tau(\U)=1$) or
charged/non\ndash Hermitian ($\U^\dag\not=\U$, $\tau(\U)=0$) and ensures
correct commutation relations in the Hermitian case, when
$a_{s}^{\dag\,\pm}(\bk)=a_{s}^{\pm}(\bk)$.

	Let us emphasize once again, the commutation relations~\eref{8.2} are
equivalent to the field equations~\eref{7.4} and, consequently, to the
initial system of equations~\eref{3.15}--\eref{3.16} under the made
hypotheses.  If by some reason one or more of these additional to the
Lagrangian formalism conditions is rejected, the trilinear system of
equations~\eref{7.4}, which is more general than~\eref{8.2}, should be
considered.

	A feature of~\eref{8.2} is that, in the massless case, the operators
$a_{3}^{\pm}(\bk)$ and $a_{3}^{\dag\,\pm}(\bk)$, \ie the polarization modes
with $s=3$ (along the vector $\bk$), do not enter in it and hence, these
operators remain completely arbitrary. In that sense, these modes remain not
`second quantized', \ie the Lagrangian formalism does not give any
information about the (anti)commutation relations between themselves or
between them and other creation and annihilation operators.

	As we have noted in~\cite{bp-QFTinMP-scalars}, the concepts of a
distribution (generalized function) and operator-valued distribution appear
during the derivation of the commutation relations~\eref{8.2}. In
particular, the canonical commutation relations~\eref{8.2} have a sense iff
the commutators of the creation and/or annihilation operators are
operator-valued distributions (proportional to $\id_\Hil$), which is
\emph{not} the case if the fields considered are described via
ordinary operators acting on $\Hil$. These facts point to inherent
contradiction of quantum field theory if the field variables are considered
as operators acting on a Hilbert space. The rigorous mathematical setting
requires the fields variables to be regarded as operator\ndash valued
distributions. However, such a setting is out of the scope of the present
work and the reader is referred to books
like~\cite{Streater&Wightman,Jost,
Bogolyubov&et_al.-AxQFT,Bogolyubov&et_al.-QFT}
for more details and realization of that program. In what follows, the
distribution character of the quantum fields will be encoded in the Dirac's
delta function, which will appear in relations like~\eref{7.4} and~\eref{8.2}.

	As an application of the commutation relations~\eref{8.2}, we shall
calculate the commutators between the components of the spin operator and
between them and the charge operator. For the purpose, we shall apply the
following  commutation relations between quadratic combinations of creation
and/or annihilation operators:
	\begin{equation}	\label{8.3}
	\begin{split}
[ a_{s}^{\dag\,\pm}(\bk)  \circ a_{s'}^{\mp}(\bk)
&,
  a_{t}^{\dag\,\pm}(\bs p)\circ a_{t'}^{\mp} (\bs p) ]_{\_}
=
\{
\mp \delta_{st'} a_{t}^{\dag\,\pm}(\bs p)  \circ a_{s'}^{\mp}(\bk)
\pm \delta_{s't} a_{s}^{\dag\,\pm}(\bs k)  \circ a_{t'}^{\mp}(\bs p)
\} \delta^3(\bk-\bs p)
\\
[ a_{s}^{\dag\,\pm}(\bk)  \circ a_{s'}^{\mp}(\bk)
&,
  a_{t}^{\mp}(\bs p)\circ a_{t'}^{\dag\,\pm} (\bs p) ]_{\_}
=
\{
\mp \delta_{st} a_{s'}^{\mp}(\bs k)  \circ a_{t'}^{\dag\,\pm}(\bs p)
\pm \delta_{s't'} a_{t}^{\mp}(\bs p)  \circ a_{s}^{\dag\,\pm}(\bs k)
\} \delta^3(\bk-\bs p)
\\
[ a_{s}^{\pm}(\bk)  \circ a_{s'}^{\dag\,\mp}(\bk)
&,
  a_{t}^{\pm}(\bs p)\circ a_{t'}^{\dag\,\mp} (\bs p) ]_{\_}
=
\{
\mp \delta_{st'} a_{t}^{\pm}(\bs p) \circ a_{s'}^{\dag\,\mp}(\bk)
\pm \delta_{s't} a_{s}^{\pm}(\bs k) \circ a_{t'}^{\dag\,\mp}(\bs p)
\} \delta^3(\bk-\bs p)
\\
[ a_{s}^{\dag\,\pm}(\bk) \circ a_{s'}^{\mp}(\bk)
&,
  a_{t}^{\dag\,\mp}(\bs p) \circ  a_{t'}^{\pm}  (\bs p) ]_{\_}
\negthickspace = \negthickspace
\tau(\U) \{
\mp \delta_{st} a_{s'}^{\mp}(\bs k) \negthinspace \circ \negthinspace
		a_{t'}^{\pm}(\bs p)
\pm \delta_{s't'} a_{t}^{\dag\,\mp}(\bs p) \negthinspace \circ \negthinspace
		  a_{s}^{\dag\,\pm}(\bs k)
\} \delta^3(\bk-\bs p)
\\
[ a_{s}^{\dag\,\pm}(\bk) \circ a_{s'}^{\mp}(\bk)
&,
  a_{t}^{\pm}(\bs p) \circ a_{t'}^{\dag\,\mp}  (\bs p) ]_{\_}
\negthickspace = \negthickspace
\tau(\U) \{
\mp \delta_{st'} a_{t}^{\pm}(\bs p) \negthinspace \circ \negthinspace
		 a_{s'}^{\mp}(\bs k)
\pm \delta_{s't} a_{s}^{\dag\,\pm}(\bs k) \negthinspace \circ \negthinspace
		 a_{t'}^{\dag\,\mp}(\bs p)
\} \delta^3(\bk-\bs p)
\\
[ a_{s}^{\pm}(\bk) \circ a_{s'}^{\dag\,\mp}(\bk)
&,
  a_{t}^{\mp}(\bs p)  \circ   a_{t'}^{\dag\,\pm}  (\bs p) ]_{\_}
\negthickspace = \negthickspace
\tau(\U) \{
\mp \delta_{st} a_{s'}^{\dag\,\mp}(\bs k) \negthinspace \circ \negthinspace
		a_{t'}^{\dag\,\pm}(\bs p)
\pm \delta_{s't'} a_{t}^{\mp}(\bs p) \negthinspace \circ \negthinspace
		  a_{s}^{\pm}(\bs k)
\} \delta^3(\bk-\bs p)  ,
	\end{split}
	\end{equation}
where the polarization indices  $s,\ s',\ t $, and $t'$ take the values 1, 2
and 3 for $m\not=0$ and 1 and 2 for $m=0$. These equalities are simple
corollaries of the identities
\(
[A,B\circ C]_{\_} = [A,B]_{\_}\circ C + B\circ [A,C]_{\_}
\)
and
\(
[B\circ C,A]_{\_} = [B,A]_{\_}\circ C + B\circ [C,A]_{\_}
\),
applied in this order to the left-hand-sides of~\eref{8.3}, and~\eref{8.2}.

	Applying~\eref{6.27},~\eref{7.19} and~\eref{8.3}, we find:
	\begin{multline}	\label{8.4}
[ \tope{S}_{\mu\nu},\tope{S}_{\varkappa\lambda} ]_{\_}
=
[ \ope{S}_{\mu\nu},\ope{S}_{\varkappa\lambda} ]_{\_} \big|_{m\not=0}
 =
\frac{\hbar^2}{(1+\tau(\U))^2}
\\ \times
\sum_{s,s',t=1}^{3-\delta_{0m}} \int\Id^3\bk
\bigl\{
\bigl( \sigma_{\varkappa\lambda}^{ss'}(\bk) \sigma_{\mu\nu}^{s't}(\bk)
     - \sigma_{\mu\nu}^{ss'}(\bk) \sigma_{\varkappa\lambda}^{s't}(\bk)
\bigr)
\bigl( a_{s}^{\dag\,+}(\bk) \circ a_{t}^{-}(\bk)
     - a_{s}^{\dag\,-}(\bk) \circ a_{t}^{+}(\bk)
\bigr)
\\ +
\tau(\U) \sigma_{\mu\nu}^{ss'}(\bk) \sigma_{\varkappa\lambda}^{s't}(\bk)
\bigl( a_{s}^{-}(\bk) \circ a_{t}^{+}(\bk)
     - a_{s}^{+}(\bk) \circ a_{t}^{-}(\bk)
\bigr)
\\ +
\tau(\U) \sigma_{\mu\nu}^{ts'}(\bk) \sigma_{\varkappa\lambda}^{ss'}(\bk)
\bigl( a_{s}^{\dag\,+}(\bk) \circ a_{t}^{\dag\,-}(\bk)
     - a_{s}^{\dag\,-}(\bk) \circ a_{t}^{\dag\,+}(\bk)
\bigr)
\bigr\}
+ \delta_{0m} \tilde{f}_{\mu\nu\varkappa\lambda}(a_3^{\pm},a_3^{\dag\,\pm})
 ,
	\end{multline}
where
 $\tilde{f}_{\mu\nu\varkappa\lambda}(a_3^{\pm},a_3^{\dag\,\pm})$ is a term
whose integrand is a homogeneous expression in
$a_3^{\pm}(\bk)$ and $a_3^{\dag\,\pm}(\bk)$ and which term is set equal to
zero for $m\not=0$.
	The summation over $s'$ in~\eref{8.4} can be performed explicitly by
means of~\eref{6.28}:
	\begin{gather}
			\label{8.4-1}
\sum_{s'=1}^{3-\delta_{0m}}
\bigl\{
\sigma_{\mu\nu}^{ss'}(\bk) \sigma_{\varkappa\lambda}^{s't}(\bk)
\bigr\}
=
v_{\nu\varkappa}(\bk) v_\mu^s(\bk) v_\lambda^t(\bk)
- (\mu\leftrightarrow\nu) - (\varkappa\leftrightarrow\lambda)
\\			\label{8.4-2}
\sum_{s'=1}^{3-\delta_{0m}}
\bigr\{
   \sigma_{\mu\nu}^{ss'}(\bk) \sigma_{\varkappa\lambda}^{s't}(\bk)
-  \sigma_{\varkappa\lambda}^{ss'}(\bk) \sigma_{\mu\nu}^{s't}(\bk)
\bigr\}
=
- v_{\mu\varkappa}(\bk) \sigma_{\nu\lambda}^{st}(\bk)
- (\mu\leftrightarrow\nu) - (\varkappa\leftrightarrow\lambda) ,
\intertext{where}
			\label{8.4-3}
v_{\mu\nu}(\bk)
:=
\sum_{s=1}^{3-\delta_{0m}} v_{\mu}^s(\bk) v_{\nu}^s(\bk)
	\end{gather}
is given via~\eref{4.22} and~\eref{4.23} and the symbol
$- (\mu\leftrightarrow\nu)$ means that we have to subtract the previous terms
by making the change $\mu\leftrightarrow\nu$, \ie an antisymmetrization over
 $\mu$ and $\nu$ must be performed.

	Similar calculations, based on~\eref{6.27} and~\eref{6.6}, show that%
\footnote{~%
This result can formally be obtained from~\eref{8.4} with
$-\iih(1+\tau(\U)) q\delta^{st}$ for $\sigma_{\varkappa\lambda}^{st}(\bk)$.%
}
	\begin{multline*}
[ \tope{S}_{\mu\nu},\tope{Q} ]_{\_}
=
\frac{\ih q \tau(\U)}{1+\tau(\U)} \sum_{s,t=1}^{3-\delta_{0m}} \int\Id^3\bk
\sigma_{\mu\nu}^{ts}(\bk)
\bigl\{
  a_{s}^{-}(\bk) \circ a_{t}^{+}(\bk)
- a_{s}^{+}(\bk) \circ a_{t}^{-}(\bk)
\\
+ a_{s}^{\dag\,+}(\bk) \circ a_{t}^{\dag\,-}(\bk)
- a_{s}^{\dag\,-}(\bk) \circ a_{t}^{\dag\,+}(\bk)
\bigr\}
+ \delta_{0m} \tilde{f}_{\mu\nu}(a_3^{\pm},a_3^{\dag\,\pm})
,
	\end{multline*}
where
 $\tilde{f}_{\mu\nu}(a_3^{\pm},a_3^{\dag\,\pm})$ is a term
whose integrand is a homogeneous expression in
$a_3^{\pm}(\bk)$ and $a_3^{\dag\,\pm}(\bk)$ and which term is set equal to
zero for $m\not=0$.
Hence, recalling that $q=0$ for a Hermitian field and $\tau(\U)=0$ for a
non\ndash Hermitian one and, consequently, $q \tau(\U)\equiv0$, we get
	\begin{equation}	\label{8.5}
[ \tope{S}_{\mu\nu},\tope{Q} ]_{\_}
=
\delta_{0m} \tilde{f}_{\mu\nu}(a_3^{\pm},a_3^{\dag\,\pm})
	\end{equation}
in Heisenberg picture. Therefore, we obtain
	\begin{equation}	\label{8.6}
[ \ope{S}_{\mu\nu},\ope{Q} ]_{\_}
=
\delta_{0m} f_{\mu\nu}(a_3^{\pm},a_3^{\dag\,\pm})
	\end{equation}
in momentum picture (see~\eref{12.114}). In particular, we have
	\begin{equation}	\label{8.7}
[ \ope{S}_{\mu\nu},\ope{Q} ]_{\_}  = 0
\qquad\text{ if } m\not=0
\text{ or if } m=0 \text{ and } a_3^\pm(\bk) = a_3^{\dag\,\pm}(\bk) = 0 ;
	\end{equation}
so, for a massive vector field, the spin and charge operators commute.

	As other corollary of~\eref{8.2}, we shall establish the equations
	\begin{subequations}	\label{8.8}
	\begin{align}
			\label{8.8a}
& [\U_\lambda, \ope{M}_{\mu\nu}(x,x_0)]_{\_}
=
x_\mu [\U_\lambda ,\ope{P}_\nu]_{\_} - x_\nu [\U_\lambda ,\ope{P}_\mu]_{\_}
+ \ih ( \U_\mu \eta_{\nu\lambda} - \U_\nu \eta_{\mu\lambda} )
\\			\label{8.8b}
& [\U_\lambda^\dag, \ope{M}_{\mu\nu}(x,x_0)]_{\_}
=
  x_\mu [\U_\lambda^\dag ,\ope{P}_\nu]_{\_}
- x_\nu [\U_\lambda^\dag ,\ope{P}_\mu]_{\_}
+ \ih ( \U_\mu^\dag \eta_{\nu\lambda} - \U_\nu^\dag \eta_{\mu\lambda} ) ,
	\end{align}
	\end{subequations}
where $m\not=0$,
which are part of the conditions ensuring the relativistic covariance of the
theory considered~\cite{Bjorken&Drell-2}. The condition $m\not=0$ is
essential one. For $m=0$, additional terms, depending on $a_s^{\pm}(\bk)$ and
$a_s^{\dag\,\pm}(\bk)$, $s=1,2,3$, should be added to the right hand sides
of~\eref{8.8}; these terms are connected with the gauge symmetry of the
massless case --- for some details about that situation for an
electromagnetic field, see~\cite[\S~8.4]{Bjorken&Drell-2}.

	We shall prove the equalities
	\begin{subequations}	\label{8.9}
	\begin{align}
			\label{8.9a}
[\U_\lambda, \ope{S}_{\mu\nu}(x,x_0)]_{\_}
= &
\ih \int\Id^3\bs p
\Bigl\{
\sum_{t=1}^{3} v_\lambda^t(\bs p) v_\mu^t(\bs p)
			\bigl(\U_\nu^+(\bs p) + \U_\nu^-(\bs p) \bigr)
- (\mu\leftrightarrow\nu)
\Bigr\}
\\	\notag
 [\U_\lambda, \ope{L}_{\mu\nu}(x,x_0)]_{\_}
= &
  x_\mu [\U_\lambda ,\ope{P}_\nu]_{\_}
- x_\nu [\U_\lambda ,\ope{P}_\mu]_{\_}
- \ih \int\Id^3\bs p
\Bigl\{
\Bigl( \eta_{\lambda\mu} +
\sum_{t=1}^{3} v_\lambda^t(\bs p) v_\mu^t(\bs p)
\Bigr)
\\			\label{8.9b}
& \times
\bigl(\U_\nu^+(\bs p) + \U_\nu^-(\bs p) \bigr)
- (\mu\leftrightarrow\nu)
\Bigr\}
\qquad \text{ for } m\not=0
	\end{align}
	\end{subequations}
and similar ones with $\U^{\dag}_\lambda$ for $\U_\lambda$, from which the
equations~\eref{8.8} immediately follow, due to~\eref{5.5}, \eref{5.6}
and~\eref{3.20}. Here and below $p_0:=\sqrt{m^2c^2+{\bs p}^2}$ and the symbol
$-(\mu\leftrightarrow\nu)$ means that we have to subtract the previous terms
by making the change $\mu\leftrightarrow\nu$, \ie the previous expression has
to be antisymmetrized relative to $\mu$ and $\nu$.

	Equation~\eref{8.9a} is a simple corollary of~\eref{7.1},~\eref{8.2}
and~\eref{6.28}. To derive~\eref{8.9b}, we substitute~\eref{7.1}
and~\eref{6.7} in its l.h.s.\ and then, after an integration by parts of the
terms proportional to $\frac{\pd a_s^\pm(\bs p)}{\pd p^\nu}$ and
using~\eref{6.5} and~\eref{6.8}, we obtain
	\begin{multline*}
 [\U_\lambda, \ope{L}_{\mu\nu}(x,x_0)]_{\_}
=
  x_\mu [\U_\lambda ,\ope{P}_\nu]_{\_}
- x_\nu [\U_\lambda ,\ope{P}_\mu]_{\_}
+ \ih \sum_{s=1}^{3} \int\Id^3\bs p
			\{ 2c(2\pi\hbar)^3\sqrt{m^2c^2+{\bs p}^2} \}^{1/2}
\\ \times
\Bigl\{
\Bigl( \eta_{\lambda\sigma} +
\sum_{t=1}^{3} v_\lambda^t(\bs p) v_\sigma^t(\bs p)
\Bigr)
p_\mu \frac{\pd v^{\sigma,s}(\bs p)}{\pd p^\nu}
\bigl(a_s^+(\bs p) + a_s^-(\bs p) \bigr)
- (\mu\leftrightarrow\nu)
\Bigr\} .
	\end{multline*}
Since from~\eref{4.22a} with $m\not=0$~%
\footnote{~%
This is the place where the supposition $m\not=0$ is essentially used and the
proof brakes down if $m=0$.%
}
and~\eref{4.14} it follows
	\begin{multline*}
\Bigl( \eta_{\lambda\sigma} +
\sum_{t=1}^{3} v_\lambda^t(\bs p) v_\sigma^t(\bs p)
\Bigr)
p_\mu \frac{\pd v^{\sigma,s}(\bs p)}{\pd p^\nu}
=
\frac{1}{m^2c^2} p_\lambda p_\sigma p_\mu
\frac{\pd v^{\sigma,s}(\bs p)}{\pd p^\nu}
\\ =
- \frac{1}{m^2c^2} p_\lambda p_\mu \eta_{\sigma\nu} v^{\sigma,s}(\bs p)
=
- \Bigl( \eta_{\lambda\mu} +
\sum_{t=1}^{3} v_\lambda^t(\bs p) v_\mu^t(\bs p)
\Bigr)
v_{\nu}^s(\bs p)
\qquad \text{ for } m\not=0,
	\end{multline*}
the last equality implying~\eref{8.9b}, due to~\eref{5.20} and~\eref{5.4-1}.
\QED

	As it was said above, the relations~\eref{8.8} are not valid for
$m=0$ in the theory considered, unless some additional terms are taken into
account. This fact is connected with the quantization method adopted in the
present work for massless vector fields. Other such methods may restore the
validity of~\eref{8.8} for $m=0$; for example, such is the Gupta\ndash
Bleuler quantization of electromagnetic
field~\cite{Bogolyubov&Shirkov,Akhiezer&Berestetskii}, as it is proved
in~\cite[\S~19.1]{Akhiezer&Berestetskii} (for interacting electromagnetic
and spin $\frac{1}{2}$ fields).

	An interesting result is that equation~\eref{8.8}, regardless of the
condition $m\not=0$, implies the relation
	\begin{equation}	\label{8.13}
[ \ope{M}_{\varkappa\lambda} , \ope{M}_{\mu\nu} ]_{\_}
=
- \ih \bigl\{
\eta_{\varkappa\mu}	\ope{M}_{\lambda\nu} -
\eta_{\lambda\mu}	\ope{M}_{\varkappa\nu} -
\eta_{\varkappa\nu}	\ope{M}_{\lambda\mu} +
\eta_{\lambda\nu}	\ope{M}_{\varkappa\mu}
\bigr\} .
	\end{equation}
To prove this, we notice that~\eref{8.8}, in momentum representation in
Heisenberg picture, is equivalent to (see~\eref{6.46})
	\begin{equation}	\label{8.14}
	\begin{split}
[ \tope{U}_\lambda^\pm(\bk) , \tope{M}_{\mu\nu} ]_{\_}
& =
\ih \Bigl(
k_\mu \frac{\pd}{\pd k^\nu} -  k_\nu \frac{\pd}{\pd k^\mu}
\Bigr) \tope{U}_\lambda^\pm(\bk)
+
\ih \bigl(
\tope{U}_\mu^\pm(\bk) \eta_{\nu\lambda} -
\tope{U}_\nu^\pm(\bk) \eta_{\mu\lambda}
\bigr)
\\
[ \tope{U}_\lambda^{\dag\,\pm}(\bk) , \tope{M}_{\mu\nu} ]_{\_}
& =
\ih \Bigl(
k_\mu \frac{\pd}{\pd k^\nu} -  k_\nu \frac{\pd}{\pd k^\mu}
\Bigr) \tope{U}_\lambda^{\dag\,\pm}(\bk)
+
\ih \bigl(
\tope{U}_\mu^{\dag\,\pm}(\bk) \eta_{\nu\lambda} -
\tope{U}_\nu^{\dag\,\pm}(\bk) \eta_{\mu\lambda}
\bigr) ,
	\end{split}
	\end{equation}
where $k_0=\sqrt{m^2c^2+\bk^2}$. Now, applying~\eref{6.3},~\eref{6.26} and
the identity
$[A\circ B,C]_{\_} = A\circ[B,C]_{\_} + [A,C]_{\_}\circ B$, one can prove,
after a trivial but long calculations, that
	\begin{equation}	\label{8.15}
	\begin{split}
[ \tope{S}_{\varkappa\lambda} , \tope{M}_{\mu\nu} ]_{\_}
& =
- \ih \bigl\{
\eta_{\varkappa\mu}	\tope{S}_{\lambda\nu} -
\eta_{\lambda\mu}	\tope{S}_{\varkappa\nu} -
\eta_{\varkappa\nu}	\tope{S}_{\lambda\mu} +
\eta_{\lambda\nu}	\tope{S}_{\varkappa\mu}
\bigr\}
\\
[ \tope{L}_{\varkappa\lambda} , \tope{M}_{\mu\nu} ]_{\_}
& =
- \ih \bigl\{
\eta_{\varkappa\mu}	\tope{L}_{\lambda\nu} -
\eta_{\lambda\mu}	\tope{L}_{\varkappa\nu} -
\eta_{\varkappa\nu}	\tope{L}_{\lambda\mu} +
\eta_{\lambda\nu}	\tope{L}_{\varkappa\mu}
\bigr\} ,
	\end{split}
	\end{equation}
from where equation~\eref{8.13} follows in Heisenberg picture as
 $\tope{M}_{\mu\nu} = \tope{L}_{\mu\nu} + \tope{S}_{\mu\nu}$. Notice, this
derivation of~\eref{8.13} demonstrates that~\eref{8.13}  is a consequence of
the validity of~\eref{8.8} regardless of the fulfillment of the commutation
relations~\eref{8.2}. Similarly, equations~\eref{8.11} imply~\eref{8.12}
regardless of the validity of~\eref{8.2}.

	By virtue of the identity
 $[A,B\circ C]_{\_}=[A,B]_{\_}\circ C + B\circ [A,C]_{\_}$, the relations
	\begin{equation}	\label{8.10}
[a_s^{\pm}(\bk),\ope{Q}]_{\_} = q a_s^{\pm}(\bk)
\quad
[a_s^{\dag\,\pm}(\bk),\ope{Q}]_{\_} = - q a_s^{\dag\,\pm}(\bk)
\qquad
s = \big\{
	\begin{smallmatrix}
1,2,3		&\text{if } m\not=0	\\
1,2\hphantom{,3}&\text{if } m=0
	\end{smallmatrix}
	\end{equation}
are trivial corollaries from~\eref{6.6} and the commutation
relations~\eref{8.2}. From here and~\eref{7.1}, we get
	\begin{equation}	\label{8.11}
[\U_\mu, \ope{Q}]_{\_} =  q \U_\mu + \delta_{0m}(\cdots)
\quad
[\U_\mu^\dag, \ope{Q}]_{\_} = - q \U_\mu^\dag  + \delta_{0m}(\cdots)_\dag ,
	\end{equation}
with $(\cdots)$ and $(\cdots)_\dag$ denoting expressions which are linear and
homogeneous in $a_3^{\pm}(\bk)$ and $a_3^{\dag\,\pm}(\bk)$. So, the
equations~\eref{3.30} are consequences of the Lagrangian formalism under
consideration if $m\not=0$ or if $m=0$ and
$a_3^{\pm}(\bk)=a_3^{\dag\,\pm}(\bk)=0$. Moreover, the relations~\eref{8.10}
entail the commutativity of bilinear functions/functionals of
 $a_s^{\pm}(\bk)$ and $a_s^{\dag\,\pm}(\bk)$,
\(
s = \big\{
	\begin{smallmatrix}
1,2,3		&\text{if } m\not=0	\\
1,2\hphantom{,3}&\text{if } m=0
	\end{smallmatrix}
,
\)
with the charge operator $\ope{Q}$. In particular, we have
(see~\eref{6.5}--\eref{6.7} and~\eref{6.27}):
	\begin{equation}	\label{8.12}
	\begin{split}
&
[\ope{P}_\mu, \ope{Q}]_{\_} = 0 \quad
[\ope{Q}, \ope{Q}]_{\_} = 0
\\
&
[\ope{S}_{\mu\nu}, \ope{Q}]_{\_} = \delta_{0m}(\cdots)
\quad
[\ope{L}_{\mu\nu}, \ope{Q}]_{\_} =  \delta_{0m}(\cdots)
\quad
[\ope{M}_{\mu\nu}, \ope{Q}]_{\_} =  \delta_{0m}(\cdots) .
	\end{split}
	\end{equation}
So, if $m\not=0$ or if $m=0$ and $a_3^{\pm}(\bk)=a_3^{\dag\,\pm}(\bk)=0$, the
spin, orbital and total angular momentum operators commute with the charge
operator, as it is stated by~\eref{8.7}; the momentum and charge operators
always commute.


\section {Vacuum and normal ordering}
\label{Sect9}

	For a general motivation regarding the introduction of the concepts of
vacuum and normal ordering, the reader is referred, e.g.,
to~\cite{Bogolyubov&Shirkov,Roman-QFT,Bjorken&Drell} (see
also~\cite{bp-QFTinMP-scalars,bp-QFTinMP-spinors}). Below we shall
concentrate on their formal aspects in an extend enough for the purposes of
the present work.
	\begin{Defn}	\label{Defn9.1}
The vacuum of a free vector field $\U$ (satisfying the Lorenz condition) is
its physical state that contains no particles and has vanishing 4\ndash
momentum, (total) charge and (total) angular momentum. It is described by a
state vector, denoted by $\ope{X}_0$ (in momentum picture) and called also
the vacuum (of the field), such that:
	\begin{subequations}	\label{9.1}
	\begin{align}
				\label{9.1a}
& \ope{X}_0 \not= 0
\\				\label{9.1b}
& \ope{X}_0 = \tope{X}_0
\\				\label{9.1c}
& a_s^-(\bs k) (\ope{X}_0)
= a_s^{\dag\,-}(\bs k) (\ope{X}_0) = 0
\\				\label{9.1d}
& \langle\ope{X}_0|\ope{X}_0\rangle = 1
	\end{align}
	\end{subequations}
where $\langle\cdot | \cdot\rangle\colon\Hil\times\Hil\to\field[C]$ is the
(Hermitian) scalar product of system's (field's) Hilbert space of states and
$s=1,2,3$.
	\end{Defn}

	It is known, this definition is in contradiction with the expressions
for the dynamical variables, obtained in Sect.~\ref{Sect6}, as the latter
imply infinite, instead of vanishing, characteristics for the vacuum;
see~\eref{6.5}--\eref{6.7}, \eref{6.27} and~\eref{8.2}. To overcome this
problem, one should redefine the dynamical variables of a free vector field
via the so\ndash called \emph{normal ordering} of operator products
(compositions) of creation and/or annihilation operators. In short, this
procedure, when applied to free vector fields (satisfying the Lorenz
condition), says
that~\cite{Bogolyubov&Shirkov,Bjorken&Drell,Roman-QFT,Wick}:\\
	\indent	(i) The Lagrangian and the field's dynamical variables,
obtained from it and containing the field operators $\U_\mu$ and
$\U_\mu^\dag$, should be written in terms of the creation and annihilation
operators via~\eref{5.1}--\eref{5.6}.\\
	\indent	(ii) Any composition (product) of creation and/or
annihilation operators, possibly appearing under some integral sign(s), must
be changed so that all creation operators to stand to the left of all
annihilation operators.%
\footnote{~%
The relative order of the creation/annihilation operators is insignificant as
they commute according to~\eref{8.2}.%
}

	The just described procedure is known as \emph{normal ordering  (of
products)} and the result of its application on some operator is called its
normal form; in particular, its application on a product of creation and/or
annihilation operators is called their normal product. The mapping assigning
to an operator its normal form, obtained from it according to the above
procedure, will be denoted by $\ope{N}$ and it is called \emph{normal
ordering operator} and its action on a product of creation and/or
annihilation operators is defined according to the rule~(ii) given above. The
action of $\ope{N}$ on polynomials or convergent power series of creation
and/or annihilation operators is extended by linearity. Evidently, the order
of the creation and/or annihilation operators in some expression does not
influenced the result of the action of $\ope{N}$ on it. The dynamical
variables after normal ordering are denoted by the same symbols as before
it.

	It should be noticed, the normal ordering procedure, as introduced
above, concerns all degrees of freedom, \ie the ones involved in the field
equations~\eref{7.4} and the operators
$a_3^{\pm}(\bk)|_{m=0}$ and $a_3^{\dag\,\pm}(\bk)|_{m=0}$, in the massless
case. This simplifies temporary the consideration of massless vector fields,
but does not remove the problems it contains -- see Sect.~\ref{Sect11}.
Moreover, in Sect.~\ref{Sect11} arguments will be presented that the
afore\ndash given definition of normal ordering agrees with the description
of electromagnetic field and that the mentioned operators should anticommute
with the other ones. In principle, one can consider the normal ordering
operation only for creation and annihilation operators involved in the field
equations~\eref{7.4}, \ie by excluding its action on the operators
$a_3^{\pm}(\bk)|_{m=0}$ and $a_3^{\dag\,\pm}(\bk)|_{m=0}$. This will add new
problems with the spin and orbital angular momentum of the vacuum as it will
be, possibly, finite, but completely undetermined.

	From the evident equalities
	\begin{equation}	\label{9.5}
	\begin{split}
    \ope{N} \bigl( a_s^{-}(\bk)\circ a_t^{\dag\,+}(\bk) \bigr)
& = \ope{N} \bigl( a_t^{\dag\,+}(\bk)\circ a_s^{-}(\bk) \bigr)
  = a_t^{\dag\,+}(\bk)\circ a_s^{-}(\bk)
\\
    \ope{N} \bigl( a_s^{\dag\,-}(\bk)\circ a_t^{+}(\bk) \bigr)
& = \ope{N} \bigl( a_t^{+}(\bk)\circ a_s^{\dag\,-}(\bk) \bigr)
  = a_t^{+}(\bk)\circ a_s^{\dag\,-}(\bk)
\\
\ope{N} \bigl( b^\pm \xlrarrow{k_\mu\frac{\pd}{\pd k^\nu}} \circ b^\mp \bigr)
& =
\pm b^+ \xlrarrow{k_\mu\frac{\pd}{\pd k^\nu}} \circ b^-
\quad b^\pm=a_{s}^{\pm}, a_{s}^{\dag\,\pm}
	\end{split}
	\end{equation}
and the equations~\eref{6.5}--\eref{6.7},~\eref{6.27},
and~\eref{7.18}--\eref{7.20}, we see that the dynamical variables of a free
vector field (satisfying the Lorenz condition) take the following form after
normal ordering:
	\begin{align}	\label{9.6}
\ope{P}_\mu
& =
\frac{1}{1+\tau(\U)} \sum_{s=1}^{3-\delta_{0m}} \int
  k_\mu |_{ k_0=\sqrt{m^2c^2+{\bs k}^2} }
\{
a_s^{\dag\,+}(\bk)\circ a_s^-(\bk) +
a_s^+(\bk) \circ a_s^{\dag\,-}(\bk)
\}
\Id^3\bk
\displaybreak[1]\\	\label{9.7}
\ope{Q}
& =
q \sum_{s=1}^{3-\delta_{0m}}\int
\{
a_s^{\dag\,+}(\bk)\circ a_s^-(\bk) -
a_s^+(\bk) \circ a_s^{\dag\,-}(\bk)
\} \Id^3\bk
	\end{align}
\vspace{-4ex}
	\begin{multline}	\label{9.8}
\ope{L}_{\mu\nu}
=
\frac{1}{1+\tau(\U)} \sum_{s=1}^{3-\delta_{0m}} \!\! \int\! \Id^3\bk
 ( x_{\mu}k_\nu - x_{\nu}k_\mu ) |_{ k_0=\sqrt{m^2c^2+{\bs k}^2} }
\{
a_s^{\dag\,+}(\bk)\circ a_s^-(\bk) +
a_s^+(\bk) \circ a_s^{\dag\,-}(\bk)
\}
\\
+
\frac{\ih}{1+\tau(\U)}  \sum_{s,s'=1}^{3} \int \Id^3\bk \,
l_{\mu\nu}^{ss'}(\bk)
\bigl\{
 a_s^{\dag\,+}(\bk) \circ a_{s'}^-(\bk) -
 a_{s'}^+(\bk) \circ a_s^{\dag\,-}(\bk)
\bigr\}
\displaybreak[2]\\
+
\frac{\ih}{2(1+\tau(\U))}  \sum_{s=1}^{3-\delta_{0m}} \int \Id^3\bk
\Bigl\{
a_s^{\dag\,+}(\bk)
\Bigl( \xlrarrow{ k_\mu \frac{\pd}{\pd k^\nu} }
     - \xlrarrow{ k_\nu \frac{\pd}{\pd k^\mu} } \Bigr)
\circ a_s^-(\bk)
\\ +
 a_s^+(\bk)
\Bigl( \xlrarrow{ k_\mu \frac{\pd}{\pd k^\nu} }
     - \xlrarrow{ k_\nu \frac{\pd}{\pd k^\mu} } \Bigr)
\circ a_s^{\dag\,-}(\bk)
\Bigr\} \Big|_{ k_0=\sqrt{m^2c^2+{\bs k}^2} }
+
\delta_{0m} \iih(x^\lambda-x_0^\lambda)
			\ope{N}( \lindex[C]{}{L} {}_{\mu\nu\lambda} )
	\end{multline}
\vspace{-4ex}
	\begin{multline}	\label{9.9}
\ope{S}_{\mu\nu}
=
  \frac{\ih }{1+\tau(\U)} \sum_{s,s'=1}^{3} \int \Id^3\bk
\sigma_{\mu\nu}^{ss'}(\bk)
\bigl\{
  a_s^{\dag\,+}(\bk) \circ a_{s'}^{-}(\bk) -
  a_{s'}^{+}(\bk) \circ a_s^{\dag\,-}(\bk)
\bigr\}
\\ -
\delta_{0m} \iih(x^\lambda-x_0^\lambda)
			\ope{N}( \lindex[C]{}{S} {}_{\mu\nu\lambda} ) ,
	\end{multline}
where
 $\ope{N}( \lindex[C]{}{L} {}_{\mu\nu\lambda} )$ and
 $\ope{N}( \lindex[C]{}{S} {}_{\mu\nu\lambda} )$
can easily be found by means of~\eref{9.5}, \eref{7.10-1} and~\eref{7.6},
but we shall not need the explicit form of these operators. Similarly, the
vectors of the spin~\eref{6.36} and~\eref{6.37} take the following form after
normal ordering (see also~\eref{7.19}):
	\begin{multline}	\label{9.10}
\tope{R}_a
 =
\frac{\ih}{1+\tau(\U)} \varepsilon_{abc} \int \Id^3\bk
r^b (\bk)
\bigl\{
 \bs a^{\dag\,+}(\bk) \ocross \bs a^{-}(\bk) +
 \bs a^{+}(\bk) \ocross \bs a^{\dag\,-}(\bk)
\bigr\}^c
\\
 -
\delta_{0m} \iih(x^\lambda-x_0^\lambda)
			\ope{N}( \lindex[C]{}{S} {}_{0a\lambda} )
	\end{multline}
\vspace{-4ex}
	\begin{align}
			\label{9.11}
\bs{\ope{S}}
& =
\frac{\ih}{1+\tau(\U)}  \int \Id^3\bk
\bigl\{
 \bs a^{\dag\,+}(\bk) \ocross \bs a^{-}(\bk) +
 \bs a^{+}(\bk) \ocross \bs a^{\dag\,-}(\bk)
\bigr\}^c
 -
\delta_{0m} \bs{\ope{\Hat{S}}}
	\end{align}
with
\(
\ope{\Hat{S}}^a
:=
\iih \varepsilon^{abc} (x^\lambda-x_0^\lambda)
			\ope{N}( \lindex[C]{}{S} {}_{bc\lambda} ) .
\)
In particular, the third component of $\bs{\ope{S}}$ is (cf.~\eref{6.40})
	\begin{equation}	\label{9.11-1}
	\begin{split}
\ope{S}^3
 =
\frac{\ih}{1+\tau(\U)}  \int \Id^3\bk
\bigl\{
& a_1^{\dag\,+}(\bk) \circ a_2^{-}(\bk) -
  a_2^{\dag\,+}(\bk) \circ a_1^{-}(\bk)
\\ -
&  a_2^{+}(\bk) \circ a_1^{\dag\,-}(\bk) +
   a_1^{+}(\bk) \circ a_2^{\dag\,-}(\bk)
\bigr\} .
	\end{split}
	\end{equation}
The substitutions~\eref{6.41} transform the integrand in the last equality
into a `diagonal' form and preserves the ones into~\eref{9.6} and~\eref{9.7},
\ie
	\begin{align}	\label{9.11-2}
\ope{P}_\mu
& =
\frac{1}{1+\tau(\U)} \sum_{s=1}^{3-\delta_{0m}} \int
  k_\mu |_{ k_0=\sqrt{m^2c^2+{\bs k}^2} }
\{
b_s^{\dag\,+}(\bk)\circ b_s^-(\bk) +
b_s^+(\bk) \circ b_s^{\dag\,-}(\bk)
\}
\Id^3\bk
\displaybreak[1]\\	\label{9.11-3}
\ope{Q}
& =
q \sum_{s=1}^{3-\delta_{0m}}\int
\{
b_s^{\dag\,+}(\bk)\circ b_s^-(\bk) -
a_s^+(\bk) \circ b_s^{\dag\,-}(\bk)
\} \Id^3\bk
\displaybreak[1]\\	\label{9.11-4}
\ope{S}^3
& =
\frac{\hbar}{1+\tau(\U)} \sum_{s=1}^{2} \int
(-1)^{s+1}
\bigl\{
b_s^{\dag\,+}(\bk)\circ b_s^-(\bk) -
b_s^+(\bk) \circ b_s^{\dag\,-}(\bk)
\bigr\}
 \Id^3\bk .
	\end{align}

	From the just written expressions for the dynamical variables after
normal ordering is evident that
	\begin{equation}	\label{9.12}
\ope{P}_\mu(\ope{X}_0) =0 \quad
\ope{Q}(\ope{X}_0) =0 \quad
\ope{M}_{\mu\nu}(\ope{X}_0) =
\ope{L}_{\mu\nu}(\ope{X}_0) =
\ope{S}_{\mu\nu}(\ope{X}_0) = 0
	\end{equation}
and, consequently, the conserved quantities of the vacuum, the 4\ndash
momentum, charge, spin and orbital angular momenta, vanish, as required by
definition~\ref{Defn9.1}.

	Besides the dynamical variables, the normal ordering changes the field
equations~\eref{7.4} too. As the combinations quadratic in creation and
annihilation, operators in the commutators in~\eref{7.4} come from the
momentum operator (see~\eref{7.2}), the field equations~\eref{7.4} after
normal ordering, by virtue of~\eref{9.6}, read:
	\begin{subequations}	\label{9.13}
	\begin{align}
			\label{9.13a}
	\begin{split}
& \bigl[ a_s^{\pm}(\bk) ,
 a_t^{\dag\,+}(\bs q) \circ a_t^{-}(\bs q) +
 a_t^{+}(\bs q) \circ a_t^{\dag\,-}(\bs q)
\bigr]_{\_}
 \pm (1+\tau(\U)) a_s^{\pm}(\bk) \delta_{st} \delta^3(\bk-\bs q)
= f_{st}^\pm(\bk,\bs q)
	\end{split}
\\			\label{9.13b}
	\begin{split}
& \bigl[ a_s^{\dag\,\pm}(\bk) ,
 a_t^{\dag\,+}(\bs q) \circ a_t^{-}(\bs q) +
 a_t^{+}(\bs q) \circ a_t^{\dag\,-}(\bs q)
\bigr]_{\_}
\pm (1+\tau(\U))  a_s^{\dag\,\pm}(\bk) \delta_{st} \delta^3(\bk-\bs q)
= f_{st}^{\dag\,\pm}(\bk,\bs q) ,
	\end{split}
\\			\label{9.13c}
& s,t =
	\begin{cases}
1,2,3	&\text{for } m\not=0	\\
1,2	&\text{for } m=0
	\end{cases}
	\end{align}
\vspace{-4ex}
	\begin{equation}	\label{9.13d}
	\begin{split}
\sum_{t=1}^{3-\delta_{0m}} \int q_\mu\big|_{q_0=\sqrt{m^2c^2+{\bs q}^2}}
f_{st}^{\pm}(\bk,\bs q) \Id^3\bs{q}
= 0
\qquad
\sum_{t=1}^{3-\delta_{0m}} \int q_\mu\big|_{q_0=\sqrt{m^2c^2+{\bs q}^2}}
f_{st}^{\dag\,\pm}(\bk,\bs q) \Id^3\bs{q}
& = 0 .
	\end{split}
	\end{equation}
	\end{subequations}
However, one can verify that~\eref{9.13} hold identically due to the
commutation relations~\eref{8.2}. Once again, this demonstrates
that~\eref{8.2} play a role of field equations under the suppositions made
for their derivation.

	As a result of~\eref{9.9},~\eref{7.19} and~\eref{8.3}, we see that
the commutation relations~\eref{8.4} after normal ordering transform into
	\begin{multline}	\label{9.14}
[ \tope{S}_{\mu\nu},\tope{S}_{\varkappa\lambda} ]_{\_}
=
[ \ope{S}_{\mu\nu},\ope{S}_{\varkappa\lambda} ]_{\_} \big|_{m\not=0}
\\ =
 \frac{\hbar^2}{(1+\tau(\U))^2}
 \sum_{s,s',t=1}^{3-\delta_{0m}} \int\Id^3\!\bk
\bigl\{ \sigma_{\mu\nu}^{s't}(\bk) \sigma_{\varkappa\lambda}^{ss'}(\bk)
      - \sigma_{\mu\nu}^{ss'}(\bk) \sigma_{\varkappa\lambda}^{s't}(\bk)
\bigr\}
\bigl\{ a_{s}^{\dag\,+}(\bk) \circ a_{t}^{-}(\bk)
      + a_{s}^{+}(\bk) \circ a_{t}^{\dag\,-}(\bk)
\\ -
\tau(\U)  a_{s}^{+}(\bk) \circ a_{t}^{-}(\bk) +
\tau(\U)  a_{s}^{\dag\,+}(\bk) \circ a_{t}^{\dag\,-}(\bk)
\bigr\}
 +
\delta_{0m} f_{\mu\nu\varkappa\lambda}^{\ope{N}}(a_3^{\pm},a_3^{\dag\,\pm})
,
	\end{multline}
with
$f_{\mu\nu\varkappa\lambda}^{\ope{N}}(a_3^{\pm},a_3^{\dag\,\pm})$
being a term whose integrand is a homogeneous expression in
$a_3^{\pm}(\bk)$ and $a_3^{\dag\,\pm}(\bk)$ and which term is set equal to
zero for $m\not=0$.

	The commutation relations between the spin angular momentum operator
and the charge one, expressed by~\eref{8.5}--\eref{8.7} before normal
ordering, take the following form after normal ordering
	\begin{align}	\label{9.15}
[ \tope{S}_{\mu\nu},\tope{Q} ]_{\_}
& =
\delta_{0m} \tilde{f}_{\mu\nu}^{\ope{N}}(a_3^{\pm},a_3^{\dag\,\pm})
\\		\label{9.16}
[ \ope{S}_{\mu\nu},\ope{Q} ]_{\_}
& =
\delta_{0m} f_{\mu\nu}^{\ope{N}}(a_3^{\pm},a_3^{\dag\,\pm})
\\		\label{9.17}
[ \ope{S}_{\mu\nu},\ope{Q} ]_{\_} \big|_{m\not=0}
& = 0.
	\end{align}
These relations can be checked by means of~\eref{9.9}, \eref{9.7}
and~\eref{8.3}; alternatively, they are corollaries of~\eref{9.14} with
$\sigma_{\varkappa\lambda}^{st}(\bk)$ replaced by
$-\iih(1+\tau(\U))q\delta^{st}$.


\section {State vectors and particle interpretation}
\label{Sect10}

	The description of the state vectors of a free vector field
satisfying the Lorenz condition is practically identical with the one of a
free spinor field, presented in~\cite{bp-QFTinMP-spinors}. To be obtained the
former case from the latter one, the following four major changes should be
made:
	(i)~the polarization indices should run over the range~1,~2 and~3, if
$m\not=0$, or~1 and~2, if $m=0$;
	(ii)~the commutation relations~\eref{8.2} must replace the
corresponding anticommutation ones for a free spinor field;
	(iii)~the replacements
$\sigma_{\mu\nu}^{st,\pm}(\bk) \mapsto \sigma_{\mu\nu}^{st}(\bk)$ and
$l_{\mu\nu}^{st,\pm}(\bk) \mapsto l_{\mu\nu}^{st}(\bk)$
of the spin and orbital momentum coefficients should be made;
	(vi)~the additional terms, depending on the polarization $s=3$ should
be taken into account, when the spin and orbital angular momentum operators
of a massless field are considered.
	According to these changes, we present below a \emph{mutatis
mutandis} version of the corresponding considerations
in~\cite{bp-QFTinMP-spinors} (see also~\cite{bp-QFTinMP-scalars}.

	In momentum picture, in accord with the general theory of
Sect.~\ref{Sect2}, the state vectors of a vector field are spacetime\ndash
dependent, contrary to the field operators and dynamical variables
constructed from them. In view of~\eref{12.118}, the spacetime\ndash
dependence of a state vector $\ope{X}(x)$ is
	\begin{equation}	\label{10.1}
\ope{X}(x) = \ope{U}(x,x_0) (\ope{X}(x_0))
	\end{equation}
where $x_0$ is an arbitrarily fixed spacetime point and the \emph{evolution
operator} $\ope{U}(x,x_0)\colon\Hil\to\Hil$ is
	\begin{equation}	\label{10.2}
	\begin{split}
\ope{U}(x,x_0)
=
\exp
	\Bigl\{
\iih (x^\mu-x_0^\mu)
\! \sum_{s} \! \int \!
k_\mu |_{ k_0=\sqrt{m^2c^2+{\bs k}^2} }
\{
a_s^{\dag\,+}(\bs k)\circ a_s^{-}(\bs k)
 +
a_s^{+}(\bs k) \circ a_s^{\dag\,-}(\bs k)
\}
\Id^3\bs k
\Bigr\}  .
	\end{split}
	\end{equation}
due to~\eref{12.112} and~\eref{9.6} (see also~\eref{12.116}--\eref{12.118}).
The operator~\eref{10.2} plays also a role of an `$S$\ndash matrix'
determining the transition amplitudes between any initial and final states,
say $\ope{X}_i(x_i)$ and $\ope{X}_f(x_f)$ respectively. In fact, we have
	\begin{equation}	\label{10.3}
S_{fi}(x_f,x_i) := \langle\ope{X}_f(x_f) | \ope{X}_i(x_i)\rangle
=
\langle\ope{X}_f(x_f^{(0)}) | \ope{U}(x_i,x_f)
	(\ope{X}_i(x_i^{(0)}))\rangle .
	\end{equation}
For some purposes, the following expansion of $\ope{U}(x_i,x_f)$ into a power
series may turn to be useful:
	\begin{equation}	\label{10.4}
\ope{U}(x_i,x_f) = \id_\Hil + \sum_{n=1}^{\infty} \ope{U}^{(n)}(x_i,x_f)
	\end{equation}
\vspace{-2.4ex}
	\begin{multline}	\label{10.5}
\ope{U}^{(n)}(x_i,x_f)
:=
\frac{1}{n!} (x_i^{\mu_1}-x_f^{\mu_1}) \dots (x_i^{\mu_n}-x_f^{\mu_n})
\sum_{s_1,\dots,s_n=1}^{3-\delta_{0m}}
\int \Id^3\bs k^{(1)}\dots\Id^3\bs k^{(n)}
k_{\mu_1}^{(1)}\dotsb k_{\mu_n}^{(n)}
\\ \times
\bigl\{  a_{s_1}^{\dag\,+}(\bs k^{(1)})\circ a_{s_1}^-(\bs k^{(1)})
+
 a_{s_1}^+(\bs k^{(1)})\circ a_{s_1}^{\dag\,-}(\bs k^{(1)}) \bigr\}
\\
\circ\dotsb\circ
\bigl\{  a_{s_n}^{\dag\,+}(\bs k^{(n)})\circ a_{s_n}^-(\bs k^{(n)})
+
 a_{s_n}^+(\bs k^{(n)})\circ a_{s_n}^{\dag\,-}(\bs k^{(n)}) \bigr\}
	\end{multline}
where $k_0^{(a)}=\sqrt{m^2c^2+ ({\bs k}^{(a)})^2}$, $a=1,\dots,n$.

	According to~\eref{12.120} and the considerations in
Sect.~\ref{Sect5}, a state vector of a state containing $n'$
particles and $n^{\prime\prime}$ antiparticles,
$n^{\prime},n^{\prime\prime}\ge0$, such that
the $i^{\prime\,\text{th}}$ particle has 4\ndash momentum
$p'_{i'}$ and polarization $s'_{i'}$ and
the $i^{\prime\prime\,\text{th}}$ antiparticle has 4\ndash momentum
$p^{\prime\prime}_{i^{\prime\prime}}$ and polarization
$s^{\prime\prime}_{i^{\prime\prime}}$, where $i'=0,1,\dots,n'$ and
$i^{\prime\prime}=0,1,\dots,n^{\prime\prime}$, is given by the equality
	\begin{multline}	\label{10.6}
\ope{X}(
x;p'_1,s'_1;\ldots;p'_{n'},s'_{n'};
p^{\prime\prime}_1,s^{\prime\prime}_1;\ldots;
p^{\prime\prime}_{n^{\prime\prime}}, s^{\prime\prime}_{n^{\prime\prime}}
)
\\
=
\frac{1}{\sqrt{n^{\prime}! n^{\prime\prime}!}}
\exp\Bigl\{
\iih (x^{\mu}-x_0^\mu) \sum_{i'=1}^{n'} (p'_{i'})_\mu
+
\iih (x^{\mu}-x_0^\mu) \sum_{i^{\prime\prime}=1}^{n^{\prime\prime}}
     (p^{\prime\prime}_{i^{\prime\prime}})_\mu
\Bigr\}
\\ \times
\bigl(
 a_{s'_1}^+(\bs p'_1)\circ\dots\circ a_{s'_{n'}}^+(\bs p'_{n'})
\circ
 a_{s^{\prime\prime}_1}^{\dag\,+}(\bs p^{\prime\prime}_1)\circ\dots\circ
	a_{s^{\prime\prime}_{n^{\prime\prime}}}^{\dag\,+}
		(\bs p^{\prime\prime}_{n^{\prime\prime}})
\bigr)
(\ope{X}_0) ,
	\end{multline}
where, in view of the commutation relations~\eref{8.2}, the order  of
the creation operators is inessential. If $n'=0$ (resp.\
$n^{\prime\prime}=0$), the particle (resp.\ antiparticle) creation operators
and the first (resp.\ second) sum in the exponent should be absent. In
particular, the vacuum corresponds to~\eref{10.6} with
$n'=n^{\prime\prime}=0$. The state vector~\eref{10.6} is
	an eigenvector of the momentum operator~\eref{9.6}  with
eigenvalue (4\ndash momentum)
\(
\sum_{i'=1}^{n'} p'_{i'}
+
\sum_{i^{\prime\prime}=1}^{n^{\prime\prime}}
	p^{\prime\prime}_{i^{\prime\prime}}
\)
	and is also an eigenvector of the charge operator~\eref{9.7} with
eigenvalue  $(-q)(n'-n^{\prime\prime})$.%
\footnote{~%
Recall (see Sect.~\ref{Sect5}), the operator $ a_{s}^{+}(\bs k)$
creates a particle with 4\ndash momentum $k_\mu$ and charge $-q$, while
$ a_{s}^{\dag\,+}(\bs k)$ creates a particle with 4\ndash momentum $k_\mu$
and charge $+q$, where, in the both cases, $k_0=\sqrt{m^2c^2+{\bs k}^2}$. See
also equations~\eref{10.2-1}--\eref{10.2-5} below.%
}

	The reader may verify, using~\eref{8.2} and~\eref{5.21}, that
the transition amplitude between two states of a vector field,
like~\eref{10.6}, is:
	\begin{multline}	\label{10.7}
\langle
\ope{X}(
y;q'_1, t'_1;\ldots;q'_{n'},t'_{n'};
  q^{\prime\prime}_1, t^{\prime\prime}_1;\ldots;
  q^{\prime\prime}_{n^{\prime\prime}}, t^{\prime\prime}_{n^{\prime\prime}})
\\
|
\ope{X}(
x;p'_1, s'_1;\ldots;p'_{m'}, s'_{m'};
  p^{\prime\prime}_1, s^{\prime\prime}_1;\ldots;
  p^{\prime\prime}_{m^{\prime\prime}}, s^{\prime\prime}_{m^{\prime\prime}})
\rangle
\\
=
\frac{1}{n^{\prime}! n^{\prime\prime}!}
\delta_{m'n'} \delta_{m^{\prime\prime}n^{\prime\prime}}
\exp\Bigl\{
\iih (x^{\mu} - y^{\mu}) \sum_{i'=1}^{n'} (p'_{i'})_\mu
+
\iih (x^{\mu} - y^{\mu}) \sum_{i^{\prime\prime}=1}^{n^{\prime\prime}}
			 (p^{\prime\prime}_{i^{\prime\prime}})_\mu
\Bigr\}
\\ \times
\sum_{(i'_1,\dots,i'_{n'})}
\delta_{s'_{n'} t'_{i'_1} }
	\delta^3(\bs p'_{n'} - \bs q'_{i'_1})
\delta_{s'_{n'-1}  t'_{i'_2} }
	\delta^3(\bs p'_{n'-1} - \bs q'_{i'_2})
\dots
\delta_{s'_{1}  t'_{i'_{n'}} }
	\delta^3(\bs p'_{1} - \bs q'_{i'_{n'}})
\\ \times
\!
\sum_{(i^{\prime\prime}_1,\dots,i_{n^{\prime\prime}}^{\prime\prime})}
\!
	\delta_{s^{\prime\prime}_{n^{\prime\prime}}
	  			t^{\prime\prime}_{i^{\prime\prime}_1} }
	\delta^3(\bs p^{\prime\prime}_{n^{\prime\prime}} -
		\bs q^{\prime\prime}_{i^{\prime\prime}_1})
	\delta_{s^{\prime\prime}_{n^{\prime\prime}-1}
				t^{\prime\prime}_{i^{\prime\prime}_2} }
	\delta^3(\bs p^{\prime\prime}_{n^{\prime\prime}-1} -
		\bs q^{\prime\prime}_{i^{\prime\prime}_2})
\dots
\delta_{s^{\prime\prime}_{1}
		 t^{\prime\prime}_{i^{\prime\prime}_{n^{\prime\prime}}} }
\delta^3(\bs p^{\prime\prime}_{1} -
\bs q^{\prime\prime}_{i^{\prime\prime}_{n^{\prime\prime}}})
	\end{multline}
where the summations are over all
permutations $(i'_1,\dots,i'_{n'})$ of $(1,\dots,n')$ and
$(i^{\prime\prime}_1,\dots,i_{n^{\prime\prime}}^{\prime\prime})$ of
$(1,\dots,n^{\prime\prime})$.
	The conclusions from this formula are similar to the ones concerning
free scalar or spinor
fields~\cite{bp-QFTinMP-scalars,bp-QFTinMP-spinors}.
For instance, the only non\ndash forbidden transition from an
$n'$\ndash particle~+~$n^{\prime\prime}$\ndash antiparticle state is into
$n'$\ndash particle~+~$n^{\prime\prime}$\ndash antiparticle state; the both
states may differ only in the spacetime positions of the (anti)particles in
them. This result is quite natural as we are dealing with free
particles/fields.

	In particular, if $\ope{X}_n$ denotes any state containing $n$
particles and/or antiparticles, $n=0,1,\dots$, then~\eref{10.7} says that
	\begin{equation}	\label{10.8}
\langle \ope{X}_n|\ope{X}_0 \rangle = \delta_{n0} ,
	\end{equation}
which expresses the stability of the vacuum.

	Consider the one (anti)particle states
$a_{t}^{+}(\bs p)(\ope{X}_0)$ and $a_{t}^{\dag\,+}(\bs p)(\ope{X}_0)$, with
$t=1,2,3$, if $m\not=0$, or $t=1,2$, if $m=0$.
Applying~\eref{9.6}--\eref{9.11-2} and~\eref{8.2}, we find
($p_0:=\sqrt{m^2c^2+{\bs p}^2}$):%
\footnote{~%
The easiest way to derive~\eref{10.2-5} is by
applying~\eref{2.14},~\eref{7.2},~\eref{7.3} and~\eref{8.2}. Notice, in
Heisenberg picture and in terms of the Heisenberg creation/annihilation
operators~\eref{6.48}, equations~\eref{10.2-5} read
$\tope{L}_{\mu\nu}\bigl( a_{t}^{+}(\bs p)(\ope{X}_0) \bigr) = 0$ and
$\tope{L}_{\mu\nu}\bigl( a_{t}^{\dag\,+}(\bs p)(\ope{X}_0) \bigr) = 0$
which is quite understandable in view of the fact that $\tope{L}_{\mu\nu}$
is, in a sense, the average orbital momentum with respect to all spacetime
points, while $\ope{L}_{\mu\nu}(x,x_0)$ is the one relative to  $x$ and
$x_0$; the dependence on $x_0$ being hidden in the  $\ope{L}_{\mu\nu}$,
$a_{t}^{+}(\bs p)$ and $a_{t}^{\dag\,+}(\bs p)$.%
}
	\begin{gather}
			\label{10.2-1}
	\begin{split}
\ope{P}_\mu\bigl( a_{t}^{+}(\bs p)(\ope{X}_0) \bigr)
& =
p_\mu a_{t}^{+}(\bs p)(\ope{X}_0)
\quad \hphantom{+}
\ope{Q}\bigl( a_{t}^{+}(\bs p)(\ope{X}_0) \bigr)
=
 - q a_{t}^{+}(\bs p)(\ope{X}_0)
\\
\ope{P}_\mu\bigl( a_{t}^{\dag\,+}(\bs p)(\ope{X}_0) \bigr)
& =
p_\mu a_{t}^{\dag\,+}(\bs p)(\ope{X}_0)
\quad
\ope{Q}\bigl( a_{t}^{\dag\,+}(\bs p)(\ope{X}_0) \bigr)
=
 + q a_{t}^{\dag\,+}(\bs p)(\ope{X}_0)
	\end{split}
\displaybreak[1]\\		\label{10.2-2}
	\begin{split}
\ope{S}_{\mu\nu}
	\bigl( a_{t}^{+}(\bs p)(\ope{X}_0) \bigr)
& =
- \ih \sum_{s=1}^{3-\delta_{0m}}
  \sigma_{\mu\nu}^{ts}(\bs p) a_{s}^{+}(\bs p)(\ope{X}_0)
- \delta_{0m} (\cdots) (\ope{X}_0)
\\
\ope{S}_{\mu\nu}
	\bigl( a_{t}^{\dag\,+}(\bs p)(\ope{X}_0) \bigr)
& =
- \ih \sum_{s=1}^{3-\delta_{0m}}
\sigma_{\mu\nu}^{ts}(\bs p) a_{s}^{\dag\,+}(\bs p)(\ope{X}_0)
- \delta_{0m} (\cdots)_\dag\, (\ope{X}_0)
	\end{split}
\displaybreak[1]\\		\label{10.2-3}
	\begin{split}
\ope{S}^3 \bigl( a_{t}^{+}(\bs p)(\ope{X}_0) \bigr)
& =
  \ih
\{ \delta_{t2} a_{1}^{+}(\bs p) - \delta_{t1} a_{2}^{+}(\bs p) \} (\ope{X}_0)
\\
\ope{S}^3 \bigl( a_{t}^{\dag\,+}(\bs p)(\ope{X}_0) \bigr)
& =
   \ih
\{ \delta_{t2} a_{1}^{\dag\,+}(\bs p) - \delta_{t1} a_{2}^{\dag\,+}(\bs p) \}
   (\ope{X}_0)
	\end{split}
\displaybreak[1]\\		\label{10.2-5}
	\begin{split}
\ope{L}_{\mu\nu}(x) \bigl( a_{t}^{+}(\bs p)(\ope{X}_0) \bigr)
= &
\Bigl\{
(x_\mu p_\nu - x_\nu p_\mu )
- \ih \Bigl( p_\mu\frac{\pd}{\pd p^\nu} - p_\nu\frac{\pd}{\pd p^\mu} \Bigr)
\Bigr\}
    \bigl( a_{t}^{+}(\bs p)(\ope{X}_0) \bigr)
\\& - \ih \sum_{s=1}^{3-\delta_{0m}}
  l_{\mu\nu}^{ts}(\bs p)
    \bigl( a_{s}^{+}(\bs p)(\ope{X}_0) \bigr)
- \delta_{0m} (\cdots) (\ope{X}_0)
\\
\ope{L}_{\mu\nu}(x) \bigl( a_{t}^{\dag\,+}(\bs p)(\ope{X}_0) \bigr)
= &
\Bigl\{
(x_\mu p_\nu - x_\nu p_\mu )
- \ih \Bigl( p_\mu\frac{\pd}{\pd p^\nu} - p_\nu\frac{\pd}{\pd p^\mu} \Bigr)
\Bigr\}
    \bigl( a_{t}^{\dag\,+}(\bs p)(\ope{X}_0) \bigr)
\\& - \ih \sum_{s=1}^{3-\delta_{0m}}
  l_{\mu\nu}^{ts}(\bs p)
    \bigl( a_{s}^{\dag\,+}(\bs p)(\ope{X}_0) \bigr)
- \delta_{0m} (\cdots)_\dag (\ope{X}_0)
	\end{split}
	\end{gather}
where $q\tau(\U)\equiv 0$ was used and $(\cdots)$ and $(\cdots)_\dag$ denote
expressions whose integrands are homogeneous with respect to
$a_{3}^{\pm}(\bs p)$ and $a_{3}^{\dag\,\pm}(\bs p)$.
	It should be remarked the agreement
of~\eref{10.2-1}--\eref{10.2-5} with~\eref{5.22}.%
\footnote{~%
If the r.h.s.\ of~\eref{2.21} is with an opposite sign, this agreement
will be lost.%
}

	If one uses the operators $b_{s}^{\pm}(\bk)$ and
$a_{s}^{\dag\,\pm}(\bk)$ instead of $a_{s}^{\pm}(\bk)$ and
$a_{s}^{\dag\,\pm}(\bk)$ (see~\eref{6.41}), the equations~\eref{10.2-1}
and~\eref{10.2-3} will read ($p_0:=\sqrt{m^2c^2+{\bs p}^2}$):
	\begin{gather}
			\label{10.20}
	\begin{split}
\ope{P}_\mu\bigl( b_{t}^{+}(\bs p)(\ope{X}_0) \bigr)
& =
p_\mu b_{t}^{+}(\bs p)(\ope{X}_0)
\quad \hphantom{+}
\ope{Q}\bigl( b_{t}^{+}(\bs p)(\ope{X}_0) \bigr)
=
 - q b_{t}^{+}(\bs p)(\ope{X}_0)
\\
\ope{P}_\mu\bigl( b_{t}^{\dag\,+}(\bs p)(\ope{X}_0) \bigr)
& =
p_\mu b_{t}^{\dag\,+}(\bs p)(\ope{X}_0)
\quad
\ope{Q}\bigl( b_{t}^{\dag\,+}(\bs p)(\ope{X}_0) \bigr)
=
 + q b_{t}^{\dag\,+}(\bs p)(\ope{X}_0)
	\end{split}
\displaybreak[1]\\		\label{10.21}
	\begin{split}
\ope{S}^3 \bigl( b_{t}^{+}(\bs p)(\ope{X}_0) \bigr)
& =
  \hbar
\{ \delta_{t1} b_{1}^{+}(\bs p) - \delta_{t2} b_{2}^{+}(\bs p) \} (\ope{X}_0)
\\
\ope{S}^3 \bigl( b_{t}^{\dag\,+}(\bs p)(\ope{X}_0) \bigr)
& =
   \hbar
\{ \delta_{t1} b_{1}^{\dag\,+}(\bs p) - \delta_{t2} b_{2}^{\dag\,+}(\bs p) \}
   (\ope{X}_0)
	\end{split}
%
%
	\end{gather}

	We should emphasize, the equations~\eref{10.2-1}--\eref{10.21} do
\emph{not} concern the massless case, $m=0$, with polarization $t=3$, \ie the
(state?) vectors $a_{3}^{+}(\bs p)\big|_{m=0}(\ope{X}_0)$ and
$a_{3}^{\dag\,+}(\bs p)\big|_{m=0}(\ope{X}_0)$ have undetermined 4\ndash
momentum, charge, spin and orbital angular momentum. This is quite
understandable as the operators $a_{3}^{\pm}(\bs p)\big|_{m=0}$ and
$a_{3}^{\dag\,\pm}(\bs p)\big|_{m=0}$ do not enter in the field
equations~\eref{7.4} and, consequently, remain completely arbitrary. (For more
details on this situation, see Sect.~\ref{Sect11}.)

	On the formulae~\eref{10.2-1}--\eref{10.21} is based the particle
interpretation of quantum field theory (in Lagrangian formalism) of free
vector field satisfying the Lorenz condition. According to them, the state
vectors produced by the operators
$a_{s}^{+}(\bs p)$ and $b_{s}^{+}(\bs p)$ from the vacuum $\ope{X}_0$ can be
interpreted as ones representing particles with 4\ndash momentum
$(\sqrt{m^2c^2+{\bs p}^2},\bs p)$ and charge $(-q)$; the spin and orbital
angular momentum of these vectors is not definite. The states
$a_{s}^{+}(\bs p)(\ope{X}_0)$ do not have a definite projection of the
vector of spin on the direction of movement, \ie along $\bs p$ (for $\bs
p\not=\bs0$), but, if $m\not=0$, the one of $b_{s}^{+}(\bs p)(\ope{X}_0)$ is
equal to $(-1)^{s+1}\hbar$, if $s=1,2$, or to zero, if $s=3$.
	Similarly, the state vectors
$a_{s}^{\dag\,+}(\bs p)(\ope{X}_0)$ and $b_{s}^{\dag\,+}(\bs p)(\ope{X}_0)$
should be interpreted as ones representing particles, called
\emph{antiparticles} (with respect to the ones created by
$a_{s}^{+}(\bs p)$ and $b_{s}^{+}(\bs p)$),
with the same characteristics but the charge, which for them is equal to
$(+q)$. For  this reason, the particles and antiparticles of a neutral
(Hermitian), $q=0$, field coincide.

	Notice, the above interpretation of the creation and annihilation
operators does not concern these operators for a massless field and
polarization along the 3\ndash momentum, \ie for $(m,s)=(0,3)$. Besides, in
the massless case, the just said about the spin projection of the states
$b_{s}^{+}(\bs p)(\ope{X}_0)$ and $b_{s}^{\dag\,+}(\bs p)(\ope{X}_0)$) is
not valid unless the last terms, proportional to $\delta_{0m}$,
in~\eref{10.2-2} vanish.


\section
[The massless case. Electromagnetic field in Lorenz gauge]
{The massless case.\\ Electromagnetic field in Lorenz gauge}
\label{Sect11}

	A simple overview of the preceding sections reveals that the
zero-mass case, $m=0$, is more or less an exception of the general
considerations. The cause for this is that the Lorenz condition, expressed
by~\eref{3.16}, is external to the Lagrangian formalism for a massless
free vector field, contrary to the massive case, $m\not=0$. However, this
condition does not contradict to the formalism and, as we demonstrated, it
can be developed to a reasonable extend.

	Practically, the only problem in the massless case, we have met,  is with
the physical meaning/interpretation of the operators
 $a_3^{\pm}(\bk)|_{m=0}$ and $a_3^{\dag\,\pm}(\bk)|_{m=0}$.%
\footnote{~%
From pure mathematical viewpoint, everything is in order and no problems
arise.%
}
The first indications for it were the last two equations in~\eref{5.22c}
(valid if $m=0$), which were derived from the external to the Lagrangian
formalism equation~\eref{2.27} and, hence, can be neglected if one follows
rigorously the Lagrangian field theory; moreover, in Sect.~\ref{Sect6}, we
proved that the equations~\eref{6.30} imply
$a_s^{\pm}(\bk)=a_s^{\dag\,\pm}(\bk)=0$ for $s=1,2$, which, in view of the
further development of the theory, is unacceptable. The really serious
problem with the operators
$a_3^{\pm}(\bk)|_{m=0}$ and $a_3^{\dag\,\pm}(\bk)|_{m=0}$
is that they have vanishing contribution to the momentum operator~\eref{6.5}
and charge operator~\eref{6.6}, but they have, generally, non\ndash vanishing
one to the orbital and spin angular momentum operators~\eref{6.7}
and~\eref{6.27}, respectively.%
\footnote{~%
To save some space, here our considerations do not take into account the
normal ordering, \ie they concern the theory before it; \emph{vide infra}.%
}
(See also the vectors of spin~\eref{6.36} and~\eref{6.37} in which these
operators enter via~\eref{6.38}.) In connection with the (possible)
interpretation in terms of particles, this means that
$a_3^{\pm}(\bk)|_{m=0}$ and $a_3^{\dag\,\pm}(\bk)|_{m=0}$
describe creation/annihilation of neutral massless particles with vanishing
4\ndash momentum and charge, but, generally, non\ndash zero spin and orbital
angular momentum. Besides, the last two characteristics of the (hypothetical)
particles with state vectors
$a_3^{+}(\bk)|_{m=0}(\ope{X}_0)$ and $a_3^{\dag\,+}(\bk)|_{m=0}(\ope{X}_0)$
can be completely arbitrary since the field equation~\eref{7.4} do not impose
on the operators
$a_3^{\pm}(\bk)|_{m=0}$ and $a_3^{\dag\,\pm}(\bk)|_{m=0}$
any restrictions.%
\footnote{~%
However, the third component of the vector of spin $\bs{\ope{S}}$ does not
depend on $a_3^{\pm}(\bk)|_{m=0}$ and $a_3^{\dag\,\pm}(\bk)|_{m=0}$ ---
see~\eref{6.40} and~\eref{10.2-3}.%
}

	The above discussion leads to the following conclusion. The
Lagrangian formalism, without further assumptions/hypotheses, cannot give any
information about the operators
$a_3^{\pm}(\bk)|_{m=0}$ and $a_3^{\dag\,\pm}(\bk)|_{m=0}$
and, consequently, leaves them as free parameters of the quantum field
theory of massless free vector field satisfying the Lorenz condition and
described by the Lagrangian~\eref{3.7}.  Thus, these operators are carries of
a completely arbitrary degrees of freedom, which have a non\ndash vanishing
contribution to the spin and orbital angular momentum operators~\eref{6.27}
and~\eref{6.7}, respectively, unless before normal ordering we have
	\begin{subequations}	\label{11.1}
	\begin{multline}	\label{11.1a}
\sum_{s=1,2} \int\Id^3\bk  \sigma_{\mu\nu}^{s3}(\bk)\big|_{m=0}
\bigl\{
a_s^{\dag\,+}(\bk) \circ a_3^{-}(\bk) -
a_s^{\dag\,-}(\bk) \circ a_3^{+}(\bk)
\\ -
a_3^{\dag\,+}(\bk) \circ a_s^{-}(\bk) +
a_3^{\dag\,-}(\bk) \circ a_s^{+}(\bk)
\bigr\} \big|_{m=0}
= 0
	\end{multline}
\nopagebreak\vspace{-4ex}
	\begin{multline}	\label{11.1b}
\sum_{s=1,2} \int\Id^3\bk  l_{\mu\nu}^{s3}(\bk)\big|_{m=0}
\bigl\{
a_s^{\dag\,+}(\bk) \circ a_3^{-}(\bk) -
a_s^{\dag\,-}(\bk) \circ a_3^{+}(\bk)
\\ -
a_3^{\dag\,+}(\bk) \circ a_s^{-}(\bk) +
a_3^{\dag\,-}(\bk) \circ a_s^{+}(\bk)
\bigr\} \big|_{m=0}
= 0 ,
	\end{multline}
	\end{subequations}
where we have used the skewsymmetry of the quantities~\eref{6.8}
and~\eref{6.28}. The just-presented considerations  concern the theory before
(second) quantization, \ie before imposing the commutation
relations~\eref{8.2}, and normal ordering. However, since the  quantization
procedure does not concern the operators
$a_3^{\pm}(\bk)|_{m=0}$ and $a_3^{\dag\,\pm}(\bk)|_{m=0}$
(see Sect.~\ref{Sect8}), the above\ndash said remains completely valid after
these procedures, provided one takes into account the
expressions~\eref{9.6}--\eref{9.11} for the dynamical variables after normal
ordering. In particular, after normal ordering, the spin and orbital angular
momentum operators~\eref{9.9} and~\eref{9.8}, respectively, will be
independent of $a_3^{\pm}(\bk)|_{m=0}$ and $a_3^{\dag\,\pm}(\bk)|_{m=0}$ iff
	\begin{subequations}	\label{11.2}
	\begin{multline}	\label{11.2a}
\sum_{s=1,2} \int\Id^3\bk  \sigma_{\mu\nu}^{s3}(\bk)\big|_{m=0}
\bigl\{
a_s^{\dag\,+}(\bk) \circ a_3^{-}(\bk) -
a_3^{+}(\bk) \circ a_s^{\dag\,-}(\bk)
\\ -
a_3^{\dag\,+}(\bk) \circ a_s^{-}(\bk) +
a_s^{+}(\bk) \circ a_3^{\dag\,-}(\bk)
\bigr\} \big|_{m=0}
= 0
	\end{multline}
\nopagebreak\vspace{-4ex}
	\begin{multline}	\label{11.2b}
\sum_{s=1,2} \int\Id^3\bk  l_{\mu\nu}^{s3}(\bk)\big|_{m=0}
\bigl\{
a_s^{\dag\,+}(\bk) \circ a_3^{-}(\bk) -
a_3^{+}(\bk) \circ a_s^{\dag\,-}(\bk)
\\ -
a_3^{\dag\,+}(\bk) \circ a_s^{-}(\bk) +
a_s^{+}(\bk) \circ a_3^{\dag\,-}(\bk)
\bigr\} \big|_{m=0}
= 0 .
	\end{multline}
	\end{subequations}

	Thus, if one wants to construct a sensible \emph{physical} theory of
a massless free vector field satisfying the Lorenz condition, new
assumptions to the Lagrangian formalism should be added. At this point, there
is a room for different kinds of speculations. Here are two such
possibilities.

	One can demand, as an additional condition, the fulfillment of the
field equations~\eref{7.4} for $m=0$ and any polarization indices, \ie for
$s,t=1,2,3$, instead only for $s,t=1,2$ obtained from the Lagrangian
formalism. This will entail the vanishment of all of the quantities
$B_{ss'}^{\mu,\mp}(\bs q)$, i.e.~\eref{7.6} will be replaced with
	\begin{gather}	\label{11.3}
B_{ss'}^{\mu,\mp}(\bs q) \equiv 0  \qquad s,s'=1,2,3
\\\intertext{which, in its turn, leads to (see~\eref{7.10-1})}
			\label{11.3-1}
  \lindex[C]{}{S} {}_{\mu\nu}^{\lambda}
= \lindex[C]{}{L} {}_{\mu\nu}^{\lambda}
= 0
	\end{gather}
and, consequently, to the commutativity of the spin angular momentum and
momentum operators, etc.\ (see~\eref{7.10}--\eref{7.20}). Other consequence
of the above assumption will be the validity of the commutation
relations~\eref{8.2} for arbitrary polarization indices $s,t=1,2,3$ in the
massless case. As a result of them and the normal ordering procedure, the
vectors
$a_3^{+}(\bk)|_{m=0}(\ope{X}_0)$ and $a_3^{\dag\,+}(\bk)|_{m=0}(\ope{X}_0)$
will describe states with vanishing 4\ndash momentum and charge and,
generally, non\ndash vanishing spin and orbital angular momentum. It seems,
states/particles with such characteristics have not been observed until now.
This state of affairs can be improved by adding to the integrands
in~\eref{9.6} and~\eref{9.7} terms proportional to
$a_3^{\dag\,+}(\bk)\circ a_3^{-}(\bk)$ and
$a_3^{+}(\bk)\circ a_3^{\dag\,-}(\bk)$ in the massless case, but such a game
with adjustment of theory's parameters is out of the scope of the present
work.

	The second possible solution of the problem(s) with the zero-mass
case, we would like to explore, does not require drastical changes of the
formalism as the preceding one. In it to the Lagrangian formalism are added,
as subsidiary conditions, the equations~\eref{11.2} or~\eref{11.1}, depending
if the normal ordering is or is not taken into account, respectively. In this
way only the operators of spin and orbital angular momentum are changed, \viz
before normal ordering they read (see~\eref{6.27}, \eref{6.7}, \eref{7.19},
\eref{7.20} and notice that the above assumption entails~\eref{11.3-1})
	\begin{equation}	\label{11.4}
\ope{S}_{\mu\nu}
= \tope{S}_{\mu\nu}
=
  \frac{\ih }{1+\tau(\U)} \sum_{s,s'=1}^{3-\delta_{0m}} \int \Id^3\bk
\sigma_{\mu\nu}^{ss'}(\bk)
\bigl\{
  a_s^{\dag\,+}(\bk) \circ a_{s'}^{-}(\bk)
- a_s^{\dag\,-}(\bk) \circ a_{s'}^{+}(\bk)
\bigr\}
	\end{equation}
\vspace{-4ex}
	\begin{multline}	\label{11.5}
\ope{L}_{\mu\nu}
- (x_\mu - x_{0\,\mu}) \ope{P}_\nu
+ (x_\nu - x_{0\,\nu}) \ope{P}_\mu
= \tope{L}_{\mu\nu}
\\ =
\frac{1}{1+\tau(\U)} \sum_{s=1}^{3-\delta_{0m}} \!\! \int\! \Id^3\bk
 ( x_{0\,\mu}k_\nu - x_{0\,\nu}k_\mu ) |_{ k_0=\sqrt{m^2c^2+{\bs k}^2} }
\{
a_s^{\dag\,+}(\bk)\circ a_s^-(\bk) +
a_s^{\dag\,-}(\bk)\circ a_s^+(\bk)
\}
\\
+
\frac{\ih}{1+\tau(\U)}  \sum_{s,s'=1}^{3-\delta_{0m}} \int \Id^3\bk \,
l_{\mu\nu}^{ss'}(\bk)
\bigl\{
 a_s^{\dag\,+}(\bk) \circ a_{s'}^-(\bk) -
 a_s^{\dag\,-}(\bk) \circ a_{s'}^+(\bk)
\bigr\}
\displaybreak[2]\\
+
\frac{\ih}{2(1+\tau(\U))}  \sum_{s=1}^{3-\delta_{0m}} \int \Id^3\bk
\Bigl\{
a_s^{\dag\,+}(\bk)
\Bigl( \xlrarrow{ k_\mu \frac{\pd}{\pd k^\nu} }
     - \xlrarrow{ k_\nu \frac{\pd}{\pd k^\mu} } \Bigr)
\circ a_s^-(\bk)
\\ -
a_s^{\dag\,-}(\bk)
\Bigl( \xlrarrow{ k_\mu \frac{\pd}{\pd k^\nu} }
     - \xlrarrow{ k_\nu \frac{\pd}{\pd k^\mu} } \Bigr)
\circ a_s^+(\bk)
\Bigr\} \Big|_{ k_0=\sqrt{m^2c^2+{\bs k}^2} }
	\end{multline}
and, after normal ordering, they take the form (see~\eref{9.9}
and~\eref{9.8})
	\begin{equation}	\label{11.6}
\ope{S}_{\mu\nu}
=
  \frac{\ih }{1+\tau(\U)} \sum_{s,s'=1}^{3-\delta_{0m}} \int \Id^3\bk
\sigma_{\mu\nu}^{ss'}(\bk)
\bigl\{
  a_s^{\dag\,+}(\bk) \circ a_{s'}^{-}(\bk) -
  a_{s'}^{+}(\bk) \circ a_s^{\dag\,-}(\bk)
\bigr\}
	\end{equation}
\vspace{-4ex}
	\begin{multline}	\label{11.7}
\ope{L}_{\mu\nu}
=
\frac{1}{1+\tau(\U)} \sum_{s=1}^{3-\delta_{0m}} \!\! \int\! \Id^3\bk
 ( x_{\mu}k_\nu - x_{\nu}k_\mu ) |_{ k_0=\sqrt{m^2c^2+{\bs k}^2} }
\{
a_s^{\dag\,+}(\bk)\circ a_s^-(\bk) +
a_s^+(\bk) \circ a_s^{\dag\,-}(\bk)
\}
\\
+
\frac{\ih}{1+\tau(\U)}  \sum_{s,s'=1}^{3-\delta_{0m}} \int \Id^3\bk \,
l_{\mu\nu}^{ss'}(\bk)
\bigl\{
 a_s^{\dag\,+}(\bk) \circ a_{s'}^-(\bk) -
 a_{s'}^+(\bk) \circ a_s^{\dag\,-}(\bk)
\bigr\}
\displaybreak[2]\\
+
\frac{\ih}{2(1+\tau(\U))}  \sum_{s=1}^{3-\delta_{0m}} \int \Id^3\bk
\Bigl\{
a_s^{\dag\,+}(\bk)
\Bigl( \xlrarrow{ k_\mu \frac{\pd}{\pd k^\nu} }
     - \xlrarrow{ k_\nu \frac{\pd}{\pd k^\mu} } \Bigr)
\circ a_s^-(\bk)
\\ +
 a_s^+(\bk)
\Bigl( \xlrarrow{ k_\mu \frac{\pd}{\pd k^\nu} }
     - \xlrarrow{ k_\nu \frac{\pd}{\pd k^\mu} } \Bigr)
\circ a_s^{\dag\,-}(\bk)
\Bigr\} \Big|_{ k_0=\sqrt{m^2c^2+{\bs k}^2} }  .
	\end{multline}
So, formally, the replacement
$\sum_{s,s'=1}^{3}\mapsto\sum_{s,s'=1}^{3-\delta_{0m}}$ should be made and the
terms proportional to $\delta_{0m}$ should be deleted. As a result of these
changes, all terms proportional to $\delta_{0m}$ in all equations, starting
from~\eref{7.10} onwards, will disappear, i.e., for any $m$, we have
	\begin{equation}	\label{11.8}
\delta_{0m}\times (\cdots) = 0 ,
	\end{equation}
where the dots stand for some expressions, which depend on
$a_3^{\pm}(\bk)|_{m=0}$ and $a_3^{\dag\,\pm}(\bk)|_{m=0}$
 for $m=0$ and are set to zero for $m\not=0$; in particular, the
equations~\eref{11.3-1} hold (see~\eref{7.10-1} and~\eref{7.5}),
but~\eref{11.3} do not.

	In this way, the operators
$a_3^{\pm}(\bk)|_{m=0}$ and $a_3^{\dag\,\pm}(\bk)|_{m=0}$
disappear from all dynamical variables. So, if we extend the particle
interpretation on them,%
\footnote{~%
Such an extension requires the fulfillment of the commutation
relations~\eref{8.2} for $s=3$ and/or $t=3$ in the massless case.%
}
these operators will describe creation/annihilation of massless particles
with vanishing 4\ndash momentum, charge, spin and orbital angular momentum.
Naturally, such `particles' are completely unobservable. Thus, the properties
of the states
$a_3^{+}(\bk)|_{m=0}(\ope{X}_0)$ and $a_3^{\dag\,+}(\bk)|_{m=0}(\ope{X}_0)$
are similar to the ones of the vacuum (see~\eref{9.12}), but their
identification with the vacuum $\ope{X}_0$ requires additional and, in a
sense, artificial hypotheses for a self\ndash consistent development of the
theory.%
\footnote{~%
For instance, in a case of a neutral field, when
$a_s^{\dag\,\pm}(\bk)=a_s^{\pm}(\bk)$, one can satisfy~\eref{11.1} by
requiring
\(
  [ a_3^{+}(\bk) , a_s^{-}(\bk) ]_{+}|_{m=0}
= [ a_3^{-}(\bk) , a_s^{+}(\bk) ]_{+}|_{m=0},
\)
with $s=1,2$ and $[A,B]_+:=A\circ B + B\circ A$; in particular, this will be
valid if we assume $[ a_3^{\pm}(\bk) , a_s^{\mp}(\bk) ]_{+}|_{m=0}= 0$ for
$s=1,2$.%
}

	So, assuming the validity of~\eref{11.1} or~\eref{11.2}, we see that
these equations are the only place in the theory, where the operators
$a_3^{\pm}(\bk)|_{m=0}$ and $a_3^{\dag\,\pm}(\bk)|_{m=0}$
essentially appear.%
\footnote{~%
The initial operators $\U_\mu$ and $\U_\mu^\dag$ depend on these operators too
--- \emph{vide infra}.%
}
In fact, these equations should be regarded as equations of motion for the
mentioned operators, which operators do not enter in the field
equations~\eref{7.4} or in the commutation relations~\eref{8.2}. The
equations~\eref{11.1} or~\eref{11.2} possess always  the trivial solution
	\begin{equation}	\label{11.9}
a_3^{\pm}(\bk)|_{m=0} = 0 \quad a_3^{\dag\,\pm}(\bk)|_{m=0} = 0 ,
	\end{equation}
which agrees with the definition~\ref{Defn9.1} of the vacuum, but they may
have and other solutions. Since, at the moment, it seems that the operators
$a_3^{\pm}(\bk)|_{m=0}$ and $a_3^{\dag\,\pm}(\bk)|_{m=0}$
cannot lead to some physically measurable results, we shall not investigate
the problem for existence of solutions of~\eref{11.1} or~\eref{11.2},
different from~\eref{11.9}.

	Regardless of the fact that equations~\eref{11.1} or~\eref{11.2}
exclude a contribution of
$a_3^{\pm}(\bk)|_{m=0}$ and $a_3^{\dag\,\pm}(\bk)|_{m=0}$
from the dynamical variables, the initial operators $\U_\mu$ and
$\U_\mu^\dag$ depend on them via the operators
$\U_\mu^{\pm}(\bk)$ and $\U_\mu^{\dag\,\pm}(\bk)$ -- see~\eref{5.20new}
and~\eref{5.5}--\eref{5.6}.  As a consequence of this, the commutation
relations between different combinations of
$\U_\mu^{\pm}(\bk)$ and $\U_\mu^{\dag\,\pm}(\bk)$
also depend on
$a_3^{\pm}(\bk)|_{m=0}$ and $a_3^{\dag\,\pm}(\bk)|_{m=0}$
in the massless case. Moreover, one cannot calculate these relations without
additional assumptions, like~\eref{11.9} or the validity of~\eref{8.2} for
any $s,t=1,2,3$ in the massless case. Besides, the result depends essentially
on the additional condition(s) one assumes in the massless case. For
instance, if we assume~\eref{8.2} to hold for any $s,t=1,2,3$, when $m=0$,
then, for $(\bk,m)\not=(\bs0,0)$, we get
	\begin{equation}	\label{11.10}
[ \U_\mu^{\varepsilon}(\bk) , \U_\mu^{\varepsilon'}(\bk') ]_{\_}
=
f(\varepsilon,\varepsilon')
\bigl\{ 2c(2\pi\hbar)^3\sqrt{m^2c^2+\bk^2} \bigr\}^{-1} \delta^3(\bk-\bk')
\sum_{s=1}^{3} \{ v_\mu^s(\bk) v_\nu^s(\bk) \} ,
	\end{equation}
where
 $\varepsilon,\varepsilon'=\pm,\mp,\dag\,\pm,\dag\,\mp$
and we have applied~\eref{5.20} and~\eref{8.2} in the form
	\begin{subequations}	\label{11.11}
	\begin{gather}	\label{11.11a}
[ a_s^{\varepsilon}(\bk) , a_t^{\varepsilon'}(\bk') ]_{\_}
=
f(\varepsilon,\varepsilon') \delta_{st} \delta^3(\bk-\bk')
\\			\label{11.11b}
f(\varepsilon,\varepsilon')
:=
	\begin{cases}
0		&\text{for } \varepsilon,\varepsilon' = \pm, \dag\,\pm	\\
\pm\tau(U)	&\text{for }
		(\varepsilon,\varepsilon') = (\mp,\pm),(\dag\,\mp,\dag\,\pm)\\
\pm 1		&\text{for }
		(\varepsilon,\varepsilon') = (\mp,\dag\,\pm),(\dag\,\mp,\pm)
	\end{cases}
\ \ .
	\end{gather}
	\end{subequations}
Combining~\eref{11.10} and~\eref{4.22}, we obtain ($(\bk,m)\not=(\bs0,0)$)
	\begin{subequations}	\label{11.12}
	\begin{multline}	\label{11.12a}
[ \U_\mu^{\varepsilon}(\bk) , \U_\mu^{\varepsilon'}(\bk') ]_{\_}
\big|_{\substack{m\not=0}}
=
f(\varepsilon,\varepsilon')
\bigl\{ 2c(2\pi\hbar)^3\sqrt{m^2c^2+\bk^2} \bigr\}^{-1} \delta^3(\bk-\bk')
\\ \times
\bigg\{
	\begin{subarray}
\hphantom{} - \eta_{\mu\nu} + \frac{k_\mu k_\nu}{m^2c^2}
		\quad \text{for } \bk\not=\bs0			\\
\delta_{\mu\nu}
\hphantom{-}	\hphantom{+ \frac{k_\mu k_\nu}{m^2c^2}}
		\quad \text{for $\bk=\bs0$ and $\mu,\nu=1,2,3$}	\\
0
\hphantom{-}	\hphantom{\delta_{\mu} + \frac{k_\mu k_\nu}{m^2c^2}}
		\quad \text{otherwise }
	\end{subarray}
	\end{multline}
\vspace{-2.5ex}
	\begin{multline}
			\label{11.12b}
[ \U_\mu^{\varepsilon}(\bk) , \U_\mu^{\varepsilon'}(\bk') ]_{\_}
\big|_{\substack{m=0}}
=
f(\varepsilon,\varepsilon')
\bigl\{ 2c(2\pi\hbar)^3\sqrt{m^2c^2+\bk^2} \bigr\}^{-1} \delta^3(\bk-\bk')
\\ \times
\biggl(
- \eta_{\mu\nu} + \frac{k_\mu k_\nu}{\bk^2}
\times \bigg\{
	\begin{subarray}{l}
2      \quad \text{for } \mu=\nu=0		\\
0      \quad \text{for } \mu,\nu=1,2,3	\\
1      \quad \text{otherwise }
	\end{subarray}
\biggr)
\ .
	\end{multline}
	\end{subequations}

	On the other hand, if we assume~\eref{11.9}, then~\eref{11.10} must be
replaced with
	\begin{equation}	\label{11.12-1}
[ \U_\mu^{\varepsilon}(\bk) , \U_\mu^{\varepsilon'}(\bk') ]_{\_}
=
f(\varepsilon,\varepsilon')
\bigl\{ 2c(2\pi\hbar)^3\sqrt{m^2c^2+\bk^2} \bigr\}^{-1} \delta^3(\bk-\bk')
\sum_{s=1}^{3-\delta_{0m}} \{ v_\mu^s(\bk) v_\nu^s(\bk) \} ,
	\end{equation}
as a consequence of~\eref{5.20},~\eref{11.11} with
\(
s,t= \big\{
	\begin{smallmatrix}
1,2,3		&\text{if } m\not=0	\\
1,2\hphantom{,3}&\text{if } m=0
	\end{smallmatrix} \, ,
\)
and~\eref{11.9}. So, for $m\not=0$, the relation~\eref{11.12-1} reduces
to~\eref{11.12a} (see~\eref{4.22}),%
\footnote{~%
For $m\not=0$ and $\bk\not=\bs0$, as one can expect, the commutation
relations~\eref{8.2} and~\eref{11.12a} reproduce, due to~\eref{6.48}, the
known ones for a massive free vector field in Heisenberg
picture~\cite{Bogolyubov&Shirkov,Bjorken&Drell,Itzykson&Zuber}.%
}
but for $m=0$ it reads ($(\bk,m)\not=(\bs0,0)$)
	\begin{equation}
			\label{11.13}
[ \U_\mu^{\varepsilon}(\bk) , \U_\mu^{\varepsilon'}(\bk') ]_{\_}
\big|_{\substack{m=0}}
=
f(\varepsilon,\varepsilon')
\bigl\{ 2c(2\pi\hbar)^3\sqrt{m^2c^2+\bk^2} \bigr\}^{-1} \delta^3(\bk-\bk')
\times
\big\{
	\begin{subarray}
\hphantom{} \delta_{\mu\nu}		\quad \text{for } \mu,\nu=1,2	\\
0
\hphantom{_{\mu\nu}}
\quad \text{otherwise}
	\end{subarray}
\ ,
	\end{equation}
due to~\eref{4.23}. Evidently, the equations~\eref{11.12a} and~\eref{11.13}
coincide for $\mu,\nu=1,2$ but otherwise are, generally, different.

	It is now time to be paid special attention to the
\emph{electromagnetic field}.%
\footnote{~%
In fact, the description of electromagnetic field was the primary main
reason for the inclusion of the massless case in the considerations in the
preceding sections.%
}
As it is well
known~\cite{Bogolyubov&Shirkov,Bjorken&Drell,Itzykson&Zuber,Ryder-QFT}, this
field is a massless neutral vector field whose operators, called the
electromagnetic potentials, are usually denoted by $\ope{A}_\mu$ and are such
that
	\begin{equation}	\label{11.14}
\ope{A}_\mu^\dag = \ope{A}_\mu .
	\end{equation}
The (second) quantization of electromagnetic field meets some difficulties,
described in \emph{loc.\ cit.}, the causes for which are well\ndash described
in~\cite[\S~82]{Bjorken&Drell-2} (see also~\cite{Akhiezer&Berestetskii}).
The closest to our approach is the so\ndash called Gupta\ndash Bleuler
quantization~\cite{Bogolyubov&Shirkov,Akhiezer&Berestetskii,Itzykson&Zuber,
Bjorken&Drell-2} in the way it is described in~\cite{Bogolyubov&Shirkov}.
However, our method is quite different from it as we quantize only the
independent degrees of freedom, as a result of which there is no need of
considering indefinite metric, `time' (`scalar') photons and similar
objects.%
\footnote{~%
The only such a problem we meet is connected with the `longitudinal' photons
--- \emph{vide infra}.%
}
The idea of most such methods is to be started from some Lagrangian, to be
applied the standard canonical quantization
procedure~\cite{Roman-QFT,Bjorken&Drell-2}, and, then, to the electromagnetic
potentials to be imposed some subsidiary conditions, called \emph{gauge
conditions}, by means of which is (partially) fixed the freedom in the field
operators left by the field equations.

	In our scheme, the free electromagnetic field is described via~4
Hermitian operators $\ope{A}_\mu$ (for which $\tau(A)=1$ --- see~\eref{3.2}),
the Lagrangian~\eref{3.7} with $m=0$, \ie
	\begin{equation}	\label{11.15}
\ope{L}
 =
\frac{1}{2} c^2
[ \ope{A}_\nu,\ope{P}_\mu ]_{\_} \circ [ \ope{A}^\nu,\ope{P}^\mu ]_{\_}
- \frac{1}{2}c^2
[ \ope{A}^{\mu},\ope{P}_\mu ]_{\_} \circ [ \ope{A}^\nu,\ope{P}_\nu ]_{\_}
	\end{equation}
and the Lorenz conditions~\eref{3.16} with $\U=\ope{A}$, \ie
	\begin{equation}	\label{11.16}
[ \ope{A}_\mu , \ope{P}^\mu ]_{\_} = 0 .
	\end{equation}
It should be emphasized, the Lorenz condition~\eref{11.16} is imposed
directly on the field operators, not on the `physical' states etc.\ as in the
Gupta\ndash Bleuler formalism. So,~\eref{11.14}--\eref{11.16} describe an
electromagnetic field in \emph{Lorenz gauge}.

	Thus, to specialize the general theory form the preceding sections to
the case of electromagnetic field, one should put in it (see~\eref{11.14} and
sections~\ref{Sect5} and~\ref{Sect6})
	\begin{equation}	\label{11.17}
m=0\quad\!
\U=\ope{A}\quad
\tau(\U)=1\quad\!
\ope{A}_\mu^\dag=\ope{A}_\mu \quad
\ope{A}_\mu^{\dag\,\pm}=\ope{A}_\mu^{\pm} \quad
\ope{A}_\mu^{\dag\,\pm}(\bk)=\ope{A}_\mu^{\pm}(\bk) \quad
a_s^{\dag\,\pm}(\bk)=a_s^{\pm}(\bk) .
	\end{equation}

	It is important to be emphasized, the equations~\eref{11.17}
reduce~\eref{11.1} to
	\begin{subequations}	\label{11.18}
	\begin{align}	\label{11.18a}
\sum_{s=1,2} \int\Id^3\bk  \sigma_{\mu\nu}^{s3}(\bk)
\bigl\{
[ a_s^{+}(\bk) , a_3^{-}(\bk) ]_{+} -
[ a_s^{-}(\bk) , a_3^{+}(\bk) ]_{+}
\bigr\}
& = 0
\\			\label{11.18b}
\sum_{s=1,2} \int\Id^3\bk  l_{\mu\nu}^{s3}(\bk)
\bigl\{
[ a_s^{+}(\bk) , a_3^{-}(\bk) ]_{+} -
[ a_s^{-}(\bk) , a_3^{+}(\bk) ]_{+}
\bigr\}
& = 0,
	\end{align}
	\end{subequations}
where $[A,B]_+:=A\circ B + B\circ A$ is the anticommutator of operators $A$
and $B$.
Since in~\eref{11.18} enter the anticommutators
$[a_s^{\pm}(\bk),a_3^{\mp}(\bk) ]_{+}$ and $a_3^{\mp}(\bk)$ are actually
free parameters, the contributions of $a_3^{\pm}(\bk)$ in the spin and
orbital momentum operators can be eliminated via the following change in the
theory. Redefine the normal products of creation and/or annihilation
operators by assigning to them an additional (with respect to the definition
in Sect.~\ref{Sect9}) multiplier (sign) equal to $(-1)^f$, where $f$ is
equal to the number of transpositions of the operators $a_3^{\pm}(\bk)$,
relative to $a_1^{\mp}(\bk)$ and $a_2^{\mp}(\bk)$, required to be obtained
the normal form of a product, \ie all creation operators to be to the left of
all annihilation ones. Evidently, such a (redefined) normal ordering
procedure transforms~\eref{11.18} into identities and, consequently, after it
all dynamical variables become independent of the operators $a_3^{\pm}(\bk)$.
	A similar result will be valid if the normal ordering procedure, as
defined in Sect.~\eref{Sect9}, holds only for the operators $a_1^{\pm}(\bk)$
and $a_2^{\pm}(\bk)$ and the operators $a_3^{\pm}(\bk)$ anticommute with
them,
	\begin{equation}	\label{11.20}
[ a_3^{\mp}(\bk), a_s^{\pm}(\bk) ]_{+} = 0 \qquad\text{for } s=1,2 .
	\end{equation}
Notice, if we put (cf.~\eref{11.9})
	\begin{equation}	\label{11.21}
a_3^{\pm}(\bk) = 0 ,
	\end{equation}
the definitions of vacuum, normal ordering procedure, and
equations~\eref{11.20} (and hence~\eref{11.1} and~\eref{11.2}) will be
satisfied. Besides, these choices will naturally exclude from the theory the
`longitudinal' photons, represented in our theory by the vector
$a_3^{+}(\bk)(\ope{X}_0)$, which have identically vanishing dynamical
characteristics.

	The above discussion shows that the operators $a_3^{\pm}(\bk)$ can
naturally be considered as fermi or bose operators that anticommute with
$a_s^{\pm}(\bk)$, $s=1,2$, \ie with anomalous commutation relations between
both sets of operators~\cite[appendix~F]{Ohnuki&Kamefuchi}.%
\footnote{~%
Since this excludes the operators $a_3^{\pm}(\bk)$ from all physically
significant quantities, one is free to choose bilinear or other commutation
relations between $a_3^{\pm}(\bk)$.%
}
Besides, this agrees with the above-modified normal ordering procedure, which
will be accepted below in the present section.

	As a result of~\eref{11.17}, the commutation relations~\eref{8.2} for
an electromagnetic field (in Lorenz gauge) read%
\footnote{~%
According to our general considerations, the relations~\eref{11.22} are
equivalent to the Maxwell(\ndash Lorentz) equations, but written in
terms of creation and annihilation operators.%
}
	\begin{equation}	\label{11.22}
[ a_s^{\pm}(\bk), a_t^{\pm}(\bk') ]_{\_} = 0 \quad
[ a_s^{\mp}(\bk), a_t^{\pm}(\bk') ]_{\_} = \pm \delta_{st} \delta^3(\bk-\bk')
\qquad\text{for } s,t=1,2
	\end{equation}
and the operators of the dynamical variables, after the (redefined) normal
ordering is performed, for this field are (see~\eref{11.18}
and~\eref{9.6}--\eref{9.11} with $m=0$ and $\tau(\U)=1$)
	\begin{align}	\label{11.23}
\ope{P}_\mu
& =
\sum_{s=1,2} \int
  k_\mu |_{ k_0=\sqrt{\bk^2} } \,
a_s^{+}(\bk)\circ a_s^-(\bk)
\Id^3\bk
\displaybreak[1]\\	\label{11.24}
\ope{Q} & = 0
	\end{align}
\vspace{-4ex}
	\begin{multline}	\label{11.25}
\ope{L}_{\mu\nu}
=
\sum_{s=1,2} \int \Id^3\bk
 ( x_{\mu}k_\nu - x_{\nu}k_\mu ) |_{ k_0=\sqrt{\bk^2} } \,
a_s^{+}(\bk)\circ a_s^-(\bk)
+
\ih \sum_{s,s'=1,2} \int \Id^3\bk \,
l_{\mu\nu}^{ss'}(\bk)
 a_s^{+}(\bk) \circ a_{s'}^-(\bk)
\displaybreak[2]\\
+
\ih \frac{1}{2}  \sum_{s=1,2} \int \Id^3\bk
\Bigl\{
a_s^{+}(\bk)
\Bigl( \xlrarrow{ k_\mu \frac{\pd}{\pd k^\nu} }
     - \xlrarrow{ k_\nu \frac{\pd}{\pd k^\mu} } \Bigr)
\circ a_s^-(\bk)
\Bigr\} \Big|_{ k_0=\sqrt{\bk^2} }
	\end{multline}
\vspace{-2ex}
	\begin{align}	\label{11.26}
\ope{S}_{\mu\nu}
& =
  \ih \sum_{s,s'=1,2} \int \Id^3\bk
\sigma_{\mu\nu}^{ss'}(\bk)
  a_s^{+}(\bk) \circ a_{s'}^{-}(\bk)
\\			\label{11.27}
\bs{\ope{R}} & = \bs0
\\			\label{11.28}
   \ope{S}^1 & = \ope{S}^2  = 0 \quad
\ope{S}^3 =
  \ih \int \Id^3\bk
\bigl\{
a_1^{+}(\bk) \circ a_{2}^{-}(\bk) -
a_2^{+}(\bk) \circ a_{1}^{-}(\bk)
\bigr\} ,
	\end{align}
where, for the derivation of~\eref{11.27} and~\eref{11.28}, we have
used~\eref{6.39}, \eref{4.19}, \eref{4.21}, and that, formally,
\eref{11.26} corresponds to~\eref{9.9} with $a_3^\pm(\bk)=0$. Thus, the
operators $a_3^\pm(\bk)$ do not enter in all of the ;dynamical variables.%
\footnote{~%
This situation should be compared with similar one in the Gupta\ndash Bleuler
formalism. In it the contribution of the `time' (`scalar') and `longitudinal'
photons, the last corresponding to our states $a_3^\pm(\bk)(\ope{X}_0)$, is
removed from the \emph{average 4\ndash momentum of the admissible states},
but, for example, the `longitudinal' photons have a generally \emph{non\ndash
vanishing} part in the vector of spin $\bs{\ope{S}}$ --- see,
e.g.,~\cite[eq.~(12.19)]{Bogolyubov&Shirkov}.%
}

	From~\eref{11.23}--\eref{11.28}, it is evident that the particles of
an electromagnetic field, called photons, coincide with their antiparticles,
which agrees with the general considerations in Sect.~\ref{Sect10}. Besides,
the state vectors $b_1^+(\bk)(\ope{X}_0)$ and $b_2^+(\bk)(\ope{X}_0)$, with
$b_s^\pm(\bk)$ given by~\eref{6.41}, describe photons with 4\ndash momentum
$(\sqrt{\bk^2},\bk)$, zero charge, and vectors of spin $\bs{\ope{R}}=\bs0$
and $\bs{\ope{S}}=(0,0,+\hbar)$ or $\bs{\ope{S}}=(0,0,-\hbar)$
(see~\eref{10.21}), \ie their spin vector $\bs{\ope{S}}$ is collinear with
$\bk$ with projection value $+\hbar$ or $-\hbar$, respectively, on its
direction.

	It is worth to be mentioned, the commutation
relations~\eref{7.8}--\eref{7.11} for an electromagnetic field take their
`ordinary' form, \ie
	\begin{align}
			\label{11.29}
[ \ope{P}_\mu , \ope{P}_\nu ]_{\_} & = 0
\\			\label{11.30}
[ \tope{Q} , \ope{P}_\mu ]_{\_} & = 0
\\			\label{11.31}
[ \tope{S}_{\mu\nu} , \ope{P}_\lambda ]_{\_} & = 0
\\			\label{11.32}
[ \tope{L}_{\mu\nu} , \ope{P}_\lambda ]_{\_}
& =
- \ih \{ \eta_{\lambda\mu}\ope{P}_\nu - \eta_{\lambda\nu}\ope{P}_\mu \} ,
	\end{align}
as a consequence of which the dynamical variables in momentum picture are
(see~\eref{7.18}--\eref{7.20})
	\begin{align}
			\label{11.33}
\ope{Q} & = \tope{Q}
\\			\label{11.34}
\ope{S}_{\mu\nu}
&= \tope{S}_{\mu\nu}
\\			\label{11.35}
\ope{L}_{\mu\nu}
& =
\tope{L}_{\mu\nu}
+ (x_\mu - x_{0\,\mu}) \ope{P}_\nu -
  (x_\nu - x_{0\,\nu}) \ope{P}_\mu  .
	\end{align}

	At last, as we said above, the commutators
 $[A_\mu^\pm(\bk),A_\nu^\pm(\bk)]_{\_}$ and
 $[A_\mu^\mp(\bk),A_\nu^\pm(\bk)]_{\_}$
cannot be computed without knowing the explicit form of
 $[a_s^\pm(\bk),a_3^\pm(\bk)]_{\_}$ and
 $[a_s^\mp(\bk),a_3^\pm(\bk)]_{\_}$
for $s=1,2,3$. For instance, the additional conditions~\eref{11.21} lead
to~\eref{11.13} with $\ope{A}$ for $\U$ and
$\varepsilon,\varepsilon'=\pm,\mp$.

	The so-obtained quantization rules for electromagnetic field,
\ie equations~\eref{11.22} together with~\eref{11.21}, coincide with the ones
when it is quantized in \emph{Coulomb}
gauge~\cite{Bjorken&Drell-2,Ryder-QFT}, in which is assumed

	\begin{equation}	\label{11.36}
\ope{A}_0=0 \quad \sum_{a=1}^{3}[\ope{P}^a,\ope{A}_a]_{\_}=0
	\end{equation}
in momentum picture. This is not accidental as~\eref{11.36} is a special case
of~\eref{3.16}. In fact, by virtue of~\eref{4.5}, it is equivalent to
$k^a\ope{A}_a=0$ (with $k^2=k_0^2-\bk^2=0$) which is tantamount to
	\begin{equation}	\label{11.37}
0 =
( a_3^+(\bk) + a_3^-(\bk) ) ( k^a v_a^3(\bk)\big|_{m=0} )
=
	\begin{cases}
a_3^+(\bk) + a_3^-(\bk)	&\text{for } \bk\not=\bs0 \\
0			&\text{for } \bk=\bs0
	\end{cases}
\ ,
	\end{equation}
due to~\eref{5.5},~\eref{5.6},~\eref{5.20} and~\eref{4.19}--\eref{4.20}.
Therefore any choice of $a_3^\pm(\bk)$ such that
$a_3^+(\bk) + a_3^-(\bk) = 0$ reduces the Lorenz gauge to the Coulomb one.
However, the particular choice~\eref{11.21} completely reduces our
quantization method to the one in Coulomb gauge, as a little more derailed
comparison of the both methods reveals. We shall end this discussion with the
remark that the choice~\eref{11.21} is external to the Lagrangian formalism
and, of course, it is not the only possible one in that scheme.


\section{On the choice of Lagrangian}
\label{Sect12}

	Our previous exploration of free vector fields was based on the
Lagrangian (see~\eref{3.1})
	 \begin{equation}	\label{12.1}
\tope{L}'= \tope{L}
=
\frac{m^2c^4}{1+\tau(\tU)} \tU_\mu^\dag \circ \tU^\mu
+
\frac{c^2\hbar^2}{1+\tau(\tU)}
\bigl\{ - (\pd_\mu\tU_\nu^\dag) \circ (\pd^\mu\tU^\nu)
        + (\pd_\mu\tU^{\mu\dag}) \circ (\pd_\nu\tU^\nu) \bigr\}
	\end{equation}
in Heisenberg picture. In it the field operators $\tU_\mu$ and their Hermitian
conjugate $\tU_\mu^\dag$ do not enter on equal footing: in a sense,
$\tU_\mu^\dag$ are `first' and $\tU_\mu$ are `second' in order (counting from
left to right) unless the field is neutral/Hermitian. Since $\tU_\mu$
and $\tU_\mu^\dag$ are associated with the operators $a_s^{\pm}$ and
$a_s^{\dag\,\pm}$ (see Sect.~\ref{Sect5}), which create/annihilate field's
particles and antiparticles, respectively, the Lagrangian~\eref{12.1}
describes the particles and antiparticles in a non\ndash symmetric way, which
is non\ndash desirable for a free field as for it what should be called a
particle or antiparticle is more a convention than a natural distinction.
This situation is usually corrected via an additional  condition in the
theory, such as the charge symmetry, spin\ndash statistics theorem etc. Its
sense is the inclusion in the theory of the symmetry
particle~$\leftrightarrow$~antiparticle, which in terms of the creation and
annihilation operators should be expressed via theory's invariance under the
change $a_s^{\pm}(\bk)\leftrightarrow a_s^{\dag\,\pm}(\bk)$. As we demonstrated
in~\cite{bp-QFTinMP-scalars,bp-QFTinMP-spinors} for free scalar and
spin~$\frac{1}{2}$ fields, this symmetry/invariance can be incorporated in
the initial Lagrangian, from which the theory is constructed. Below we shall
show how this can be achieved for free vector fields satisfying the Lorenz
condition.

	As an alternative to the Lagrangian~\eref{12.1}, one can consider
	 \begin{equation}	\label{12.2}
\tope{L}^{\prime\prime}
\! = \!
\frac{m^2c^4}{1+\tau(\tU)} \tU_\mu \circ \tU^{\mu\dag}
+
\frac{c^2\hbar^2}{1+\tau(\tU)}
\bigl\{ - (\pd_\mu\tU_\nu) \circ (\pd^\mu\tU^{\nu\dag})
        + (\pd_\mu\tU^{\mu}) \circ (\pd_\nu\tU^{\nu\dag}) \bigr\}
\! = \!
\tope{L}'\big|_{\tU_\mu\leftrightarrow\tU_\mu^\dag}
	\end{equation}
in which the places, where the operators $\tU_\mu$ and $\tU_\mu^\dag$ are
situated in~\eref{12.1}, are interchanged. In terms of particles and
antiparticles, this means that we call fields particles antiparticles and
\emph{vice versa}, or, equivalently, that the change
$a_s^{\pm}(\bk)\leftrightarrow a_s^{\dag\,\pm}(\bk)$
has been made. Obviously, the Lagrangian~\eref{12.2} suffers from the same
problems as~\eref{12.1}. However, judging by our experience
in~\cite{bp-QFTinMP-scalars} and partially in~\cite{bp-QFTinMP-spinors}, we
can expect that the half\ndash sum of the Lagrangians~\eref{12.1}
and~\eref{12.2}, \ie
	 \begin{multline}	\label{12.3}
\tope{L}^{\prime\prime\prime}
 =
\frac{m^2c^4}{2(1+\tau(\tU))}
\bigl\{\tU_\mu^\dag \circ \tU^\mu + \tU_\mu \circ \tU^{\mu\dag} \bigr\}
+
\frac{c^2\hbar^2}{1+\tau(\tU)}
\bigl\{ - (\pd_\mu\tU_\nu^\dag) \circ (\pd^\mu\tU^\nu)
        + (\pd_\mu\tU^{\mu\dag}) \circ (\pd_\nu\tU^\nu)
\\	- (\pd_\mu\tU_\nu) \circ (\pd^\mu\tU^{\nu\dag})
        + (\pd_\mu\tU^{\mu}) \circ (\pd_\nu\tU^{\nu\dag}) \bigr\}
 =
\frac{1}{2} (\tope{L}^{\prime} + \tope{L}^{\prime\prime} ) ,
	\end{multline}
is one of the Lagrangians we are looking for, as it is invariant under the
change $\tU_\mu\leftrightarrow\tU_\mu^\dag$.

	To any one of the Lagrangians~\eref{12.1}--\eref{12.3}, we add the
Lorenz conditions
	\begin{equation}	\label{12.4}
\pd^\mu\tU_\mu = 0 \quad \pd^\mu\tU_\mu^\dag = 0 ,
	\end{equation}
which are symmetric under the transformation
$\tU_\mu\leftrightarrow\tU_\mu^\dag$ and for $m=0$ are additional conditions
for the Lagrangian formalism, but for $m\not=0$ they are consequences from
the field equations (see Sect.~\ref{Sect3} and below).

	According to the general rules of Sect.~\ref{Sect2}
(see~\eref{12.114} and~\eref{2.5}), the Lagrangians~\eref{12.1}--\eref{12.3}
and the Lorenz conditions~\eref{12.4} in momentum picture respectively are:
	\begin{align}	\label{12.5}
\ope{L}^{\prime}
& =
\frac{m^2c^4}{1+\tau(\U)} \U_\mu^\dag\circ\U^\mu
+
\frac{c^2}{1+\tau(\U)}
\bigl\{
[ \U_\nu^\dag , \ope{P}_\mu ]_{\_} \circ [ \U^\nu , \ope{P}^\mu ]_{\_}
-
[ \U^{\mu\dag} , \ope{P}_\mu ]_{\_} \circ [ \U^\nu , \ope{P}_\nu ]_{\_}
\bigr\}
\\			\label{12.6}
\ope{L}^{\prime\prime}
& =
\frac{m^2c^4}{1+\tau(\U)} \U_\mu\circ\U^{\mu\dag}
+
\frac{c^2}{1+\tau(\U)}
\bigl\{
[ \U_\nu , \ope{P}_\mu ]_{\_} \circ [ \U^{\nu\dag}  , \ope{P}^\mu ]_{\_}
-
[ \U^{\mu} , \ope{P}_\mu ]_{\_} \circ [ \U^{\nu\dag} , \ope{P}_\nu ]_{\_}
\bigr\}
	\end{align}
\vspace{-4ex}
	\begin{multline}	\label{12.7}
\ope{L}^{\prime\prime\prime}
=
\frac{m^2c^4}{2(1+\tau(\U))}
\bigl\{ \U_\mu^\dag\circ\U^\mu + \U_\mu\circ\U^{\mu\dag} \bigr\}
 +
\frac{c^2}{2(1+\tau(\U))}
\bigl\{
[ \U_\nu^\dag , \ope{P}_\mu ]_{\_} \circ [ \U^\nu , \ope{P}^\mu ]_{\_}
\\ -
[ \U^{\mu\dag} , \ope{P}_\mu ]_{\_} \circ [ \U^\nu , \ope{P}_\nu ]_{\_}
 +
[ \U_\nu , \ope{P}_\mu ]_{\_} \circ [ \U^{\nu\dag}  , \ope{P}^\mu ]_{\_}
-
[ \U^{\mu} , \ope{P}_\mu ]_{\_} \circ [ \U^{\nu\dag} , \ope{P}_\nu ]_{\_}
\bigr\}
	\end{multline}
\vspace{-4ex}
	\begin{equation}	\label{12.8}
[ \U_{\mu} , \ope{P}^\mu ]_{\_} = 0 \quad
[ \U_\mu^\dag , \ope{P}^\mu ]_{\_} = 0  .
	\end{equation}

	The derivatives of the above Lagrangians happen to coincide and are
as follows:%
\footnote{~%
This assertion is valid if the derivatives are calculated according to the
classical rules of analysis of commuting variables, as it is done below. Such
an approach requires additional rules for ordering of the operators entering
into the expressions for the dynamical variables, as the ones presented below.
Both of these assumptions, in the particular cases we are considering here,
have their rigorous explanation in a different way for computing derivatives
of \emph{non}\ndash commuting variables, as it is demonstrated
in~\cite{bp-QFT-action-principle}, to which paper the reader is referred for
further details.%
}
	\begin{equation}	\label{12.9}
	\begin{split}
\frac{\pd\ope{L}^{\prime}}{\pd\U^\mu} =
\frac{\pd\ope{L}^{\prime\prime}}{\pd\U^\mu} =
\frac{\pd\ope{L}^{\prime\prime\prime}}{\pd\U^\mu}
& =
m^2c^4 \U_\mu^\dag
\quad
\frac{\pd\ope{L}^{\prime}}{\pd\U^{\mu\dag}} =
\frac{\pd\ope{L}^{\prime\prime}}{\pd\U^{\mu\dag}} =
\frac{\pd\ope{L}^{\prime\prime\prime}}{\pd\U^{\mu\dag}}
=
m^2c^4 \U_\mu
\\
\pi_{\mu\lambda}
= \frac{\pd\ope{L}^{\prime}}{\pd y^{\mu\lambda}}
= \frac{\pd\ope{L}^{\prime\prime}}{\pd y^{\mu\lambda}}
= \frac{\pd\ope{L}^{\prime\prime\prime}}{\pd y^{\mu\lambda}}
& =
  \ih c^2 [\U_\mu^\dag,\ope{P}_\lambda]_{\_}
- \ih c^2 \eta_{\mu\lambda} [\U^{\nu\dag},\ope{P}_\nu]_{\_}
\\
\pi_{\mu\lambda}^\dag
= \frac{\pd\ope{L}^{\prime}}{\pd y^{\mu\lambda\dag}}
= \frac{\pd\ope{L}^{\prime\prime}}{\pd y^{\mu\lambda\dag}}
= \frac{\pd\ope{L}^{\prime\prime\prime}}{\pd y^{\mu\lambda\dag}}
& =
  \ih c^2 [\U_\mu,\ope{P}_\lambda]_{\_}
- \ih c^2 \eta_{\mu\lambda} [\U^{\nu},\ope{P}_\nu]_{\_}
	\end{split}
	\end{equation}
with
\(
y_{\mu\lambda} := \iih [\U_\mu,\ope{P}_\lambda]_{\_}
\text{ and }
y_{\mu\lambda}^\dag := \iih [\U_\mu^\dag,\ope{P}_\lambda]_{\_} .
\)

	As a consequence of~\eref{12.9}, the field equations of the
Lagrangians~\eref{12.5}--\eref{12.7} coincide and are given by~\eref{3.12}
(and~\eref{12.8} as an additional conditions/equation for $m=0$) which, for
$m\not=0$, split into the Klein\ndash Gordon equations~\eref{3.15} and the
Lorenz conditions~\eref{12.8}. From here it follows that the material of
sections~\ref{Sect4} and~\ref{Sect5} remains valid without any changes for
the Lagrangian theories arising from any one of the
Lagrangians~\eref{12.5}--\eref{12.7} (under the Lorenz conditions in the
massless case).

	The densities of the operators of the dynamical variables for the
Lagrangian~\eref{12.1} are given via~\eref{3.18}--\eref{3.22}. Similarly,
the energy\ndash momentum tensor, (charge) current and spin angular momentum
operators for the Lagrangians~\eref{12.2} and~\eref{12.3} respectively are:%
\footnote{~%
Excluding the spin operators, the other density operators can be obtained, by
virtue of~\eref{12.4}, as sums of similar ones corresponding to $\U_0$,
$\U_1$, $\U_2$ and $\U_3$ and considered as free scalar fields ---
see~\cite{bp-QFTinMP-scalars}. For a rigorous derivation,
see~\cite{bp-QFT-action-principle}.%
}
	\begin{subequations}	\label{12.10}
	\begin{align}
			\label{12.10a}
	\begin{split}
\tope{T}_{\mu\nu}^{\prime\prime}
& =
\frac{1}{1+\tau(\tU)}
\bigl\{
(\pd_\nu\tU^\lambda)\circ\tope{\pi}_{\lambda\mu} +
\tope{\pi}_{\lambda\mu}^\dag\circ(\pd_\nu\tU^{\lambda\dag})
\bigr\}
- \eta_{\mu\nu}\tope{L}^{\prime\prime}
\\ & =
- \frac{c^2\hbar^2}{1+\tau(\tU)}
\bigl\{
(\pd_\mu\tU_\lambda)\circ(\pd_\nu\tU^{\lambda\dag}) +
(\pd_\nu\tU_\lambda)\circ(\pd_\mu\tU^{\lambda\dag})
\bigr\}
- \eta_{\mu\nu}\tope{L}^{\prime\prime}
	\end{split}
\displaybreak[1]\\			\label{12.10b}
	\begin{split}
\tope{J}_{\mu}^{\prime\prime}
& =
\frac{q}{\ih}
\bigl\{
\tU^\lambda	\circ\tope{\pi}_{\lambda\mu} -
\tope{\pi}_{\lambda\mu}^\dag\circ\tU^{\lambda\dag}
\bigr\}
=
\ih q c^2
\bigl\{
- ( \pd_\mu\tU_\lambda)\circ\tU^{\lambda\dag} +
\tU_\lambda\circ(\pd_\mu\tU^{\lambda\dag})
\bigr\}
	\end{split}
\displaybreak[2]\\			\label{12.10c}
	\begin{split}
\tope{S}_{\mu\nu}^{\prime\prime\lambda}
& : =
\frac{1}{1+\tau(\tU)}
\bigl\{
( I_{\rho\mu\nu}^{\sigma}\tU_\sigma ) \circ \tope{\pi}^{\rho\lambda} +
\tope{\pi}^{\rho\lambda\dag}\circ( I_{\rho\mu\nu}^{\dag\, \sigma}\tU_\sigma^\dag )
\bigr\}
\\ & =
\frac{\hbar^2c^2}{1+\tau(\tU)}
\bigl\{
(\pd^\lambda \tU_\mu) \circ \tU_\nu^\dag -
(\pd^\lambda \tU_\nu) \circ \tU_\mu^\dag -
\tU_\mu \circ (\pd^\lambda\tU_\nu^\dag) +
\tU_\mu \circ (\pd^\lambda\tU_\mu^\dag)
\bigr\}
	\end{split}
	\end{align}
	\end{subequations}
\vspace{-4ex}
	\begin{subequations}	\label{12.11}
	\begin{align}
			\label{12.11a}
	\begin{split}
\tope{T}_{\mu\nu}^{\prime\prime\prime}
& =
\frac{1}{2(1+\tau(\tU))}
\bigl\{
\tope{\pi}_{\lambda\mu}\circ(\pd_\nu\tU^\lambda) +
(\pd_\nu\tU^{\lambda\dag})\circ\tope{\pi}_{\lambda\mu}^\dag
\\ &\hphantom{=}
+
(\pd_\nu\tU^\lambda)\circ\tope{\pi}_{\lambda\mu} +
\tope{\pi}_{\lambda\mu}^\dag\circ(\pd_\nu\tU^{\lambda\dag})
\bigr\}
- \eta_{\mu\nu}\tope{L}^{\prime\prime\prime}
 =
 \frac{1}{2}
\bigl\{ \tope{T}_{\mu\nu}^{\prime} + \tope{T}_{\mu\nu}^{\prime\prime} \bigr\}
	\end{split}
\displaybreak[1]\\			\label{12.11b}
	\begin{split}
\tope{J}_{\mu}^{\prime\prime\prime}
& =
\frac{q}{2\ih}
\bigl\{
\tope{\pi}_{\lambda\mu}\circ\tU^\lambda -
\tU^{\lambda\dag}\circ\tope{\pi}_{\lambda\mu}^\dag +
\tU^\lambda	\circ\tope{\pi}_{\lambda\mu} -
\tope{\pi}_{\lambda\mu}^\dag\circ\tU^{\lambda\dag}
\bigr\}
=
\frac{1}{2}
\bigl\{ \tope{J}_{\mu}^{\prime} + \tope{J}_{\mu}^{\prime\prime} \bigr\}
	\end{split}
\displaybreak[2]\\			\label{12.11c}
	\begin{split}
\tope{S}_{\mu\nu}^{\prime\prime\prime\lambda}
& : =
\frac{1}{2(1+\tau(\tU))}
\bigl\{
\tope{\pi}^{\rho\lambda} \circ ( I_{\rho\mu\nu}^{\sigma}\tU_\sigma ) +
( I_{\rho\mu\nu}^{\dag\, \sigma}\tU_\sigma^\dag )\circ\tope{\pi}^{\rho\lambda\dag}
\\ &\hphantom{=}
+
( I_{\rho\mu\nu}^{\sigma}\tU_\sigma ) \circ \tope{\pi}^{\rho\lambda} +
\tope{\pi}^{\rho\lambda\dag}\circ( I_{\rho\mu\nu}^{\dag\, \sigma}\tU_\sigma^\dag )
\bigr\}
 =
\frac{1}{2}
\bigl\{ \tope{S}_{\mu\nu}^{\prime\lambda}
+ 	\tope{S}_{\mu\nu}^{\prime\prime\lambda}
\bigr\} .
	\end{split}
	\end{align}
	\end{subequations}

	Thus, we see that the dynamical variables derived from
$\ope{L}^{\prime\prime}$ can be obtained from the ones for
$\ope{L}^{\prime}=\ope{L}$ by making the change
 $\U_\mu\leftrightarrow\U_\mu^\dag$ and reversing the current's sign.
Besides, the dynamical variables for $\ope{L}^{\prime\prime\prime}$ are equal
to the half\ndash sum of the corresponding ones for
$\ope{L}^{\prime}=\ope{L}$ and $\ope{L}^{\prime\prime}$. So, symbolically we
can write (see also~\eref{2.0}, \eref{2.10}--\eref{2.12} and~\eref{12.114})
	\begin{equation}	\label{12.12}
\ope{D}^{\prime\prime}
= \pm \ope{D}^{\prime} \big|_{\U_\mu\leftrightarrow\U_\mu^\dag}
\quad
\ope{D}^{\prime\prime\prime} =
\frac{1}{2}( \ope{D}^{\prime}  + \ope{D}^{\prime\prime} ),
	\end{equation}
where
\(
\ope{D}=
\ope{T}_{\mu\nu},\ope{J}_{\mu}, \ope{S}_{\mu\nu}^{\lambda},
	\ope{L}_{\mu\nu}^{\lambda},
\ope{P}_{\mu},\ope{Q}, \ope{S}_{\mu\nu}, \ope{L}_{\mu\nu}
\)
and the minus sign in the first equality stand only for
 $\ope{D}=\ope{J}_\mu,\ope{Q}$. If we express the dynamical variables in
terms of creation and annihilation operators, which are identical for the
Lagrangians we consider (\emph{vide infra}), then~\eref{12.12} takes the form
	\begin{equation}	\label{12.13}
\ope{D}^{\prime\prime}
= \pm \ope{D}^{\prime}
\big|_{a_s^{\pm}(\bk)\leftrightarrow a_s^{\dag\,\pm}(\bk)} \quad
\ope{D}^{\prime\prime\prime} =
\frac{1}{2}( \ope{D}^{\prime}  + \ope{D}^{\prime\prime} ) ,
	\end{equation}
with
\(
\ope{D}
= \ope{P}_{\mu},\ope{Q}, \ope{S}_{\mu\nu}, \ope{L}_{\mu\nu}.
\)%
\footnote{~%
For $\ope{D} = \ope{P}_{\mu},\ope{Q}$, the first equation in~\eref{12.13} is
evident (see~\eref{6.1} and~\eref{6.2}), but for
$\ope{D} = \ope{S}_{\mu\nu}, \ope{L}_{\mu\nu}$ some simple manipulations are
required for its proof --- see~\eref{6.8} and~\eref{6.28}.%
}
To save some space, we shall write explicitly only the conserved operator
quantities for the Lagrangian~\eref{12.6}.
Combining~\eref{6.5}--\eref{6.7} and~\eref{6.27} with the rule~\eref{12.13},
we get in Heisenberg picture (and before normal ordering):
	\begin{subequations}	\label{12.14}
	\begin{align}	\label{12.14a}
\ope{P}_\mu^{\prime\prime}
& =
\frac{1}{1+\tau(\U)} \sum_{s=1}^{3-\delta_{0m}} \int
  k_\mu |_{ k_0=\sqrt{m^2c^2+{\bs k}^2} }
\{
a_s^{+}(\bk)\circ a_s^{\dag\,-}(\bk) +
a_s^{-}(\bk)\circ a_s^{\dag\,+}(\bk)
\}
\Id^3\bk
\displaybreak[1]\\	\label{12.14b}
\tope{Q}^{\prime\prime}
& =
- q \sum_{s=1}^{3-\delta_{0m}}\int
\{
a_s^{+}(\bk)\circ a_s^{\dag\,-}(\bk) -
a_s^{-}(\bk)\circ a_s^{\dag\,+}(\bk)
\} \Id^3\bk
\displaybreak[1]\\	\label{12.14c}
\tope{S}_{\mu\nu}^{\prime\prime}
& =
  \frac{\ih }{1+\tau(\U)} \sum_{s,s'=1}^{3} \int \Id^3\bk
\sigma_{\mu\nu}^{ss'}(\bk)
\bigl\{
  a_s^{+}(\bk) \circ a_{s'}^{\dag\,-}(\bk)
- a_s^{-}(\bk) \circ a_{s'}^{\dag\,+}(\bk)
\bigr\}
	\end{align}
\vspace{-4ex}
	\begin{multline}	\label{12.14d}
\tope{L}_{\mu\nu}^{\prime\prime}
=
\frac{1}{1+\tau(\U)} \sum_{s=1}^{3-\delta_{0m}} \!\! \int\! \Id^3\bk
 ( x_{0\,\mu}k_\nu - x_{0\,\nu}k_\mu ) |_{ k_0=\sqrt{m^2c^2+{\bs k}^2} }
\{
a_s^{+}(\bk)\circ a_s^{\dag\,-}(\bk) +
a_s^{-}(\bk)\circ a_s^{\dag\,+}(\bk)
\}
\\
+
\frac{\ih}{1+\tau(\U)}  \sum_{s,s'=1}^{3} \int \Id^3\bk \,
l_{\mu\nu}^{ss'}(\bk)
\bigl\{
 a_s^{+}(\bk) \circ a_{s'}^{\dag\,-}(\bk) -
 a_s^{-}(\bk) \circ a_{s'}^{\dag\,+}(\bk)
\bigr\}
\displaybreak[2]\\
+
\frac{\ih}{2(1+\tau(\U))}  \sum_{s=1}^{3-\delta_{0m}} \int \Id^3\bk
\Bigl\{
a_s^{+}(\bk)
\Bigl( \xlrarrow{ k_\mu \frac{\pd}{\pd k^\nu} }
     - \xlrarrow{ k_\nu \frac{\pd}{\pd k^\mu} } \Bigr)
\circ a_s^{\dag\,-}(\bk)
\\ -
a_s^{-}(\bk)
\Bigl( \xlrarrow{ k_\mu \frac{\pd}{\pd k^\nu} }
     - \xlrarrow{ k_\nu \frac{\pd}{\pd k^\mu} } \Bigr)
\circ a_s^{\dag\,+}(\bk)
\Bigr\} \Big|_{ k_0=\sqrt{m^2c^2+{\bs k}^2} } \ ,
	\end{multline}
	\end{subequations}

	Let us turn now our attention to the field equations in terms of
creation and annihilation operators for the Lagrangians $\ope{L}^{\prime}$,
$\ope{L}^{\prime\prime}$ and $\ope{L}^{\prime\prime\prime}$. For
$\ope{L}^{\prime}=\ope{L}$ they are given by~\eref{7.4}. To derive them for
$\ope{L}^{\prime\prime}$ and $\ope{L}^{\prime\prime\prime}$, one should
repeat the derivation of~\eref{7.4} from~\eref{7.2} with
$\ope{P}_\mu^{\prime\prime}$ and
\(
\ope{P}_\mu^{\prime\prime\prime}
= \frac{1}{2}(\ope{P}_\mu^{\prime} + \ope{P}_\mu^{\prime\prime} ),
\)
respectively, for $\ope{P}_\mu$. In this way, from~\eref{7.2} with
 $\ope{P}_\mu=\ope{P}_\mu^{\prime},\ope{P}_\mu^{\prime\prime}$, \eref{12.14a}
and~\eref{12.13} with $\ope{D}=\ope{P}_\mu$, we obtain the field equations
derived from the Lagrangians
$\ope{L}^{\prime\prime}$ and $\ope{L}^{\prime\prime\prime}$
respectively as:
	\begin{subequations}	\label{12.15}
	\begin{align}
			\label{12.15a}
	\begin{split}
& \bigl[ a_s^{\pm}(\bk) ,
a_t^{+}(\bs q) \circ a_t^{\dag\,-}(\bs q)
 +
a_t^{-}(\bs q) \circ a_t^{\dag\,+}(\bs q)
\bigr]_{\_}
 \pm (1+\tau(\U)) a_s^{\pm}(\bk) \delta_{st} \delta^3(\bk-\bs q)
=
{\lindex[\mspace{-3mu}f]{}{\prime\prime}}_{st}^{\pm}(\bk,\bs q)
	\end{split}
\\			\label{12.15b}
	\begin{split}
& \bigl[ a_s^{\dag\,\pm}(\bk) ,
a_t^{+}(\bs q) \circ a_t^{\dag\,-}(\bs q)
 +
a_t^{-}(\bs q) \circ a_t^{\dag\,+}(\bs q)
\bigr]_{\_}
\pm (1+\tau(\U))  a_s^{\dag\,\pm}(\bk) \delta_{st} \delta^3(\bk-\bs q)
=
{\lindex[\mspace{-3mu}f]{}{\prime\prime}}_{st}^{\dag\,\pm}(\bk,\bs q)
	\end{split}
	\end{align}
	\end{subequations}
	\begin{subequations}	\label{12.16}
	\begin{multline}
			\label{12.16a}
 \bigl[ a_s^{\pm}(\bk) ,
a_t^{\dag\,+}(\bs q) \circ a_t^{-}(\bs q)
 +
a_t^{\dag\,-}(\bs q) \circ a_t^{+}(\bs q)
\bigr]_{\_}
 +
 \bigl[ a_s^{\pm}(\bk) ,
a_t^{+}(\bs q) \circ a_t^{\dag\,-}(\bs q)  +
a_t^{-}(\bs q) \circ a_t^{\dag\,+}(\bs q)
\bigr]_{\_}
\\
 \pm 2 (1+\tau(\U)) a_s^{\pm}(\bk) \delta_{st} \delta^3(\bk-\bs q)
=
{\lindex[\mspace{-3mu}f]{}{\prime\prime\prime}}_{st}^{\pm}(\bk,\bs q)
	\end{multline}
\vspace{-4ex}
	\begin{multline}
			\label{12.16b}
 \bigl[ a_s^{\dag\,\pm}(\bk) ,
a_t^{\dag\,+}(\bs q) \circ a_t^{-}(\bs q) +
a_t^{\dag\,-}(\bs q) \circ a_t^{+}(\bs q)
\bigr]_{\_}
 +
 \bigl[ a_s^{\dag\,\pm}(\bk) ,
a_t^{+}(\bs q) \circ a_t^{\dag\,-}(\bs q)
 +
a_t^{-}(\bs q) \circ a_t^{\dag\,+}(\bs q)
\bigr]_{\_}
\\
\pm 2 (1+\tau(\U))  a_s^{\dag\,\pm}(\bk) \delta_{st} \delta^3(\bk-\bs q)
=
{\lindex[\mspace{-3mu}f]{}{\prime\prime\prime}}_{st}^{\dag\,\pm}(\bk,\bs q) ,
	\end{multline}
	\end{subequations}
where the polarization indices $s$ and $t$ take the values
	\begin{equation}	\label{12.17}
s,t =
	\begin{cases}
1,2,3	&\text{for } m\not=0	\\
1,2	&\text{for } m=0
	\end{cases}
	\end{equation}
and the operator-valued (generalized) functions
$ {\lindex[\mspace{-3mu}f]{}{a}}^{\pm} (\bs k,\bs q)$ and
$ {\lindex[\mspace{-3mu}f]{}{a}}^{\dag\,\pm} (\bs k,\bs q)$, with
 $a=\prime,\prime\prime,\prime\prime\prime$, are such that
	\begin{equation}	\label{12.18}
\int q_\mu |_{q_0=\sqrt{m^2c^2+{\bs q}^2}}
	{\lindex[\mspace{-3mu}f]{}{a}}^{\pm} (\bs k,\bs q) \Id^3\bs q
=
\int q_\mu |_{q_0=\sqrt{m^2c^2+{\bs q}^2}}
	{\lindex[\mspace{-3mu}f]{}{a}}^{\dag\,\pm} (\bs k,\bs q) \Id^3\bs q
= 0 .
	\end{equation}

	Equations, similar to~\eref{7.8}--\eref{7.11}, can be derived
form~\eref{12.15} and~\eref{12.16} and, consequently, expressions for the
dynamical variables in momentum picture, similar to~\eref{7.18}--\eref{7.20},
can easily be obtained from these equations.

	As we see, the dynamical variables and the field equations in terms
of creation and annihilation operators for the
Lagrangians~\eref{12.1}--\eref{12.3} are completely different for a non\ndash
Hermitian field, $\U_\mu^\dag\not=\U_\mu$ or
$a_s^{\dag\,\pm}(\bk)\not=a_s^\pm(\bk)$, and, in this sense, the arising from
them quantum field theories of free vector field satisfying the Lorenz
condition are different. A step toward the identification of these theories
is achieved via the `second' quantization procedure, \ie by
establishing/imposing for/on the creation and annihilation operators
commutation relations, like~\eref{8.2} for the Lagrangian~\eref{12.1}. These
relations for the Lagrangians~\eref{12.2} and~\eref{12.3} can be derived
analogously to the ones for~\eref{12.1}, \ie by making appropriate changes in
the derivation of the commutation relations for an arbitrary free scalar
field, given in~\cite{bp-QFTinMP-scalars} (see
also~\cite{bp-QFTinMP-spinors}, where free spinor fields are investigates).
Without going into details, we shall say that this procedure results into the
commutation relations~\eref{8.2} for any one of the
Lagrangians~\eref{12.1}--\eref{12.3} (under the Lorenz
conditions~\eref{12.4} in the massless case). In this way, the systems of
field equations~\eref{7.4}, \eref{12.15} and~\eref{12.16} became identical
and equivalent to~\eref{8.2}.

	It should be emphasized, the derivation of~\eref{8.2} for the
Lagrangians~\eref{12.1}--\eref{12.3} is not
identical~\cite{bp-QFTinMP-scalars}: the Lagrangian~\eref{12.3} does not
admit quantization via \emph{anti}commutators, contrary to~\eref{12.1}
and~\eref{12.2}. So, the establishment of~\eref{8.2} for $\ope{L}^{\prime}$
and $\ope{L}^{\prime\prime}$ requires as an additional hypothesis the
quantization via commutators or some equivalent to it assertion, like the
charge symmetry, spin\ndash statistics theorem,
etc.~\cite{Bogolyubov&Shirkov}. Said differently, this additional assumption
is not needed  for the Lagrangian~\eref{12.3} as it entails such a hypothesis
in the framework of the Lagrangian formalism. The initial cause for this
state of affairs is that the symmetry particle~$\leftrightarrow$~antiparticle
is encoded in the Lagrangian $\ope{L}^{\prime\prime\prime}$ via its invariance
under the change $\U_\mu\leftrightarrow\U_\mu^\dag$. In particular, since for a
neutral field we, evidently, have
	\begin{equation}	\label{12.19}
\ope{L}^{\prime}
= \ope{L}^{\prime\prime}
= \ope{L}^{\prime\prime\prime}
\qquad\text{if } \U_\mu = \U_\mu^\dag,
	\end{equation}
for such a field, \eg for the electromagnetic one, the spin-statistics
theorem and other equivalent to it assertions are consequences form the
Lagrangian formalism investigated in the present paper.

	Since the commutation relations for the
Lagrangians~\eref{12.1}--\eref{12.3} are identical, we assume the normal
ordering procedures and the definitions of the vacuum for them to be
identical, respectively, and to coincide with the ones given in
Sect.~\ref{Sect9}.

	Applying the normal ordering procedure to the dynamical variables
corresponding to the Lagrangians~\eref{12.1}--\eref{12.3}
(see~\eref{6.5}--\eref{6.8}, \eref{6.27}, \eref{12.14} and~\eref{12.13}), we
see that, after this operation, they became independent of the Lagrangian we
have started, \ie symbolically we can write
	\begin{equation}	\label{12.20}
\ope{D}^{\prime}
= \ope{D}^{\prime\prime}
= \ope{D}^{\prime\prime\prime}
= \ope{D}
\qquad \ope{D}
= \ope{P}_{\mu},\ope{Q}, \ope{L}_{\mu\nu}, \ope{S}_{\mu\nu} ,
	\end{equation}
where the operators for $\ope{L}'=\ope{L}$ are given
by~\eref{9.6}--\eref{9.9}. (To prove these equations for
$ \ope{D} = \ope{L}_{\mu\nu}, \ope{S}_{\mu\nu}$,
one has to use the antisymmetry of the quantities~\eref{6.8}
and~\eref{6.28}.)

	Let us summarize at the end. The
Lagrangians~\eref{12.1}--\eref{12.3}, which are essentially different for
non\ndash Hermitian fields, generally entail quite different Lagrangian field
theories unless some additional conditions are added to the Lagrangian
formalism. In particular, these theories became identical if one assumes the
commutation relations~\eref{8.2}, the normal ordering procedure and the
definition of vacuum, as given in Sect.~\ref{Sect9}. The
Lagrangian~\eref{12.3} has the advantage that the spin\ndash statistics
theorem (or charge symmetry, etc.) is encoded in it, while, for the
Lagrangians~\eref{12.1} and~\eref{12.2} this assertion should be postulated
(imposed) as an additional condition to the Lagrangian formalism. For a
neutral free vector field satisfying the Lorenz condition, \ie for
electromagnetic field in Lorenz gauge, the spin\ndash statistics theorem is
a consequence from the Lagrangian formalism.%
\footnote{~%
Recall~\cite{bp-QFTinMP-scalars}, the proof of the spin\ndash statistics
theorem (charge symmetry, etc.) for the Lagrangian~\eref{12.3} requires as a
hypothesis, additional to the Lagrangian formalism, the assertion that the
commutators or anticommutators of all combinations of creation and/or
annihilation operators to be proportional to the identity mapping of system's
Hilbert space of states, \ie to be  $c$\ndash numbers.%
}


\section
[On the role of the Lorenz condition in the massless case]
{On the role of the Lorenz condition in the massless case}
\label{Sect13}

	Until now, in the description of free massless vector fields, we
supposed that they satisfy the Lorenz condition, \ie the equations
 $\pd^\mu\tU_\mu=0$ and $\pd^\mu\tU_\mu^\dag=0$ in Heisenberg picture
or~\eref{3.16} in momentum one, as subsidiary restrictions to the Lagrangian
formalism. Such a theory contains some physical, not mathematical, problems
which were summarized and partially analyzed in Sect.~\ref{Sect11}. The
present section is devoted to a brief exploration of a Lagrangian formalism
for free massless vector field without additional restrictions, like the
Lorenz condition. As we shall see, in this case the problems inherent to a
formalism with the Lorenz conditions remain and new ones are added to them.

	For other point of view on the topic of this section, see,
e.g.,~\cite[\S~7.1]{Ryder-QFT}.

	Most of the considerations in this section will be done in Heisenberg
picture which will prevent the exposition from new details (which are not quite
suitable for the purpose).

\subsection
[Description of free massless vector fields (without the Lorenz condition)]
{Description of free massless vector fields\\ (without the Lorenz condition)}
\label{Subsec13.1}

	The description of a free massless vector field coincides with the
one of a free massive vector field, given in Sect.~\ref{Sect3}, with the only
difference that the field's mass parameter $m$ (which is equal to the mass of
field's particles, if $m\not=0$) is set equal to zero,
	\begin{equation}	\label{13.1}
m = 0.
	\end{equation}
In particular, the Lagrangian formalism can start from the Lagrangian
(see~\eref{3.1} and~\eref{3.7})
	 \begin{gather}	\label{13.2}
\tope{L}
=
-
\frac{c^2\hbar^2}{1+\tau(\tU)}	(\pd_\mu\tU_\nu^\dag) \circ
				(\pd^\mu\tU^\nu)
+
\frac{c^2\hbar^2}{1+\tau(\tU)}	(\pd_\mu\tU^{\mu\dag}) \circ
				(\pd_\nu\tU^\nu)
\\			\label{13.3}
\ope{L}
 =
\frac{c^2}{1+\tau(\U)}
\bigl\{
[ \U_\nu^\dag , \ope{P}_\mu ]_{\_} \circ [ \U^\nu , \ope{P}^\mu ]_{\_}
-
[ \U^{\mu\dag} , \ope{P}_\mu ]_{\_} \circ [ \U^\nu , \ope{P}_\nu ]_{\_}
\bigr\}
	\end{gather}
in Heisenberg and momentum picture, respectively. The Euler-Lagrange
equations for this Lagrangian are the following massless Proca equations
(see~\eref{3.5} and~\eref{3.12})
	\begin{gather}	\label{13.4}
\tope{\square}(\tU_\mu) - \pd_\mu(\pd^\lambda\tU_\lambda) = 0
\qquad
\tope{\square}(\tU_\mu^\dag) - \pd_\mu(\pd^\lambda\tU_\lambda^\dag) =0
\\			\label{13.5}
  [ [ \U_\mu , \ope{P}_\lambda]_{\_} , \ope{P}^\lambda ]_{\_}
- [ [ \U_\nu , \ope{P}^\nu]_{\_} , \ope{P}^\mu]_{\_}
= 0
\qquad
  [ [ \U_\mu^\dag , \ope{P}_\lambda]_{\_} , \ope{P}^\lambda ]_{\_}
- [ [ \U_\nu^\dag , \ope{P}^\nu]_{\_} , \ope{P}^\mu]_{\_}
= 0
	\end{gather}
in Heisenberg and momentum picture, respectively.  (Recall,
$\tope{\square}:=\pd^\mu\pd_\mu$.) As pointed in Sect.~\ref{Sect3}, these
equations do not imply that the field operators satisfy the massless
Klein\ndash Gordon equations and the Lorenz conditions
(see~\eref{3.6}, \eref{3.12}, \eref{3.15} and~\eref{3.16} with $m=0$). The
common solutions of the massless Klein\ndash Gordon equations and the
Lorenz conditions for the field operators are solutions of the massless
Proca equation, but the opposite is not necessary, \ie the latter system of
equations is more general than the former one. This is the cause why, for
solutions of~\eref{3.5}, the Lagrangian~\eref{13.3} cannot be reduced
to~\eref{3.17} with $m=0$ in the general case (unless the Lorenz
conditions~\eref{3.16} are imposed on the solutions of~\eref{13.5} as
additional conditions).

	The general expressions for the densities of the dynamical variables
through the generalize momenta $\tope{\pi}_{\lambda\mu}$ (see the first
equalities in~\eref{3.18}--\eref{3.22}) remain, of course, valid in the
massless case too, but their particular dependence on the field operators is
different from the second equalities in~\eref{3.18}--\eref{3.22}, as
now~\eref{3.11}, without the Lorenz conditions on these operators, should be
used. Thus, the dynamical variables of a massless vector field are:
	\begin{align}
			\label{13.6}
	\begin{split}
\tope{T}_{\mu\nu}
& =
- \frac{c^2\hbar^2}{1+\tau(\tU)}
\bigl\{
(\pd_\mu\tU_\lambda^\dag)\circ(\pd_\nu\tU^\lambda) +
(\pd_\nu\tU_\lambda^\dag)\circ(\pd_\mu\tU^\lambda)
\bigr\}
- \eta_{\mu\nu}\tope{L}
\\ &\hphantom{=}
+ \frac{c^2\hbar^2}{1+\tau(\tU)}
\bigl\{
(\pd_\varkappa\tU^{\varkappa\dag})\circ(\pd_\nu\tU_\mu) +
(\pd_\nu\tU_\mu^\dag)\circ(\pd_\varkappa\tU^\varkappa)
\bigr\}
	\end{split}
\displaybreak[1]\\			\label{13.7}
	\begin{split}
\tope{J}_{\mu}
& =
\ih q c^2
\bigl\{
(\pd_\mu\tU_\lambda^\dag)\circ\tU^\lambda -
\tU_\lambda^\dag\circ(\pd_\mu\tU^\lambda)
- (\pd_\varkappa\tU^{\varkappa\dag})\circ\tU_\mu
+ \tU_\mu^\dag\circ(\pd_\varkappa\tU^\varkappa)
\bigr\}
	\end{split}
\displaybreak[1]\\			\label{13.8}
  \tope{M}_{\mu\nu}^{\lambda}
& = \tope{L}_{\mu\nu}^{\lambda} + \tope{S}_{\mu\nu}^{\lambda}
\\			\label{13.9}
\tope{L}_{\mu\nu}^{\lambda}
: & =
x_\mu\Sprindex[\tope{T}]{\nu}{\lambda} -
x_\nu\Sprindex[\tope{T}]{\mu}{\lambda}
\displaybreak[2]\\			\label{13.10}
	\begin{split}
\tope{S}_{\mu\nu}^{\lambda}
& =
\frac{\hbar^2c^2}{1+\tau(\tU)}
\bigl\{
(\pd^\lambda \tU_\mu^\dag) \circ \tU_\nu -
(\pd^\lambda \tU_\nu^\dag) \circ \tU_\mu -
\tU_\mu^\dag \circ (\pd^\lambda\tU_\nu) +
\tU_\nu^\dag \circ (\pd^\lambda\tU_\mu)
\\ & \hphantom{=}
-\delta_\mu^\lambda
\bigl( (\pd_\varkappa \tU^{\varkappa\dag}) \circ \tU_\nu +
\tU_\nu^\dag \circ (\pd_\varkappa \tU^\varkappa)
\bigr)
+
\delta_\nu^\lambda
\bigl( (\pd_\varkappa \tU^{\varkappa\dag}) \circ \tU_\mu +
\tU_\mu^\dag (\pd_\varkappa \tU^\varkappa)
\bigr)
\bigr\} .
	\end{split}
	\end{align}
Notice, in~\eref{13.6} the Lagrangian $\tope{L}$ must be replaced by its
value given by~\eref{13.2}, not by~\eref{3.17} with $m=0$. Evidently, the
Lorenz conditions~\eref{3.6b} reduce the equations~\eref{13.6}--\eref{13.10}
to~\eref{3.18}--\eref{3.22}, respectively. If needed, the reader can easily
write the above equations in momentum picture by means of the general rules
of Sect.~\ref{Sect2}.

	It should be remarked, as the energy-momentum tensor~\eref{13.6} is
non\ndash symmetric, the spin and orbital angular momentum are no longer
conserved quantities.

\subsection{Analysis of the Euler-Lagrange equations}
\label{Subsec13.2}

	Since the solutions of the Euler-Lagrange equations~\eref{13.4}
(or~\eref{13.5}) generally do not satisfy the Klein\ndash Gordon equation, we
cannot apply to free massless vector fields the methods developed for free
scalar fields. To explore the equations~\eref{13.4}, we shall transform them
into algebraic ones in the \emph{momentum representation} in Heisenberg
picture~\cite{Bogolyubov&Shirkov,Bjorken&Drell-2,Itzykson&Zuber}.

	Define the Fourier images $\tu_\mu(k)$ and $\tu_\mu^\dag(k)$,
$k\in\field[R]^4$, of the field operators via the Fourier transforms
	\begin{equation}	\label{13.16}
\tU_\mu(x)
=
\frac{1}{(2\pi)^2} \int \e^{-\iih kx} \tu_\mu(k) \Id^3k
\qquad
\tU_\mu^\dag(x)
=
\frac{1}{(2\pi)^2} \int \e^{-\iih kx} \tu_\mu^\dag(k) \Id^3k ,
	\end{equation}
where $\Id^4k:=\Id k^0\Id k^1\Id k^2\Id k^3$ and $kx:=k_\mu x^\mu$. Since
$\tU_\mu^\dag(x)$ is the Hermitian conjugate of $\tU_\mu(x)$,
$\tU_\mu^\dag(x):=(\tU_\mu(x))^\dag$, we have
	\begin{equation}	\label{13.17}
\tu_\mu^\dag(k) = (\tu_\mu(-k))^\dag .
	\end{equation}

	Substituting~\eref{13.16} into~\eref{13.4}, we find the systems of
equations
	\begin{equation}	\label{13.18}
k^2 \tu_\mu(k) - k_\mu k^\nu \tu_\nu(k) = 0
\qquad
k^2 \tu_\mu^\dag(k) - k_\mu k^\nu \tu_\nu^\dag(k) = 0 ,
	\end{equation}
which is equivalent to~\eref{13.4}. Here and below $k^2:=k_\mu k^\mu$. (From
the context, it will be clear that, in most cases, by $k^2$ we have in mind
$k_\mu k^\mu$, not the second contravariant component of $k$.)

	Let us consider the classical analogue of the equation~\eref{13.18},
\ie
	\begin{equation}	\label{13.19}
k^2 v_\mu(k) - k_\mu k^\nu v_\nu(k)
\equiv  ( k^2 \eta_{\mu\nu} - k_\mu k_\nu ) v^\nu(k)
= 0 ,
	\end{equation}
where $v_\mu(k)$ is a classical, not operator-valued, vector field over the
$k$\ndash space $\field[R]^4$. This is a linear homogeneous system of~4
equations for the~4 variables $v_0(k)$, $v_1(k)$, $v_2(k)$ and $v_3(k)$.
Since the determinant of the matrix of~\eref{13.19} is%
\footnote{~%
One can easily prove that
\(
\det\bigl[ c_{\mu\nu} - z_\mu z_\nu \bigr]_{\mu,\nu=0}^{3}
=
c_0 c_1 c_2 c_3 + z_0^2 c_1 c_2 c_3 + c_0 z_1^2 c_2 c_3 +
		  c_0 c_1 z_2^2 c_3 + c_0 c_1 c_2 z_3^2
\)
for a diagonal matrix $[c_{\mu\nu}]=\diag(c_0,c_2,c_2,c_3)$ and any 4\ndash
vector $z_\mu$. Putting here $c_{\mu\nu}=k^2\eta_{\mu\nu}$ and $z_\mu=k_\mu$,
we get the cited result.%
}
\[
\det[k^2 \eta_{\mu\nu} - k_\mu k_\nu]_{\mu,\nu=0}^{3}
=
k^2 ( - k^2 + k^2 ) \equiv 0
\]
the system of equation~\eref{13.19} possesses always a non-zero solution
relative to $v^\nu(k)$. Besides, the form of this determinant indicates that
the value $k^2=0$ is crucial for the number of linearly independent solutions
of~\eref{13.19}. A simple algebraic calculation reveals that the rank $r$ of
the matrix $[k^2 \eta_{\mu\nu} - k_\mu k_\nu]_{\mu,\nu=0}^{3}$,
as a function of $k$, is:
$r=0$ if $k_\mu=0$,
$r=1$ if $k^2=0$ and $k_\mu\not=0$ for some $\mu=0,1,2,3$ and
$r=3$ if $k^2\not=0$.
Respectively, the number of linearly independent solutions of~\eref{13.19} is
infinity if $k_\mu=0$,
three if $k^2=0$ and $k_\mu\not=0$ for some $\mu=0,1,2,3$, and
one if $k^2\not=0$.

	For $k^2=0$, the system~\eref{13.19} reduces to the
equation~\eref{4.15} with $m=0$, which was investigated in Sect.~\ref{Sect4}.
For $k^2\not=0$, it has the evident solution
	\begin{equation}	\label{13.20}
w_\mu(k) := \iu \frac{k_\mu}{\sqrt{k^2}} \qquad (\,k^2\not=0\,) ,
	\end{equation}
which is normalized to $-1$,
	\begin{equation}	\label{13.21}
w_\mu(k) w^\nu(k) = -1	\qquad (\,k^2\not=0\,) ,
	\end{equation}
and any other solution of~\eref{13.19} is proportional to $w_\mu(k)$, as
defined by~\eref{13.20}.

	Therefore, for any $k_\mu$, we can write the general solution
of~\eref{13.19} as
	\begin{equation}	\label{13.25}
v_\mu(k)
=
\delta_{0 k^2}  \sum_{s=1}^{3} \alpha_s(k) v_\mu^s(\bk)\big|_{m=0}
+
(1-\delta_{0 k^2})  \alpha_4(k) w_\mu(k) ,
	\end{equation}
where $\alpha_1(k),\dots,\alpha_4(k)$ are some functions of
$k=(k_0,\dots,k_3)$,  $v_\mu^s(\bk)$ with $s=1,2,3$ are defined
by~\eref{4.15}--\eref{4.21}, and the Kronecker delta\ndash symbol
$\delta_{0 k^2}$~($:=1$ for $k^2=0$ and~$:=0$ for $k^2\not=0$) takes care of
the number of linearly independent solutions of~\eref{13.19}.

	Returning to the operator equations~\eref{13.18}, we can express
their solutions as
	\begin{equation}	\label{13.26}
	\begin{split}
\tu_\mu(k)
& =
(2\pi)^2 \{ \iu c^2 (2\pi\hbar)^3 \}^{-1/2}
\bigl\{
\delta_{0 k^2}  \sum_{s=1}^{3} \ta_s(k) v_\mu^s(\bk)\big|_{m=0}
+
(1-\delta_{0 k^2})  \ta_4(k) w_\mu(k)
\bigr\}
\\
\tu_\mu^\dag(k)
& =
(2\pi)^2 \{ \iu c^2 (2\pi\hbar)^3 \}^{-1/2}
\bigl\{
\delta_{0 k^2}  \sum_{s=1}^{3} \ta_s^\dag(k) v_\mu^s(\bk)\big|_{m=0}
+
(1-\delta_{0 k^2})  \ta_4^\dag(k) w_\mu(k)
\bigr\} ,
	\end{split}
	\end{equation}
where $\ta_1(k),\dots, \ta_4^\dag(k)$ are some operator-valued functions of
$k$, which, by~\eref{13.17}, are such that
	\begin{equation}	\label{13.27}
\ta_\omega^\dag(k) = (\ta_\omega(-k) )^\dag
\qquad
\omega=1,2,3,4 ,
	\end{equation}
and the factor $(2\pi)^2 \{ \iu c^2 (2\pi\hbar)^3 \}^{-1/2}$ is
introduced for future convenience. (The operators
$\ta_1(k),\dots, \ta_4^\dag(k)$ are closely related to the creation and
annihilation operators, but we shall not consider this problem here.)

	At last, combining~\eref{13.27} and~\eref{13.16}, we can write the
solutions of the field equations~\eref{13.4} as
	\begin{equation}	\label{13.28}
	\begin{split}
\tU_\mu(x)
= &
(2\pi)^2 \{ \iu c^2 (2\pi\hbar)^3 \}^{-1/2}
\int\limits_{k^2=0}   \sum_{s=1}^{3}
\e^{-\iih kx} \ta_s(k) v_\mu^s(\bk)\big|_{m=0} \Id ^4k
\\& +
(2\pi)^2 \{ \iu c^2 (2\pi\hbar)^3 \}^{-1/2}
\int\limits_{k^2\not=0}
\e^{-\iih kx} \ta_4(k) w_\mu(\bk) \Id ^4k
\\
\tU_\mu^\dag(x)
= &
(2\pi)^2 \{ \iu c^2 (2\pi\hbar)^3 \}^{-1/2}
\int\limits_{k^2=0}   \sum_{s=1}^{3}
\e^{-\iih kx} \ta_s^\dag(k) v_\mu^s(\bk)\big|_{m=0} \Id ^4k
\\& +
(2\pi)^2 \{ \iu c^2 (2\pi\hbar)^3 \}^{-1/2}
\int\limits_{k^2\not=0}
\e^{-\iih kx} \ta_4^\dag(k) w_\mu(\bk) \Id ^4k .
	\end{split}
	\end{equation}

	Since the Lorenz conditions (see~\eref{3.6})
	\begin{equation}	\label{13.29}
\pd^\mu\tU_\mu(x) = 0 \qquad \pd^\mu\tU_\mu^\dag(x) = 0
	\end{equation}
in momentum representation in Heisenberg picture read (see~\eref{13.16}),
	\begin{equation}	\label{13.30}
k^\mu\tu_\mu(k) = 0 \qquad k^\mu\tu_\mu^\dag(k) = 0 ,
	\end{equation}
we see that they are equivalent to the selection of solutions of~\eref{13.18}
with
	\begin{equation}	\label{13.31}
k^2= 0.
	\end{equation}
Said differently, the Lorenz condition on the field operators is equivalent
to the imposition of the restrictions
	\begin{equation}	\label{13.32}
\ta_4(k) = 0 \qquad \ta_4^\dag(k) = 0 \qquad (\, k^2\not=0 \,) ,
	\end{equation}
due to~\eref{13.26} (or~\eref{13.28}). Here the operators
$\ta_4(k)$ and $\ta_4^\dag(k)$ may be considered as a measure of the
satisfaction of the Lorenz condition by the field operators. Therefore the
sum of the terms containing an integral over the hyperboloid $k^2=0$
in~\eref{13.28} corresponds to field operators satisfying the Lorenz
condition and, consequently, to them is valid the theory developed in the
preceding sections. In particular, up to a constant, the operators
$\ta_s(k)$, $s=1,2,3$, are sums of the creation and annihilation operators
(in Heisenberg picture --- see~\eref{6.48}) of a free massless vector field
satisfying the Lorenz condition.

\subsection{Dynamical variables}
\label{Subsect13.3}

	To reveal the meaning of the operators $\ta_4(k)$ and $\ta_4^\dag(k)$
in~\eref{13.28}, we shall express the field's dynamical variables in terms of
$\ta_\omega(k)$ and $\ta_\omega^\dag(k)$, $\omega=1,2,3,4$. For the purpose,
the decompositions~\eref{13.28} should be inserted into the
expressions~\eref{13.6}--\eref{13.10} and, then, the conserved
operators~\eref{2.0} , \eref{2.10}--\eref{2.12} to be calculated.  It is not
difficult to be seen, a dynamical variable $\tope{D}$, with
$\ope{D}=\ope{P}_\mu,\ope{Q},\ope{L}_{\mu\nu},\ope{S}_{\mu\nu}$ for
respectively the momentum, charge, orbital and spin angular momentum
operators, has the following structure:
	\begin{equation}	\label{13.33}
\tope{D}
=
\lindex[\mspace{-4mu}\tope{D}]{}{0} +
\lindex[\mspace{-3mu}\tope{D}]{}{0-4} +
\lindex[\mspace{-3mu}\tope{D}]{}{4} .
	\end{equation}
Here
\(
\lindex[\mspace{-4mu}\tope{D}]{}{0}
:=
\int_{k^2=0}\Id^4k \int_{{k'}^2=0} \Id^4k' \{\cdots\}
\)
is the dynamical variable under the conditions~\eref{13.32}, \ie if the field
operators were supposed to satisfy the Lorenz condition;
the second term is of the form
\(
\lindex[\mspace{-3mu}\tope{D}]{}{0-4}
:=
\int_{k^2=0}\Id^4k \int_{{k'}^2\not=0} \Id^4k' \{\cdots\}
\)
with the expression in braces being a linear combination of terms like
 $\ta_s^\dag(k)\circ\ta_4(k')$ and $\ta_4^\dag(k')\circ\ta_s(k)$,
where $s=1,2,3$;
and the structure of the last term is
\(
\lindex[\mspace{-3mu}\tope{D}]{}{4}
:=
\int_{k^2\not=0}\Id^4k \int_{{k'}^2\not=0} \Id^4k' \{\cdots\}
\)
with the expression in braces being proportional to the operator
$\ta_4^\dag(k)\circ\ta_4(k')$.

	By means of the explicit formulae~\eref{4.19}--\eref{4.21}
and~\eref{13.20}, one can prove, after simple algebraic calculations, that
	\begin{equation}	\label{13.34}
\lindex[\mspace{-3mu}\tope{D}]{}{4} = 0
\qquad
\tope{D}=\tope{P}_\mu,\tope{Q},\tope{L}_{\mu\nu},\tope{S}_{\mu\nu}.
	\end{equation}
Thus, if we regard $\ta_4(k')$ and $\ta_4^\dag(k')$ as independent degrees of
freedom (possibly connected with some particles), then their pure (`free')
contribution to the dynamical variables is vanishing. However, the second
term in~\eref{13.33} is generally non\ndash zero. Simple, but long and
tedious, algebraic calculations give the following results:
	\begin{subequations}	\label{13.35}
	\begin{multline}	\label{13.35a}
\lindex[\mspace{-3mu}\tope{P}_\mu]{}{0-4}
=
\frac{1}{1+\tau(\tU)} \sum_{s=1}^{3}
\int_{k^2=0}\Id^4k \int_{{k'}^2\not=0} \Id^4k'
\delta^3(\bk+\bk')
\e^{-\iih(k_0-k'_0)x^0}
\\ \times
\frac{k_0k'_\mu}{\sqrt{{k'_0}^2-k_0^2}}
\Big\{ \sqrt{k_0^2} \delta^{3s} - k_0 \Big\}
\{ \ta_s^\dag(k) \circ \ta_4(k') + \ta_4^\dag(k') \circ \ta_s(k) \}
	\end{multline}
\vspace{-4ex}
	\begin{multline}	\label{13.35b}
\lindex[\mspace{-3mu}\tope{Q}]{}{0-4}
=
q  \sum_{s=1}^{3}
\int_{k^2=0}\Id^4k \int_{{k'}^2\not=0} \Id^4k'
\delta^3(\bk+\bk')
\e^{-\iih(k_0-k'_0)x^0}
\\ \times
\frac{k_0}{\sqrt{{k'_0}^2-k_0^2}}
\Big\{ \sqrt{k_0^2} \delta^{3s} - k_0 \Big\}
\{ - \ta_s^\dag(k) \circ \ta_4(k') + \ta_4^\dag(k') \circ \ta_s(k) \}
	\end{multline}
\vspace{-4ex}
	\begin{multline}	\label{13.35c}
\lindex[\mspace{-3mu}\tope{S}_{\mu\nu}]{}{0-4}
=
\frac{\ih}{1+\tau(\tU)} \sum_{s=1}^{3}
\int_{k^2=0}\Id^4k \int_{{k'}^2\not=0} \Id^4k'
\delta^3(\bk+\bk')
\e^{-\iih(k_0-k'_0)x^0}
\\ \times
\frac{1}{\sqrt{{k'_0}^2-k_0^2}}
\{ \ta_s^\dag(k) \circ \ta_4(k') + \ta_4^\dag(k') \circ \ta_s(k) \}
\\ \times
	\begin{cases}
(-1)^{\delta_{0\mu}}
\Big\{
k_0(k'_0-k_0) v_a^s(\bk) \! - \!
k'_a \big( \sqrt{k_0^2} \delta^{3s} \! - \!k'_0 \big)
\Big\}
		&\text{\!\!\!for } (\mu,\nu)=(0,a),(a,0) \text{ with } a=1,2,3
\\
(k_0-k'_0) (k_\mu v_\nu^s(\bk) - k_\nu v_\mu^s(\bk))
		&\text{\!\!\!otherwise} \, .
	\end{cases}
	\end{multline}
	\end{subequations}
Notice, the expression ${k'_0}^2-k_0^2$ in~\eref{13.35} is different from
zero as
\(
  {k'_0}^2-k_0^2
= {k'_0}^2-\bk^2
= {k'_0}^2-{\bk'}^2
= {k'}^2 \not= 0,
\)
due to $0=k^2=k_0^2-{\bk}^2$ and the  $\delta$\ndash function
$\delta^3(\bk+\bk')$ in~\eref{13.35}.

\subsection{The field equations}
\label{Subsec13.4}

	Recall now that we consider quantum field theories in which the
Heisenberg relations~\eref{2.28} hold as a subsidiary restriction on the
field operators. Consequently, the system of field equations consists of the
Euler\ndash Lagrange equations~\eref{13.4}, the Heisenberg relations
	\begin{equation}	\label{13.40}
[\tU_\mu(x),\ope{P}_\nu]_{\_} = \ih \frac{\pd \tU_\mu(x)}{\pd x^\nu}
\qquad
[\tU_\mu^\dag(x),\ope{P}_\nu]_{\_} = \ih \frac{\pd \tU_\mu^\dag(x)}{\pd x^\nu}
	\end{equation}
and the explicit connection between $\ope{P}_\mu$ and the field operators,
\ie (see~\eref{13.33}--\eref{13.35})
	\begin{equation}	\label{13.41}
\tope{P}_\mu
=
\lindex[\mspace{-4mu}\tope{P}_\mu]{}{0} +
\lindex[\mspace{-3mu}\tope{P}_\mu]{}{0-4}
	\end{equation}
with $\lindex[\mspace{-4mu}\tope{P}_\mu]{}{0}$ given by the r.h.s.\
of~\eref{6.5} and $\lindex[\mspace{-3mu}\tope{P}_\mu]{}{0-4}$ defined
via~\eref{13.35a}.

	Since the expansions~\eref{13.28} take care of the Euler-Lagrange
equation~\eref{13.4}, the equations~\eref{13.40} remain the only restrictions
on the field operators. Substituting equation~\eref{13.28} into~\eref{13.40},
we get
	\begin{subequations}	\label{13.42}
	\begin{align}	\label{13.42a}
[\ta_s(k),\ope{P}_\mu]_{\_} & = - k_\mu \ta_s(k) \qquad s=1,2 \quad k^2=0
\\			\label{13.42b}
[\ta_4(k),\ope{P}_\mu]_{\_} & = - k_\mu \ta_4(k) \qquad k^2\not=0 .
	\end{align}
	\end{subequations}
One can verify that~\eref{13.42a} is equivalent to~\eref{7.2} with $m=0$ and
$\ope{P}_\mu$ given by~\eref{13.41}. Notice, the operators
$\ta_3(k)$ and $\ta_3^\dag(k)$
(or $a_3^\pm(\bk)$ and $a_3^{\dag\,\pm}(\bk)$), with $k^2=0$,
enter in the field equations~\eref{13.42} and into the dynamical variables
via~\eref{13.41} and~\eref{13.35} (see also~\eref{6.7} and~\eref{6.27}).
However, a particle interpretation of the degrees of freedom connected with
$\ta_3(k)$ and $\ta_3^\dag(k)$ fails as they enter in~\eref{13.42},
\eref{6.7} and~\eref{6.27} only in combinations/compositions with
$\ta_\omega(k)$ and $\ta_\omega^\dag(k)$ with $\omega=1,2,4$. In that sense,
the operators $\ta_3(k)$ and $\ta_3^\dag(k)$  serve as `coupling constants'
with respect to the remaining ones. Similar is the situation with the
operators $\ta_4(k)$ and $\ta_4^\dag(k)$, with $k^2\not=0$, regarding the
dynamical variables, but these operators are `more dynamical' as they must
satisfy the equations~\eref{13.42b}.

	The explicit equations of motion for
$\ta_\omega(k)$ and $\ta_\omega^\dag(k)$, $\omega=1,2,3,4$,
 can be obtained by inserting~\eref{13.41} (see also~\eref{13.35a}
and~\eref{6.5} with $m=0$) into~\eref{13.42}. The result will be similar
to~\eref{7.3} or~\eref{7.4}, with $m=0$, but additional terms, depending on
$\ta_4(k)$ and $\ta_4^\dag(k)$ with $k^2\not=0$, will be presented. We shall
not write these equations as they will not be used further and it seems that
$\ta_4(k) = \ta_4^\dag(k) =0 $, when they coincide with~\eref{7.4}, is the
only their physically meaningful solution (see Subsect.~\ref{Subsect13.5}
below).

\subsection{Discussion}
\label{Subsect13.5}

	Equations~\eref{13.34} and~\eref{13.35} show that the operators
$\ta_4(k')$ and $\ta_4^\dag(k')$ (with ${k'}^2\not=0$) do not have their own
contributions to the dynamical variables, but they do contribute to them via
the combinations
 $\ta_s^\dag(k) \circ \ta_4(k')$ and $\ta_4^\dag(k') \circ \ta_s(k)$,
with $k^2=0$, ${k'}^2\not=0$ and $s=1,2,3$. In a sense, the operators
$\ta_4(k')$ and $\ta_4^\dag(k')$
act as `operator\ndash valued coupling constants' for the operators
$\ta_s^\dag(k)$ and $\ta_s(k)$
(and hence for $\ta_s^{\dag\,\pm}(k)$ and $\ta_s^\pm(k)$), as via them they
bring an additional contribution to the dynamical variables with respect to
vector fields satisfying the Lorenz condition. In this aspect, the operators
$\ta_4(k')$ and $\ta_4^\dag(k')$ are similar to
$a_3^\pm(\bk)$ and $a_3^{\dag\,\pm}(\bk)$
(or $\ta_3(k)$ and $\ta_3^\dag(k)$) (for details, see Sect.~\ref{Sect11}).

	Consequently, the absence of the Lorenz conditions brings new
problems, in addition to the similar ones with the operators
$a_3^\pm(\bk)$ and $a_3^{\dag\,\pm}(\bk)$  (see Sect.~\ref{Sect11}). One can
say that a massless free vector field, which does not satisfy the Lorenz
condition, is equivalent to a similar field satisfying that restriction and
with self\ndash interaction determined by $\ta_4(k')$ and $\ta_4^\dag(k')$.
One can try to get rid of the contributions of the last operators in the
dynamical variables by imposing on them some subsidiary conditions, called
gauges in terms of the initial operators $\tU_\mu(x)$ and $\tU_\mu^\dag(x)$.
The problem of gauge freedom of massless vector fields is well\ndash studied
in the literature~\cite{Bjorken&Drell,Ryder-QFT,Bogolyubov&Shirkov} to which
the reader is referred. In particular, one can set the operators~\eref{13.35}
to zero by demanding

       \begin{equation}	\label{13.36}
\ta_s^\dag(k) \circ \ta_4(k') + \ta_4^\dag(k') \circ \ta_s(k) = 0
\quad
\ta_s^\dag(k) \circ \ta_4(k') - \ta_4^\dag(k') \circ \ta_s(k) = 0
	\end{equation}
for $k^2=0,\ {k'}^2\not=0$, and $s=1,2,3$, or, equivalently,
	\begin{equation}	\label{13.37}
\ta_s^\dag(k) \circ \ta_4(k') = 0
\quad
\ta_4^\dag(k') \circ \ta_s(k) = 0
\qquad
(\, k^2=0,\ {k'}^2\not=0,\ s=1,2,3 \,) .
	\end{equation}
For instance, these equalities are identically satisfied if the Lorenz
condition, \eg in the form~\eref{13.32}, is valid.

	Let us summarize. A theory of massless free vector field, based
on the Lagrangian~\eref{13.2} contains as free parameters the operators
 $a_{3}^{\pm}(\bk)$ and $a_{3}^{\dag\,\pm}(\bk)$ and, partially,
 $\ta_{4}(k')$ and $\ta_{4}^{\dag}(k')$, with $k^2=0$ and ${k'}^2\not=0$.
These operators have, generally, non\ndash vanishing contributions to the
dynamical variables only via their compositions with the (`physical')
operators
 $a_{s}^{\pm}(\bk)$ and $a_{s}^{\dag\,\pm}(\bk)$ (or
 $\ta_{s}(k)$ and $\ta_{s}^{\dag}(k)$ ), with $s=1,2$, and via the
combinations
 $\ta_{3}^\dag(k)\circ \ta_{4}(k')$ and
 $\ta_{4}^{\dag}(k')\circ\ta_{3}(k)$.
As a result, these operators describe degrees of freedom with vanishing own
dynamical characteristics and, consequently, they do not admit particle
interpretation.%
\footnote{~%
If one assigns particle interpretation of the discussed operators, then they
will have vanishing 4\ndash momentum, charge and spin and hence will be
unobservable.%
}
As we saw in Sect.~\ref{Sect11}, the operators
$a_{s}^{\pm}(\bk)$ $a_{s}^{\dag\,\pm}(\bk)$, with $s=1,2,3$ and $k^2\not=0$,
describe reasonably well massless free vector fields satisfying the Lorenz
condition, in particular the electromagnetic field.  From this point of view,
we can say that the operators
 $a_{3}^{\pm}(\bk)$, $a_{3}^{\dag\,\pm}(\bk)$,
 $\ta_{4}(k')$ and $\ta_{4}^{\dag}(k')$
 describe some selfinteraction of the preceding field, but it seems such a
selfinteracting, massless, free vector field is not known to exist in the
Nature at the moment. The easiest way for exclusion of that selfinteraction
from the theory is the pointed operators to be set equal to zero, \ie on the
field operators to be imposed the Lorenz conditions and~\eref{11.9}.
However, it is possible that other restrictions on the Lagrangian formalism
may achieve the same goal.

	At the end, the above considerations point that the Lorenz condition
should be imposed as an addition (subsidiary) condition to the Lagrangian
formalism of massless free vector fields, in particular to the quantum
theory of free electromagnetic field. Besides, the conditions~\eref{11.9} also
seems to be necessary for a satisfactory description of these fields.


\section {Conclusion}
\label{Conclusion}

	A more or less detailed Lagrangian quantum field theory of free
vector fields, massless in Lorenz gauge and massive ones, in momentum picture
was constructed in the present paper. Regardless of a common treatment of the
both types of fields, the massless case has some specific features and
problems. The Lorenz conditions are external to the Lagrangian formalism of
massless vector fields, but they are compatible with it. However, for an
electromagnetic field, which is a neutral massless vector field, in Lorenz
gauge, we have obtained a problem\ndash free description in terms of creation
and annihilation operators, \ie in terms of particles. This description is
similar to the Gupta\ndash Bleuler quantization of electromagnetic field, but
is quite different from the latter one and it is free of the problems this
formalism contains. Our formalism reproduces, under suitable additional
conditions, the quantization of electromagnetic field in Coulomb gauge.

	Between the Lagrangians, considered for a suitable description of
free vector fields satisfying the Lorenz conditions, we have singled out the
Lagrangian~\eref{12.3}. It is invariant under the transformation
particle$\leftrightarrow$antiparticle, described in appropriate variables, so
that in it is encoded the charge symmetry (or spin\ndash statistics theorem).
The field equations in terms of creation and annihilation operators for this
Lagrangian are~\eref{12.16} (under the conditions~\eref{12.17}
and~\eref{12.18}). They can equivalently be rewritten as
	\begin{subequations}	\label{c.1}
	\begin{align}
			\label{c.1a}
	\begin{split}
\bigl[ [a_t^{\dag\,+}(\bs q) , a_t^{-}(\bs q)]_{+} ,
	a_s^{\pm}(\bk) \bigr]_{\_}
& +
\bigl[ [a_t^{+}(\bs q) , a_t^{\dag\,-}(\bs q)]_{+} ,
	a_s^{\pm}(\bk) \bigr]_{\_}
\\
& =
\pm 2 (1+\tau(\U)) a_s^{\pm}(\bk) \delta_{st} \delta^3(\bk-\bs q)
-
{\lindex[\mspace{-3mu}f]{}{\prime\prime\prime}}_{st}^{\pm} (\bs k,\bs q)
	\end{split}
\\			\label{c.1b}
	\begin{split}
\bigl[ [a_t^{\dag\,+}(\bs q) , a_t^{-}(\bs q)]_{+} ,
	a_s^{\dag\,\pm}(\bk) \bigr]_{\_}
& +
\bigl[ [a_t^{+}(\bs q) , a_t^{\dag\,-}(\bs q)]_{+} ,
	a_s^{\dag\,\pm}(\bk) \bigr]_{\_}
\\
& =
\pm 2 (1+\tau(\U)) a_s^{\dag\,\pm}(\bk) \delta_{st} \delta^3(\bk-\bs q)
-
{\lindex[\mspace{-3mu}f]{}{\prime\prime\prime}}_{st}^{\dag\,\pm}
							     (\bs k,\bs q).
	\end{split}
	\end{align}
	\end{subequations}
Trilinear equations of this kind are typical for the so-called parastatistics
and parafield
theory~\cite{Green-1953,Volkov-1959,Volkov-1960,Greenberg&Messiah-1964,
Greenberg&Messiah-1965}, in which they play a role of (para)commutation
relations.  In a forthcoming paper, we intend to demonstrate how the parabose
commutation relations for free vector fields (satisfying the Lorenz
condition) can be obtained from~\eref{c.1}.




\addcontentsline{toc}{section}{References}
\bibliography{bozhopub,bozhoref}
\bibliographystyle{unsrt}
\addcontentsline{toc}{subsubsection}{This article ends at page}

\end{document}